\documentclass[11pt,a4paper]{article}
\pdfoutput=1
\usepackage[utf8]{inputenc}
\usepackage{jheppub}

\usepackage{lmodern}
\usepackage[T1]{fontenc}

\usepackage{amssymb}
\usepackage{amsmath}

\usepackage{graphicx}
\usepackage{epsfig}
\usepackage{float}

\usepackage{url}
\usepackage{color}
\usepackage{nicefrac}

\usepackage{comment}

\usepackage{physics}
\usepackage{caption}
\usepackage{subcaption}

\usepackage[normalem]{ulem} 

\newcommand{\refeq}[1]{Eq.~\eqref{#1}}

\newcommand{\mr}{\mathrm}

\newcommand{\beq}{\begin{equation}}
\newcommand{\eeq}{\end{equation}}
\newcommand{\bea}{\begin{eqnarray}}
\newcommand{\eea}{\end{eqnarray}}
\newcommand{\nn}{\nonumber}

\newcommand{\MCatNLO}{{\tt MC@NLO}}
\newcommand{\POWHEG}{{\tt POWHEG}}
\newcommand{\POWHEGBOX}{{\tt POWHEG~BOX}}
\newcommand{\BOX}{{\tt POWHEG~BOX}}

\newcommand{\POWHEGBOXR}{{\tt POWHEG~BOX~RES}}

\newcommand{\PYTHIA}{{\tt Pythia}}
\newcommand{\PYTHIAE}{{\tt Pythia8}}
\newcommand{\HERWIG}{{\tt Herwig}}
\newcommand{\HERWIGS}{{\tt Herwig7}}
\newcommand{\SHERPA}{{\tt Sherpa}}
\newcommand{\SHERPAT}{{\tt Sherpa2}}
\newcommand{\VINCIA}{{\sc Vincia}}

\newcommand\sss{\mathchoice%
{\displaystyle}%
{\scriptstyle}%
{\scriptscriptstyle}%
{\scriptscriptstyle}%
}
\newcommand{\mathd}{{\rm d}}

\newcommand{\as}{\alpha_s}

\newcommand{\GeV}{\;\mathrm{GeV}}

\newcommand{\MINLO}{\texttt{MINLO}}

\usepackage{xcolor}

\newcommand{\ydis}{y_{\sss \rm DIS}}
\newcommand{\xdis}{x_{\sss \rm B}}
\newcommand{\xB}{x_{\sss \rm B}}

\newcommand{\uk}{{\underline{k}}}

\newcommand{\xitilde}{\tilde{\xi}}
\newcommand{\ximax}{\xi_{\sss \rm max}}

\newcommand{\ck}{y_k}

\newcommand{\lint}{\mathcal{L}_\mr{int}}

\newcommand{\ptj}{p_T^\mathrm{jet}}
\newcommand{\etaj}{\eta^\mathrm{jet}}

\title{A POWHEG generator for deep inelastic scattering}

	\author[a]{Andrea Banfi,}
	\author[b]{Silvia Ferrario Ravasio,}
	\author[c]{Barbara J\"ager,}
	\author[b]{Alexander Karlberg,}
	\author[c]{Felix Reichenbach,}
	\author[d,e]{Giulia Zanderighi}

\affiliation[a]{Department of Physics and Astronomy, University of Sussex, Sussex House, Brighton, BN1 9RH,
UK}
\affiliation[b]{Theoretical Physics Department, CERN, CH-1211 Geneva 23, Switzerland}
\affiliation[c]{Institute for Theoretical Physics, University of T\"ubingen, Auf der Morgenstelle 14, 72076 T\"ubingen, Germany}

\affiliation[d]{Max-Planck-Institut f\"ur Physik, F\"ohringer Ring 6, 80805 M\"unchen, Germany}
\affiliation[e]{Physik-Department, Technische Universit\"at M\"unchen, James-Franck-Strasse 1, 85748 Garching,
Germany}

\emailAdd{a.banfi@sussex.ac.uk,
silvia.ferrario.ravasio@cern.ch,
jaeger@itp.uni-tuebingen.de,
alexander.karlberg@cern.ch,
felix.reichenbach@uni-tuebingen.de,
zanderi@mpp.mpg.de}
\preprint{CERN-TH-2023-152, MPP-2023-164}

\abstract{We present a new event generator for the simulation of both
  neutral- and charged-current deep inelastic scattering (DIS) at
  next-to-leading order in QCD matched to parton showers using the
  \POWHEG{} method. Our implementation builds on the existing
  \POWHEGBOX{} framework originally designed for hadron-hadron
  collisions, supplemented by considerable extensions to account for
  the genuinely different kinematics inherent to lepton-hadron
  collisions. In particular, we present new momentum mappings that
  conserve the special kinematics found in DIS, which we use to modify
  the \POWHEGBOX{} implementation of the Frixione-Kunszt-Signer
  subtraction mechanism. We compare our predictions to fixed-order and
  resummed predictions, as well as to data from the HERA $ep$
  collider. Finally we study a few representative distributions for
  the upcoming Electron Ion Collider.  }

\begin{document}

\maketitle
\flushbottom

\section{Introduction}
\label{sec:introduction}

Electron-proton ($ep$) colliders are powerful 
tools to perform
high-precision studies of quantum chromodynamics (QCD) and act as 
microscopes to probe the internal structure of the proton.
Particularly well suited to that end, are deep inelastic scattering
(DIS) processes where a photon or massive vector boson of high
virtuality is exchanged between the lepton and the partonic
constituents of the proton.
In fact, in such a reaction, the space-like vector boson exchanged in
the $t$-channel probes the charged constituents of the protons through
the electromagnetic and weak interaction in the cleanest possible
environment.
From the external momenta of the incoming and outgoing leptons ($p_l$
and $p_l'$) one can determine the internal hard space-like momentum
$q$ which probes the proton structure, $Q^2 = -q^2 = -(p_l-p_l')^2
>0$.

The Hadron Electron Ring Accelerator (HERA) at the Deutsches
Elektronen Synchrotron (DESY) was the first dedicated
high centre-of-mass energy $ep$ collider.
HERA operated in two phases -- HERA~I, from 1991 to 2000, and HERA~II
from 2002 to 2007, colliding protons up to energies of 920~GeV and
electrons (or positrons) at 27.5~GeV, spanning several orders of
magnitude in $Q^2$, thereby probing the proton structure at the
attometer level.
Besides measurements of exclusive reactions and diffraction, the main
legacy results from HERA collisions as measured by the H1 and ZEUS
collaborations include precise determinations of parton distribution
functions (PDFs) resulting in the HERAPDF
family~\cite{H1:2009pze,H1:2012xnw,H1:2015ubc,H1:2018flt}, a range of
precision QCD
studies~\cite{H1:1992fuc,ZEUS:1994mec,H1:1996naa,ZEUS:1997bxs,H1:1999yes,H1:2000muc,ZEUS:2002nms,ZEUS:2003xml,ZEUS:2005iex,H1:2009jxj}
and constraints on physics beyond the Standard
Model~\cite{H1:1993vsn,H1:1999dil,H1:2004rlm,ZEUS:1993vas}.
Proton PDFs from HERA played a crucial role for physics studies at the
Tevatron and at the Large Hadron Collider (LHC). In particular, the
fast discovery of the top quark at the Tevatron would not have been
possible without the knowledge of proton distribution functions
determined using data collected by the H1 and ZEUS collaborations.
HERA data are still included in global fits of PDFs, though more
recent PDF determinations rely more and more on LHC data (see
e.g.\ ref.~\cite{PDF4LHCWorkingGroup:2022cjn} and references
therein). This is particularly the case for the gluon distribution
function which is mostly probed indirectly at HERA, through the
precise measurement of the evolution of the quark distribution
functions via the
DGLAP~equations~\cite{Dokshitzer:1977sg,Gribov:1972ri,Altarelli:1977zs}.

In June 2021, the U.S.~Department of Energy has authorised the start
of the project execution phase of a new electron-ion collider (EIC),
with construction planned to start in 2024 at Brookhaven National
Laboratory (BNL).%
\footnote{See \href{https://www.energy.gov/science/articles/electron-ion-collider-achieves-critical-decision-1-approval}{https://www.energy.gov/science/articles/electron-ion-collider-achieves-critical-decision-1-approval}. } 
Other possible lepton-hadron colliders included in
the European Strategy for Particle Physics~\cite{CERN-ESU-015} are a
Large Hadron electron Collider (LHeC) at CERN and a Future Circular
electron-hadron Collider (FCC-eh).
These new-generation lepton-hadron colliders will enable
experimentalists to collect much higher luminosity compared to HERA,
and they will open up the possibility to explore an even larger range
in energy scales.

The EIC will collide 5 to 18~GeV electron beams with proton beams
spanning the energies from 41 to 275~GeV, with the possibility to have
both the electron and the proton beams polarised.
An electron-proton peak luminosity of $10^{34} {\rm cm}^{-2} {\rm
  s}^{-1}$ at 105 GeV centre-of-mass energy is foreseen.
Furthermore, a rich heavy ion program is planned, including the
possibility to have light polarised ions (such as $^3$He) with
energies up to 166~GeV and unpolarised heavy ions with energies up to
110~GeV.
For more details on the EIC, see for instance
refs.~\cite{AbdulKhalek:2021gbh,Bruning:2022hro,AbdulKhalek:2022hcn}.

From the theory side the last fifteen years, since the shutdown of
HERA, have seen considerable progress in the calculation of higher
order perturbative corrections (see,
e.g.~\cite{Heinrich:2020ybq,Gross:2022hyw} and references therein).
Although most of this progress has been in the context of automated
next-to-leading order (NLO) QCD corrections and
next-to-next-to-leading order (NNLO) corrections for two-to-two
scattering processes for hadron-hadron collisions, the DIS coefficient
functions have been computed through an impressive three loops in
QCD~\cite{Moch:2004xu,Vermaseren:2005qc,Moch:2008fj,Davies:2016ruz,Blumlein:2022gpp},
and using the projection-to-Born method~\cite{Cacciari:2015jma} fully
differential next-to-next-to-next-to-leading order (N$^3$LO)
single-jet distributions have been obtained by the NNLOJET
collaboration~\cite{Currie:2018fgr,Gehrmann:2018odt}. At fixed order
this makes DIS one of the best understood processes in QCD.

However, given that the LHC started operation around 2010, general
purpose Monte-Carlo generators have almost exclusively focused on
including higher-order corrections to hadron-hadron collisions, most
notably in the \POWHEG{}~\cite{Nason:2004rx,Frixione:2007vw} and
\MCatNLO{}~\cite{Frixione:2002ik} approaches, along with their
implementations in the \POWHEGBOX{}~\cite{Alioli:2010xd} and {\tt
  MadGraph5\_aMC@NLO}~\cite{Alwall:2014hca} frameworks.  In contrast
to the highly refined tools nowadays used per default at the LHC,
physics studies for the EIC widely rely on general-purpose event
generators that are only being adapted to the needs of an $ep$
collider.
These include the Monte-Carlo generators
\HERWIGS~\cite{Bahr:2008pv,Bellm:2019zci},
\SHERPAT~\cite{Gleisberg:2008ta,Sherpa:2019gpd}, and
\PYTHIAE~\cite{Sjostrand:2014zea,Bierlich:2022pfr}. 
Additionally, the EIC user community resorts to some generators for
more specialised issues such as the transverse-momentum dependence of
the proton or nuclear effects in collisions of electrons with a
heavy-ion beam, and on the generator {\tt
  DJANGOH}~\cite{Schuler:1991yg} that allows for a merging of QED and
QCD effects.\footnote{See, e.g.,
\href{https://eic.github.io/software/mcgen.html}{https://eic.github.io/software/mcgen.html}
for a compilation of software used by the EIC user community.}

Fixed-order programs widely used in the operation of HERA, such as
{\tt DISENT}~\cite{Catani:1996vz}, {\tt
  DISASTER++}~\cite{Graudenz:1997gv}, and {\tt
  NLOJET++}~\cite{Nagy:2001xb}, provide NLO accurate predictions for
neutral current and charged current processes with one or two jets in
the final state.
The {\tt DISResum} package, together with the {\tt Dispatch} package,
provides resummed predictions for certain event shapes at
next-to-leading logarithmic (NLL) accuracy matched to the fixed-order
programs above~\cite{Dasgupta:2002dc}. The automated NLL resummation
of event shapes in DIS can be obtained in the {\tt CAESAR}
framework~\cite{Banfi:2004yd} as was recently done for plain and
groomed 1-jettiness~\cite{Knobbe:2023ehi}.

While the internal matching functionalities of the multi-purpose
generators \HERWIG{} and \SHERPA{}~\cite{Carli:2010cg,Hoche:2018gti}
allow for DIS simulations at NLO+PS, neither the {\tt
  MadGraph5\_aMC@NLO} framework nor previous versions of the
\POWHEGBOX{} support the simulation of DIS.
The purpose of this paper is to present the first \emph{dedicated}
\POWHEG{} NLO+PS generator for DIS, and embed it in the \POWHEGBOX{}
framework. Concretely, the implementation provides results that can be
matched to a generic parton shower. This, in particluar, means that
NLO accurate events can be interfaced to \PYTHIAE{}, something which
has so far not been possible. Our code has been made publicly
available and can be downloaded following the instructions given in
the \POWHEGBOX{} webpage~\cite{POWHEGBOXWEB}.

The paper is organised as follows: 
In Sec.~\ref{sec:implementation}, we describe key changes required to
the \POWHEGBOXR{} to describe lepton-hadron collisions.
Section~\ref{sec:validation} is devoted to validation of our code and
comparisons with fixed-order results.
In Sec.~\ref{sec:pheno} we present sample phenomenological results at
HERA (Sec.~\ref{sec:hera}) and at the EIC (Sec.~\ref{sec:EIC}).
We present our summary and outlook in Sec.~\ref{sec:summary}.
Technical details regarding the phase-space parametrisation are
provided in App.~\ref{app:ps}, the generation of final- and
initial-state radiation in App.~\ref{app:genradfsr}
and~\ref{app:genradisrdis}, respectively, and the matching to the
\PYTHIA{} parton shower is described in detail in
App.~\ref{app:matching}.

\section{Details of the implementation}
\label{sec:implementation}

In this section, we provide a detailed description of the process
considered in this work and elaborate on the extensions made to the
\POWHEGBOXR{} framework for its implementation.
Specifically, we present comprehensive details regarding three key
aspects: the phase-space generation, the generation of radiation and
the treatment of real-radiation damping.

\subsection{The DIS process}

To set the stage it is useful to first recall the leading order (LO)
kinematics of DIS.
We consider the scattering of a massless (anti-)quark $q$ off a
massless \mbox{(anti-)lepton} $l$ via the exchange of a photon or
electroweak gauge boson $V$ of virtuality $Q^2$.
In our notation, the external four-momenta are given by $k_i$
(incoming lepton), $k_f$ (outgoing lepton), $p_i$ (incoming quark),
and $p_f$ (outgoing quark).

It is customary to define a set of DIS variables $\xdis$, $Q^2$, and
$\ydis$, given by
\begin{align}
  Q^2 = -q^2 = -(k_i - k_f)^2, \qquad  \xdis = \frac{Q^2}{2 P \cdot q},
  \qquad  \ydis = \frac{P \cdot q}{P \cdot k_i} = \frac{p_i \cdot q}{p_i \cdot k_i} ,
  \label{eq:dis-variables}
\end{align}
where $P$ is the proton four-momentum. At LO, neglecting the proton
mass, the Bjorken $\xdis$ variable coincides with the longitudinal
momentum fraction $x$ carried by the incoming quark, $p_i = x
P$. 
The LO
phase space is 
\begin{equation}
  \dd\Phi_2 =
\dd x \frac{\dd^4k_f}{(2\pi)^4}\frac{\dd^4p_f}{(2\pi)^4} (2\pi)\delta(k_f^2)(2\pi)\delta(p_f^2)(2\pi)^4\delta^4(k_i+p_i-k_f-p_f) = 
  \frac{\dd x \dd\ydis \dd\bar\phi}{16 \pi^2},
  \label{eq:dphi2}
\end{equation}
and the differential partonic cross section (for photon
exchange),\footnote{Our implementation includes also diagrams with $Z$
  exchange including the interference with the photon
  diagrams. Additionally the code can also handle the charged current
  process where a $W^+/W^-$ is exchanged.} after integrating over the azimuthal angle of the lepton, is given by 
\begin{equation}
  \frac{\dd^2\hat{\sigma}}{\dd\xdis \dd Q^2} = \frac{4 \pi
    \alpha^2}{Q^4}\left[1 + (1-\ydis)^2\right]\frac12 e_q^2 \delta(\xdis
  -x),
\end{equation}
where we have used that $Q^2 = \xdis \, \ydis \, S$, with $S = 2 P \cdot
k_i$ the total squared centre-of-mass energy.

At next-to-leading order (NLO) the process receives both virtual loop
corrections and real emission tree-level corrections. 
The full three-particle DIS phase
space $\dd\Phi_3$ for the real correction is given by 
\begin{align}
	\dd\Phi_3 = \dd x\, \dd\phi_3 = \dd x \frac{\dd^3 \vb k_f}{2k_f^0 (2\pi)^3} \frac{\dd^3 \vb p_f}{2p_f^0 (2\pi)^3} \frac{\dd^3 \vb p_r}{2p_r^0 (2\pi)^3} (2\pi)^4 \delta^{(4)}\qty(k_i + p_i - k_f - p_f - p_r),
	\label{eq:ps3-sec2}
\end{align}
where $x$ is the longitudinal momentum fraction of the incoming parton
and $\dd\phi_3$ is the Lorentz invariant three particle phase
space. As above, $k_{i/f}$ denote the incoming (outgoing) lepton,
$p_{i}$ the incoming parton, and $p_{f/r}$ denote now the two outgoing
QCD partons.

\subsection{Extension of the \POWHEGBOXR{}}

The \POWHEGBOX{} is a very powerful framework for matching fixed-order
NLO processes to parton shower Monte Carlos in hadron-hadron
collisions.
A large range of collider processes are implemented and have
been used in many LHC analyses.
Together with interfaces to NLO codes, the framework can in principle
be used to generate events for arbitrary hadron-collider processes.

However, in its original formulation, the \POWHEGBOX{} could not be
used to generate events for processes with lepton beams.\footnote{
The \POWHEGBOX{} can handle processes where leptons are treated as hadron constituents, 
see e.g.~\cite{Buonocore:2021bsf,Buonocore:2022msy}.} 
The \POWHEGBOXR{}~\cite{Jezo:2015aia} can, however, straightforwardly be
modified to handle processes where both incoming beams are leptons and there is no initial state radiation (as
done for example in Ref.~\cite{FerrarioRavasio:2018ubr}), as one
simply needs to replace the incoming beam PDFs with $\delta$--functions.
This approach does not work for DIS processes that involve initial state radiation (ISR), 
as the \POWHEG{} mappings for ISR would modify
the kinematics of both incoming beams, whereas, in the case of DIS, one needs to
keep the momentum of the incoming lepton fixed.
Moreover, although not necessary when performing a fixed-order
calculation, during the event generation (and the subsequent
parton-shower evolution), it is important to preserve the momentum
transfer between the incoming and outgoing leptons, to accurately
reproduce the NLO predictions for inclusive quantities.
In the following we give more technical details and we better motivate
the importance of preserving the DIS invariants at the stage of event
generation.

\subsubsection{POWHEG ingredients}
\label{sec:pwgingredients}

Before describing the modifications we made to handle DIS, we briefly
summarise the main ingredients of the \POWHEG{}
method~\cite{Nason:2004rx} as implemented in the
\POWHEGBOX{}~\cite{Frixione:2007vw}.

A building block of the \POWHEG{} cross section is the inclusive NLO cross section
\begin{align}
  \frac{\mathd \sigma_{\sss \rm NLO}}{\mathd \bar{\Phi}_n} =& \sum_{f_b} \Big[ B_{f_b}(\bar{\Phi}_b) + V_{f_b}(\bar{\Phi}_b) 
  + \sum_{f_r} \sum_{\alpha \in f_r \to f_b} \int \mathd\Phi_{\sss \rm rad}^{\alpha} R_{\alpha}(\Phi_{n+1}(\bar{\Phi}_n, \Phi_{\sss \rm rad}))\Big],
\end{align}
where $\bar{\Phi}_n$ denotes the phase space of the underlying Born
configuration, $f_b$ labels the partonic subprocess contributing at LO,
and $f_r$ is summed over the partonic subprocesses entering the real
contribution. $B_{f_b}$ corresponds to the Born matrix element
(including luminosity and flux factors), $V_{f_b}$ corresponds to the
UV-renormalised virtual corrections, and $R_{f_r}$ is the real matrix
element.  The real cross section is partitioned in several
contributions, labelled with the index $\alpha$, each of them
associated with a singular region. The notation ``$\alpha\in f_r\to
f_b$'' means that all the singular regions leading to the underlying
Born subprocess $f_b$ are considered.  This writing assumes that the
phase space for the real contribution can be written in a factorised
form
\begin{equation}
\mathd\Phi_{n+1}^{\alpha} = \mathd \bar{\Phi}_n \mathd \Phi_{\sss \rm rad}^{\alpha}.
 \end{equation}
The radiation phase space $\Phi_{\sss \rm rad}^{\alpha}$ is
parameterised  in terms of three variables, an energy fraction $\xi$,
the cosine of the angle between two partons that can become collinear $y$,
and an azimuthal angle $\phi$, according to the
Frixione-Kunszt-Signer~(FKS)~\cite{Frixione:1995ms} subtraction technique.
The exact expression of $\Phi_{\sss \rm rad}^{\alpha}$ depends on the
singular region.  In the DIS case, there are two singular regions, one
associated with initial-state radiation, one with final-state
radiation.

The \POWHEG{} cross section reads
\begin{align}
  \label{eq:sigmaPWGnaive}
\mathd\sigma_{\sss \rm PWG} =& \sum_{f_b} \bar{B}_{f_b}(\bar{\Phi}_n) \mathd \bar{\Phi}_n \Bigg[ \prod_{\alpha \in
    f_r \to f_b} \Delta^{f_b}_{\alpha}(\bar{\Phi}_n, \mu_0)\\
  &+ \sum_{\alpha \in
    f_r \to f_b} \mathd\Phi_{\sss \rm rad}^{\alpha}  \Theta(\kappa_{t}^{\alpha}(\Phi_{\sss \rm rad}^{\alpha})>\mu_0) \frac{R_{\alpha}(\Phi_{n+1}(\bar{\Phi}_n, \Phi_{\sss \rm rad}^{\alpha}))}{{B}_{f_b}(\bar{\Phi}_n)}\Delta^{f_b}_{\alpha}(\bar{\Phi}_n, \kappa_{t}^{\alpha}(\Phi_{\sss \rm rad}^{\alpha}))
  \Bigg],
\nonumber
\end{align}
where
\begin{equation}
\bar{B}_{f_b}(\bar{\Phi}_n)=B_{f_b}(\bar{\Phi}_b) + V_{f_b}(\bar{\Phi}_b) 
+ \sum_{f_r} \sum_{\alpha \in f_r \to f_b} \int \mathd\Phi_{\sss \rm rad}^{\alpha} R_{\alpha}(\Phi_{n+1}(\bar{\Phi}_n, \Phi_{\sss \rm rad})),
\label{eq:BTilde}
  \end{equation}
$\kappa_{t}^{\alpha}(\Phi_{\sss \rm rad}^{\alpha})$ is a quantity used
to measure the hardness of an emission, that depends on the radiation
variables $\xi$ and $y$, and becomes equal to the transverse momentum
of the emission in the soft-collinear limit, $\mu_0$ is an infrared scale
of the order of $1$ GeV, below which real radiation is considered
unresolved, and
\begin{equation}
  \Delta^{f_b}_{\alpha}(\bar{\Phi}_n, k_{\sss \rm T}) =\exp\left(- \int \mathd\Phi_{\sss \rm rad}^{\alpha}  \Theta(\kappa_{t}^{\alpha}(\Phi_{\sss \rm rad}^{\alpha})>k_{\sss \rm T}) \frac{R_{\alpha}(\Phi_{n+1}(\bar{\Phi}_n, \Phi_{\sss \rm rad}^{\alpha}))}{{B}_{f_b}(\bar{\Phi}_n)}\right)\, 
  \label{eq:SudakovPWG}
\end{equation}
is the Sudakov form factor.
After integrating over the radiation
phase space, the squared bracket appearing in
Eq.~\eqref{eq:sigmaPWGnaive} yields 1. 
For this reason, preserving the DIS invariants when building the radiation phase  
space ensures that one exactly reproduces the NLO distributions for $x_{\rm \sss DIS}$,
$y_{\rm \sss DIS}$ and $Q^2_{\rm \sss DIS}$, and does not introduce spurious higher-order corrections in inclusive quantities. 
In Secs.~\ref{sec:phspISR} and~\ref{sec:phspFSR} we present new
parametrisations of the radiation phase space, for ISR and final state
radiation (FSR) respectively, that enable one to preserve the DIS
invariants.
We also need to modify our definition of the hardness variable
$k_{\sss \rm T}^{\alpha}(\xi,y)$, as detailed in
Sec.~\ref{sec:genRad}.

One of the features 
of Eq.~\eqref{eq:sigmaPWGnaive}, is that it
can significantly depart from the fixed-order NLO calculation when
considering non-inclusive observables (i.e.\ observables that are
vanishing at LO) even in the limit in which the radiation is very
hard.
This is due to the ratio $\bar{B}/B$, and to higher-order effects
encoded in the Sudakov form factor of Eq.~\eqref{eq:sigmaPWGnaive} (e.g.\ related to the
treatment of the QCD coupling constant, which is modified to include
the dominant logarithmically-enhanced corrections at all
orders~\cite{Catani:1990rr}).  
To remedy this, one can introduce a monotonic function $h(k_{\sss \rm T})$, such that
\begin{equation}
  \lim_{k_{\sss \rm T} \to 0} h(k_{\sss \rm T}) = 1, \qquad \lim_{k_{\sss \rm T} \to \infty} h(k_{\sss \rm T}) = 0,
  \label{eq:hdamp}
\end{equation}
and separate the real cross section into a singular ($s$) and a finite ($f$) contribution,
\begin{align}
  R_{\alpha}^{(s)}(\Phi_{n+1}) =  h(k_{\sss \rm T}) \times R_{\alpha}(\Phi_{n+1}), \qquad
  R_{\alpha}^{(f)}(\Phi_{n+1}) = (1- h(k_{\sss \rm T})) \times R_{\alpha}(\Phi_{n+1}). 
\end{align}
One can then use $R_{\alpha}^{(s)}$ instead of $R_{\alpha}$ in the definition
of $\bar{B}_{f_b}$ in Eq.~\eqref{eq:BTilde}, of the Sudakov form factor $
\Delta^{f_b}_{\alpha}$ of Eq.~\eqref{eq:SudakovPWG} and in the \POWHEG{} cross
section $d\sigma_{\sss \rm PWG}$ in Eq.~\eqref{eq:sigmaPWGnaive}.
One then also needs to add a ``remnant'' contribution to  $\mathd \sigma_{\sss \rm PWG}$:
\begin{align}
\mathd\sigma_{\sss \rm PWG}^{\sss \rm remn} = \sum_{f_r}  \sum_{\alpha \in f_r} d\Phi_{n+1} R_{\alpha}^{(f)}(\Phi_{n+1}).
\end{align}
In the \BOX{}, this procedure is dubbed the \texttt{hdamp}
mechanism. 
In the \BOX{}, it is also possible to use the
\texttt{Bornzerodamp} mechanism, which moves to $R_{\alpha}^{(f)}$ all
the configurations where the real matrix element
departs significantly from its soft or collinear
approximation.\footnote{Practically, the code checks if the real
  matrix element is 5 times bigger or has a different sign than its
  soft or collinear approximation.}
The impact of the damping functions is discussed in App.~\ref{sec:damp}.
If regular contributions (i.e.\ those not associated with any singularity)
are present, those are also treated alongside the remnant
contributions.

\subsubsection{Phase-space parameterisation for initial-state radiation}
\label{sec:phspISR}
In order to evaluate the phase-space of Eq.~(\ref{eq:ps3-sec2}) for
the case of ISR, we write the centre-of-mass momenta in the final state as 
\begin{align}
  p_r &= \xi \frac{\sqrt{s}}{2}\qty(1,\sqrt{1-y^2} \cos\phi, \sqrt{1-y^2} \sin\phi, y)\,, \\
  k_f &= \xi_k \frac{\sqrt{s}}{2}\qty(1,\sqrt{1-y_k^2} \cos\phi_k, \sqrt{1-y_k^2} \sin\phi_k, y_k)\,,  
\label{eq:prkf-sec2}
\end{align}
where $\xi$, $y$ and $\phi$ are the FKS variables that are used to
parametrise the real-radiation phase space.  After some algebra one
may express the three-particle phase space, $\dd\Phi_3$, of
Eq.~\eqref{eq:ps3-sec2} in terms of the two-particle phase space in
Eq.~\eqref{eq:dphi2} as follows:
\begin{align}
	\dd \Phi_3 &= \frac{1}{32 \pi^3} \dd \Phi_2 \; \dd \lambda \; \dd \xi \; \dd \phi \; \dd y \; 
	\left[\delta(\lambda - \lambda_+) + \delta(\lambda-\lambda_-)\right] \notag \\
	&\qquad \times  \frac{\bar s\ydis \xi}{\lambda \left| \lambda  \ydis (\xi(1+y)-2)-\xi  \cos \left(\Delta \phi\right) \sqrt{\lambda  \left(1-y^2\right) (1-\ydis) \ydis}\right|}\,, 
        \label{eq:isr-ps3}
\end{align}
where $\Delta\phi = \phi-\phi_k$ and, as in Ref.~\cite{Frixione:2007vw}, we
use the bar to indicate underlying Born quantities, like the squared Born 
centre-of-mass energy $\bar{s} =\xB S$. 
The two $\delta$-functions arise due to energy conservation, which
gives rise to a quadratic equation in $\lambda=\bar x/x$. The two solutions are
given by
\begin{align}
  \lambda_\pm &= \frac{\pm 2 \xi  \cos \Delta \phi \sqrt{A}+\xi ^2 \left(1-y^2\right) (1-\ydis) \cos \left(2 \Delta \phi\right)+2 (1-\xi ) (2 \ydis -\xi(1+y))}{\ydis (\xi(1+y)-2)^2},
  \label{eq:lambdasol-sec2}
\end{align}
and the argument of the root, $A$, is given in Eq.~\eqref{eq:A}. As
discussed in App.~\ref{app:ps}, in the soft and collinear regions only
$\lambda_-$ is a valid solution.

The form of Eq.~\eqref{eq:isr-ps3} is not yet suitable for numerical
implementation in the \BOX{} due to the presence of the
$\delta$-functions and the additional associated integration over
$\lambda$. Schematically, the $\xi$ and $\lambda$ integrations of a generic function $f(\lambda, \xi,y,\phi)$ can
then be written as
\begin{align}
  \int \dd \lambda \dd \xi f(\lambda, \xi,y,\phi) \qty(\delta(\lambda - \lambda_+) + \delta(\lambda-\lambda_-)) &
	= \int_0^{\ximax} \dd \xi f_-(\xi) + \int_{\xi_0}^{\ximax} \dd \xi f_+(\xi),
\end{align}
where $f_\pm(\xi)=f(\lambda_\pm, \xi,y,\phi)$ and 
the limits in the $\xi$ integrations are set by requiring that
the $\lambda_\pm$ solutions are physical.  The explicit expressions for
$\xi_0$ and $\ximax$ are given in App.~\ref{app:ps}. As shown in that
appendix the integral in the above equation can then be written as
\begin{align}
& \int_0^{\ximax} \dd \xi f_-(\xi) + \int_{\xi_0}^{\ximax} \dd \xi f_+(\xi) \nn\\
  &\qquad = \int_0^{\ximax'} \dd \xi \qty(f_-(\xi) \Theta\qty(\ximax-\xi) + f_+(2\ximax-\xi) \Theta\qty(\xi-\ximax)),
\end{align}
with $\ximax' = 2 \ximax - \xi_0$. Lastly, one can make the
transformation to $\xitilde=\xi/\ximax'$, 
to obtain 
\begin{align} 
\int_0^{\ximax} \dd \xi f_-(\xi) + \int_{\xi_0}^{\ximax} \dd \xi f_+(\xi)
  = \int_0^1 \dd & \xitilde \ximax' \left(f_-(\xitilde \ximax') \Theta(\ximax-\xitilde\ximax') \right. \nonumber \\ 
  &  \left. + f_+(\ximax'(1-\xitilde)+\xi_0) \Theta(\xitilde \ximax' - \ximax)\right)\,.
    \label{eq:isr-ps3-final}
\end{align}
Since the $\lambda$ integration has been eliminated and the $\xi$
integral has been remapped into a single integral between 0 and 1, one
can evaluate the radiation phase space as usual in the \BOX{}.

\subsubsection{Phase-space parameterisation for final-state radiation}
\label{sec:phspFSR}
The starting point for the FSR phase-space derivation is the same as
the one given in Eq.~\eqref{eq:ps3-sec2}. We first introduce the
momentum sum $k = p_f + p_r$ of the two outgoing QCD partons. The
radiation variables are then given as in the \BOX{}, i.e.\
\begin{align}
  \xi = \frac{2 p_r^0}{\sqrt{s}}, \qquad y = \frac{\vec{p}_r \cdot \vec{k}_f}{p_r^0 k_f^0}, \qquad \phi = \phi\left( \vec{\eta} \times \vec{k},\vec{p}_r \times \vec{k}\right),
\end{align}
where $\vec{\eta}$ is an arbitrary direction that serves to define the
origin of the azimuthal angle. 

In this case, after some algebra, one arrives at an
expression in terms of the Born phase space and the FKS radiation
variables that can be integrated numerically, given by
\begin{align}
  \int \dd \Phi_3 &= \frac{1}{16 \pi^3} \int \dd \Phi_2 \;  \dd \xi \; \dd y \; \dd \phi \frac{(1-\xi ) \xi  \bar s}{\lambda _0^2 (2-\xi  (1-y)) (2-(2-\xi ) \xi  (1-y))}\,,
  \label{eq:fsr-ps3-final}
\end{align}
where the value of $\lambda_0$ can be found in Eq.~\eqref{eq:lambda0}.

\subsubsection{Generation of radiation}
\label{sec:genRad}
The standard \BOX{} generation of FSR radiation is discussed in
detail in Ref.~\cite{Frixione:2007vw} and recalled explicitly in
App.~\ref{app:genradfsrstandard}. In the case of DIS, since the energy
of the incoming lepton is fixed, the energy of the incoming parton is
reduced even in the case of FSR, by an amount equal to
\begin{equation}
\lambda = \frac{\bar x}{x} =1-  \frac{\xi  (1-\xi)(1-y)}{y_{\text{DIS}} (2-\xi  (1-y))}. 
\end{equation}
Since in the soft or collinear limits $\lambda \to 1$, and hence
$s\approx\bar{s}$, one can use as ordering variable
\begin{equation}
\kappa_{t}^2 = \frac{\bar s}{2}\xi^2 (1-y), 
\end{equation}
which now involves explicitly the underlying Born centre-of-mass
energy. The upper bound for $\kappa_{t}^2$ is in this case simply
$\bar s$. One can then generate a radiation phase-space point in the
usual way in the \BOX{}, and accept or reject it using the
standard hit-and-miss technique with an upper bound of the form (see
App.~C of Ref.~\cite{Alioli:2010xd}) 
\begin{equation}
  U(\xi, y) d \xi d y \propto\frac{\alpha_s(\kappa_t^2)}{\xi (1-y)} d \xi d y\,.
  \label{eq:upperbound}
\end{equation}

In the case of ISR, the standard \POWHEG{} code in the default setup
handles the two collinear regions along the beam together. In our case
instead we only have one collinear region. For example, if the collinear region is
for $y\to +1$, our upper-bound is identical to the one in Eq.~\eqref{eq:upperbound}. 
Furthermore one can use as ordering variable
\begin{equation}
\kappa_t^2 = \frac{\xi^2}{2-\xi(1+y)} \bar{s}(1-y), 
\end{equation}
which, as shown in the App.~\ref{app:genradisrdis}, is also bounded
from above by $\kappa_t^2 < \bar s$.  In the soft ($\xi \to 0$) or
collinear limit ($y \to 1$) is it easy to verify that $\kappa_t^2 \to
\frac{s}{2} \xi^2(1-y)$, i.e.\ it corresponds to the transverse momentum
of the emission.

\section{Code validation and comparison to existing predictions}
\label{sec:validation}

In the following we present a validation of our code and 
a comparison to other existing theory predictions, both for inclusive
observables, as well as for observables related to the jet kinematics.
All the \POWHEGBOX{} results presented in this section have been obtained using
the \texttt{Bornzerodamp} mechanism, described in
Sec.~\ref{sec:pwgingredients}, to separate the singular and
non-singular contributions in the real cross section.
Alternative choices of the damping functions are discussed in App.~\ref{sec:damp}.

\subsection{Inclusive observables}
One of the defining features of an NLO+PS generator is that it should
reproduce quantities that are inclusive in radiation not present at
Born level, with NLO accuracy~\cite{Nason:2012pr}. In DIS one
typically decomposes the inclusive cross section in terms of the three
proton structure functions $F_1$ (or $F_L$), $F_2$, and
$F_3$~\cite{ParticleDataGroup:2022pth},
\begin{align}
  \frac{\dd^2\sigma}{\dd\xdis\, dQ^2} = \frac{4\pi\alpha^2}{\xdis Q^4}\left[\xdis\ydis^2 F_1 + (1-\ydis)F_2 +\xdis\ydis(1-\frac12 \ydis)F_3\right]\,,
  \label{eq:sigmaF1}
\end{align}
where $\xdis$, $\ydis$, and $Q^2$ are the usual DIS variables as defined
in Eq.~\eqref{eq:dis-variables}, and $\alpha$ is the electromagnetic
fine-structure constant. The structure functions themselves depend on
both $\xdis$ and $Q^2$. At leading order $F_2 = 2 \xdis F_1$ and $F_3$
only receives contributions from diagrams with a $Z$ boson. The above
equation is therefore often recast using $F_L = F_2 - 2\xdis F_1$ instead, 
such that it reads
\begin{align}
  \frac{\dd^2\sigma}{\dd\xdis\, dQ^2} = \frac{4\pi\alpha^2}{\xdis Q^4}\left[\frac{1}{2}(1+(1- \ydis)^2)F_2 -\frac{1}{2}\ydis^2 F_L +\xdis\ydis(1-\frac12 \ydis)F_3\right]\,,
  \label{eq:sigmaFL}
\end{align}
where we have suppressed again the arguments of the structure
functions. 
One then defines the (dimensionless) reduced cross section by \cite{H1:2012qti}
\begin{align}
  \sigma_R(x,Q^2) = \frac{\xdis Q^4}{2\pi\alpha^2(1+(1- \ydis)^2)} \frac{\dd^2\sigma}{\dd\xdis\, \dd Q^2} \,, 
  \label{eq:sigmaR}
\end{align}
which has the property that at leading order it is equal to
$F_2$ when considering only photon exchange.  
In this validation section, we use the reduced cross section
to investigate inclusive predictions of the new \POWHEGBOX{}
generator.

For the numerical results presented here we will consider
positron and proton collisions at energies of $27.6$~GeV and
$920$~GeV, respectively.
We will use the NNLO~PDF set {\tt
  NNPDF30\_nnlo\_as\_0118\_hera}~\cite{NNPDF:2014otw} based on HERA
data with the associated strong coupling $\alpha_s(M_Z)=0.118$ as
implemented in {\tt LHAPDF} v6.5.3~\cite{Buckley:2014ana}. We set the
central renormalisation, $\mu_R$, and factorisation, $\mu_F$, scales
equal to $Q$, and to estimate the perturbative uncertainty we do a
standard 7-point scale variation by a factor of two around these
values. The number of active flavours is set to $N_f=5$. 

\begin{figure}[tb!]
  \centering\includegraphics[width=0.92\textwidth,page=1]{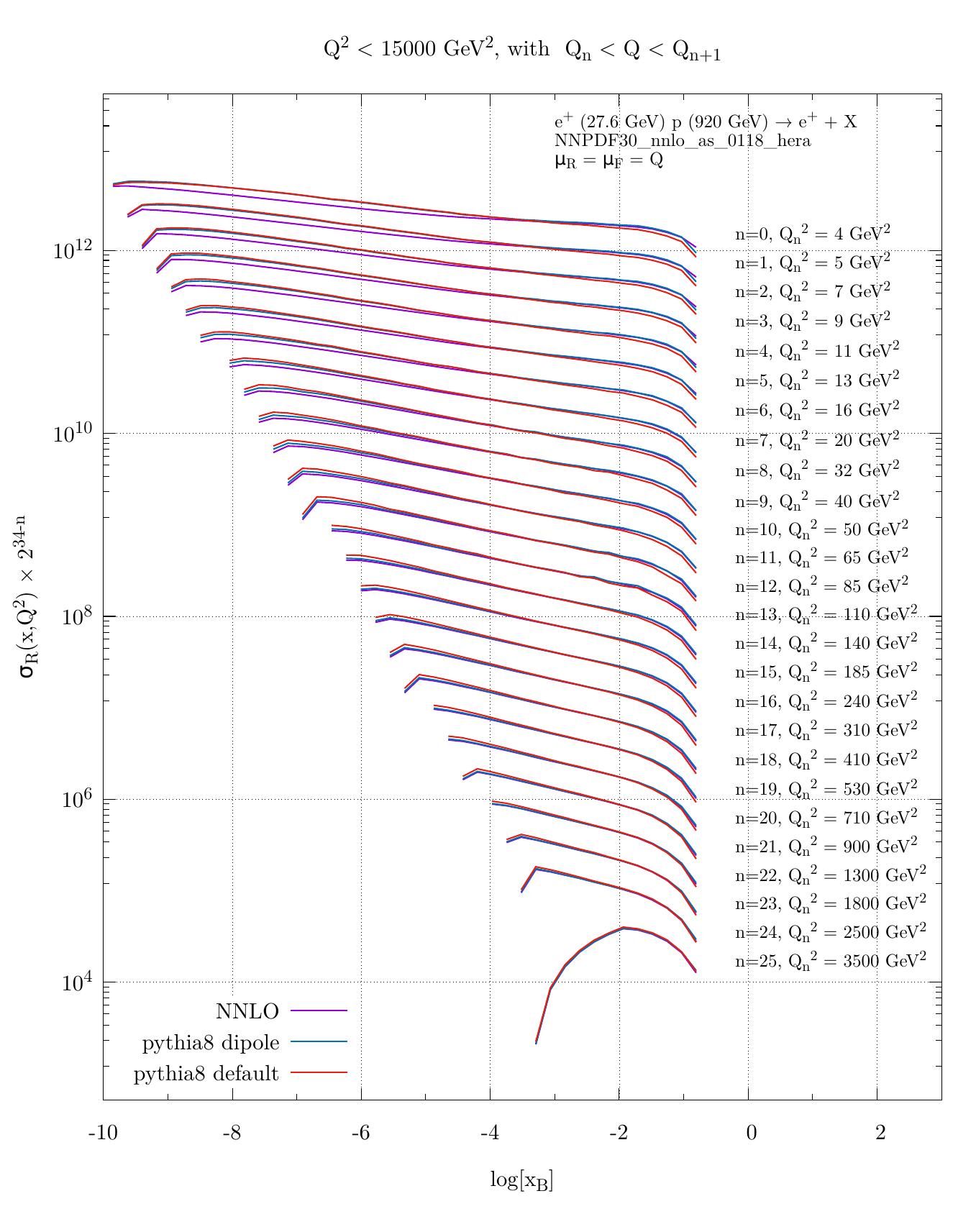}
  \caption{The reduced double differential cross section, defined in
    Eq.~\eqref{eq:sigmaR}, binned in $Q^2$ and plotted as a function
    of $\log{\xdis}$. For a given $n$ the lower bin-edge in $Q^2$ is given
    by the printed $Q_n^2$ and the upper limit is $Q_{n+1}^2$. For the
    last bin with $n=25$ the upper edge is given by $Q^2 =
    15000$~GeV$^2$. Note that in order to plot all curves in the same
    panel they have been multiplied by $2^{25-n}$. Plotted are
    fixed-order NNLO results (dark purple), \POWHEG{} events
    showered with the dipole \PYTHIAE{} shower (blue), and \POWHEG{} 
    events showered with the default \PYTHIAE{} shower (red), both at parton level.}
  \label{fig:sigmaRinlogx}
\end{figure}
\begin{figure}[tb!]
  \centering\includegraphics[width=0.92\textwidth,page=2]{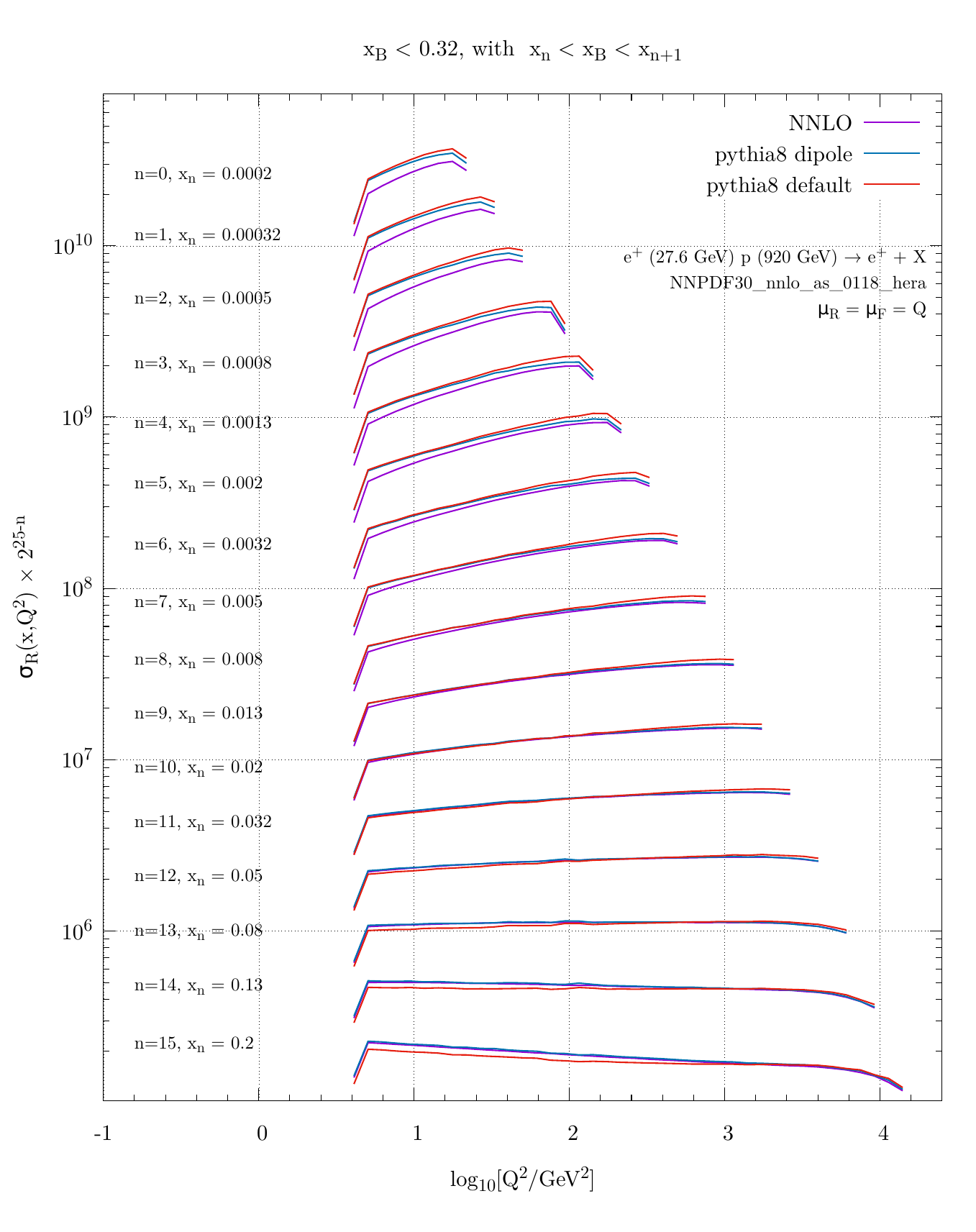}
  \caption{The reduced double differential cross section, defined in
    Eq.~\eqref{eq:sigmaR}, binned in $\xdis$ and plotted as a function of
    $\log{Q^2}$. For a given $n$ the lower bin-edge in $\xdis$ is given by
    the printed $x_n$ and the upper limit is $x_{n+1}$. For the last
    bin with $n=16$ the upper edge is given by $\xdis = 0.5$. Note that in
    order to plot all curves in the same panel they have been
    multiplied by $2^{16-n}$. Plotted are fixed-order NNLO results (dark
    purple), \POWHEG{} events showered with the dipole \PYTHIAE{}
    shower (blue), and \POWHEG{} events showered with the default
    \PYTHIAE{} shower (red), both at parton level.}
  \label{fig:sigmaRinlogQ}
\end{figure}

In Figs.~\ref{fig:sigmaRinlogx} and \ref{fig:sigmaRinlogQ} we show the
reduced double differential cross section, as defined in
Eq.~\eqref{eq:sigmaR}.  The results are either binned in $Q^2$ and
plotted in $\log{\xdis}$, or binned in $\xdis$ and plotted in $\log{Q^2}$, where $Q$ is measured in GeV.
In particular, we compare our NLO+PS result to an NNLO calculation
obtained with {\tt disorder}~\cite{disorder} that uses the DIS
structure function branch of {\tt HOPPET}~\cite{Salam:2008qg} which
was developed in the context of the {\tt proVBFH}
programs~\cite{Cacciari:2015jma,Dreyer:2016oyx,Dreyer:2018qbw,Dreyer:2018rfu}
and uses the NNLO DIS coefficient functions computed in
Refs.~\cite{vanNeerven:1999ca,vanNeerven:2000uj}. We obtain NLO+PS
results using two different versions of the
\PYTHIAE{}\footnote{Specifically we run version 8.308.}
showers~\cite{Sjostrand:2014zea}, namely the default \PYTHIAE{} and
the dipole like \PYTHIAE{} shower introduced in
Ref.~\cite{Cabouat:2017rzi}. For the purposes of the comparison in
this section, the main difference between the two is that the default
shower does not preserve the lepton kinematics, whereas the dipole
variant does. In these plots we only run the shower phase of
\PYTHIAE{}, i.e.\ we do not include hadronisation and underlying event
simulation.
More details on the matching procedure with \PYTHIAE{} are given in
App.~\ref{app:matching}.
As explained in Sec.~\ref{sec:implementation}, our \POWHEGBOX{}
implementation is constructed such that the radiation mappings
preserve the underlying DIS kinematics, i.e.\ the lepton
kinematics. For this reason the reduced cross~sections obtained at
pure NLO and at the level of the unshowered Les Houches Event
(LHE)~\cite{Alioli:2013nda} file are in perfect agreement,
i.e.\ there are no spurious higher-order terms induced by the
\POWHEG{} Sudakov for this observable.\footnote{One finds only very
small discrepancies for very large values of $\xdis$. These are due to the
fact that in the \POWHEGBOX{} the weight of events with negative PDF
values is set to zero.  We have checked that the discrepancy reduces
when one uses a PDF set exhibiting fewer negative values. Small
differences between the LHE and the NLO distributions also arise in
the very small $Q$ region due to the momentum reshuffling procedure to
introduce mass effects for final-state particles.}
We note that this would not have been the case if
our mappings did not preserve the DIS kinematics.
Furthermore, with a parton shower which preserves the DIS invariants,
such as the dipole \PYTHIAE{} shower~\cite{Cabouat:2017rzi}, the
reduced cross-sections after parton shower are also identical. For
this reason in Figs.~\ref{fig:sigmaRinlogx} and \ref{fig:sigmaRinlogQ}
we do not show explicitly the NLO and LHE curves, as they are almost
identical to the dipole (\POWHEG +)\PYTHIAE{} shower results.
On the other hand, the default \PYTHIAE{} shower does not preserve the
DIS kinematics, and therefore one can expect modifications from the
shower to the reduced cross sections, as is evident from
Figs.~\ref{fig:sigmaRinlogx} and \ref{fig:sigmaRinlogQ}, even though
the hardest emission event generated by \POWHEG{} does preserve the DIS
kinematics.  In particular,
there are significant deviations for larger values of $\xdis$ together
with small to moderate values of $Q^2$. It is interesting to note that
in this kinematic region the NNLO prediction is very close to the
dipole \PYTHIAE{} prediction (i.e.\ it is very close to the NLO
result), and hence true NNLO corrections are tiny in these
regions. The discrepancies between the two \PYTHIAE{} showers can
therefore be seen as spurious effects. This was already
pointed out in Ref.~\cite{Cabouat:2017rzi} where it was found that the
dipole \PYTHIAE{} shower correctly reproduces the singular limits of
LO DIS matrix elements, whereas the default \PYTHIAE{} shower does
not.
\begin{figure}[tb!]
  \centering\includegraphics[width=0.49\textwidth,page=1]{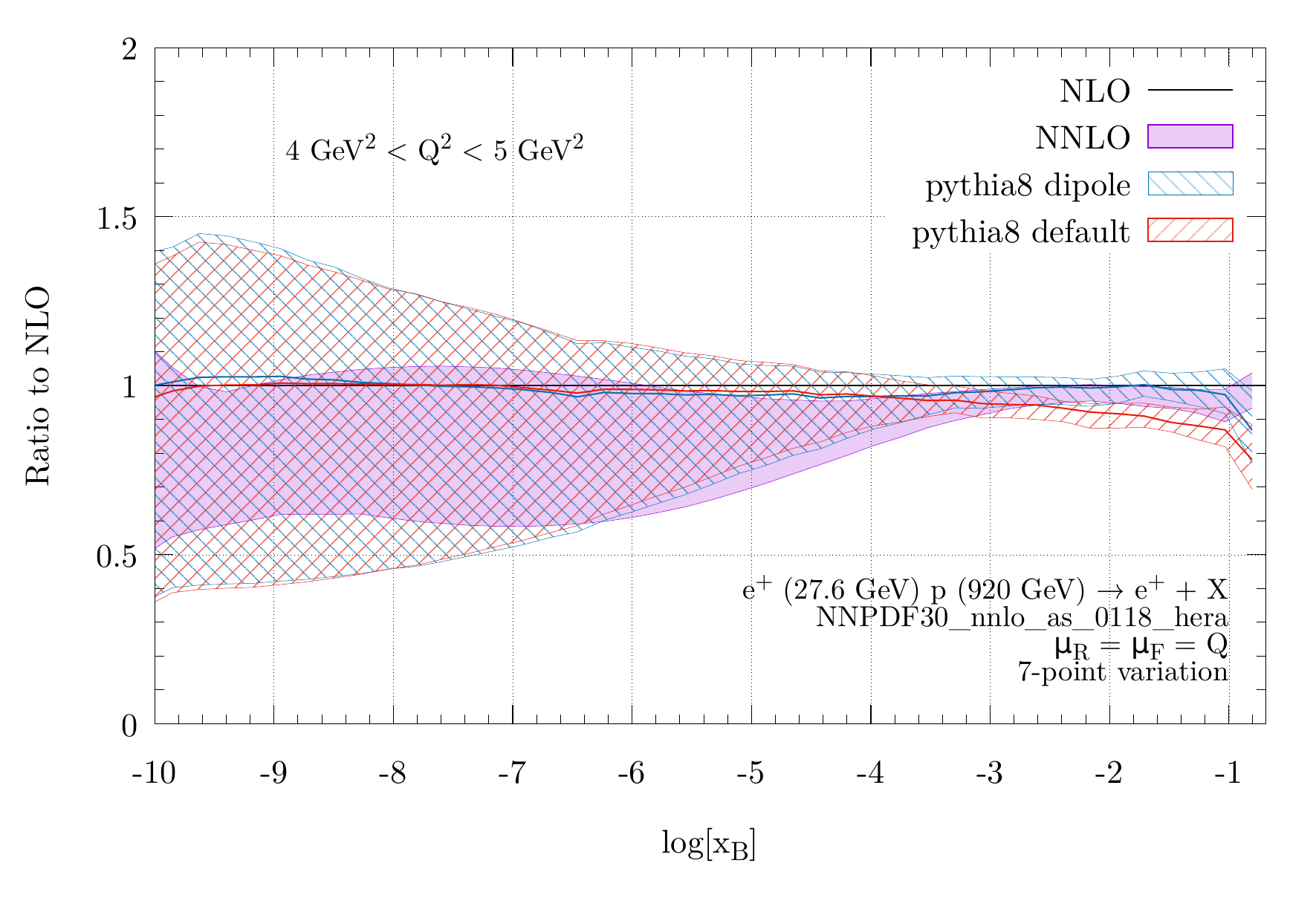}
  \centering\includegraphics[width=0.49\textwidth,page=27]{fig-temp/sigmaR-ratios.pdf}
  \caption{Ratios for the reduced cross section in
    Eq.~\eqref{eq:sigmaR} of NNLO (purple), \PYTHIAE{} dipole (blue),
    and \PYTHIAE{} default (red) both at parton level to NLO (black) for the bin
    $4\,\mathrm{GeV}^2 < Q^2 < 5\,\mathrm{GeV}^2$ as a function of
    $\log{\xdis}$ (left) and the same ratios for the bin
    $0.0002<\xdis<0.00032$ as a function of $\log_{10}{Q^2}$ (right). The
    bands represent the 7-point scale variation of $\mu_R$ and $\mu_F$
    by a factor of two around the central value $Q$.}
  \label{fig:sigmaRratio1}
\end{figure}
\begin{figure}[tb!]
  \centering\includegraphics[width=0.49\textwidth,page=13]{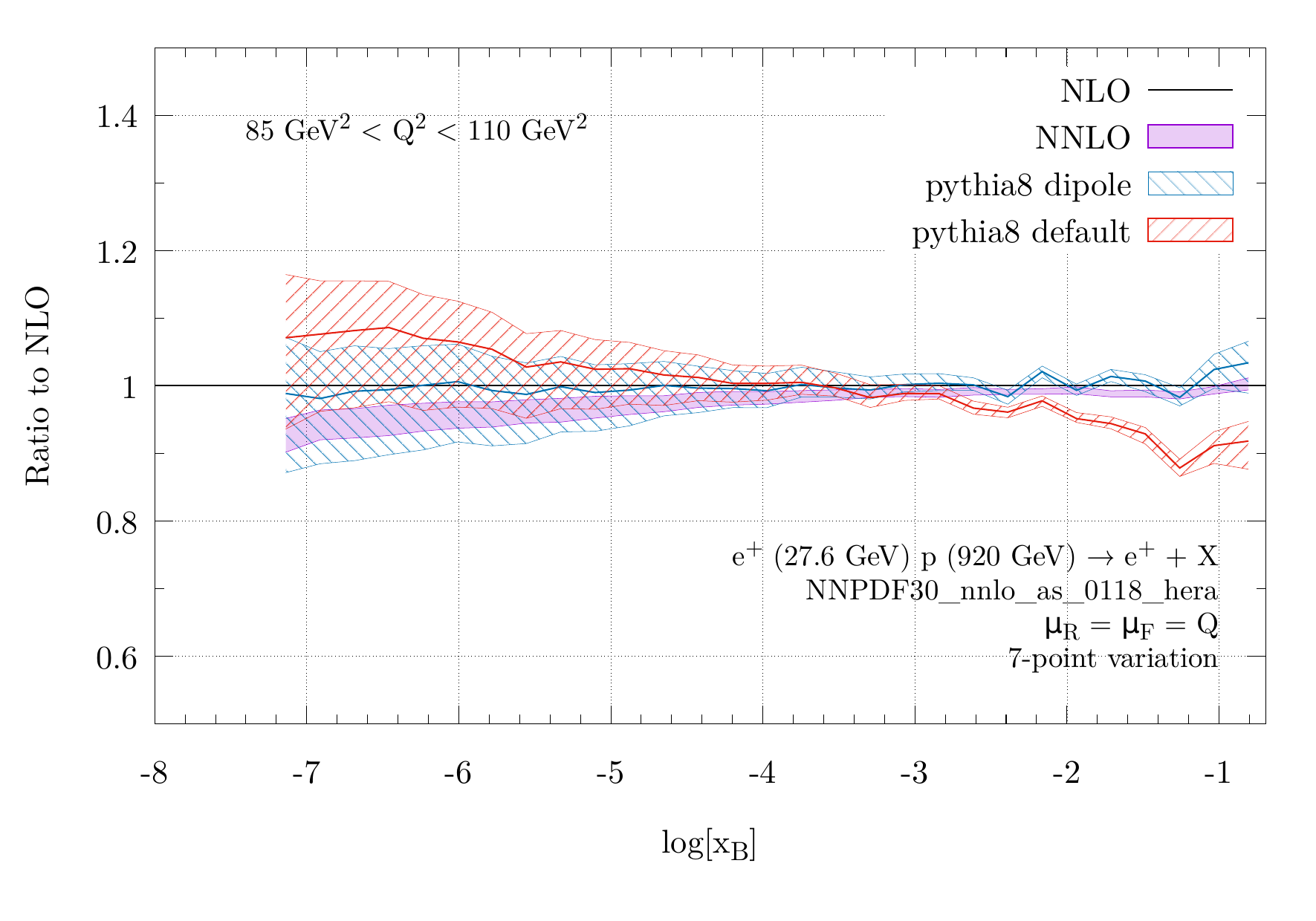}
  \centering\includegraphics[width=0.49\textwidth,page=33]{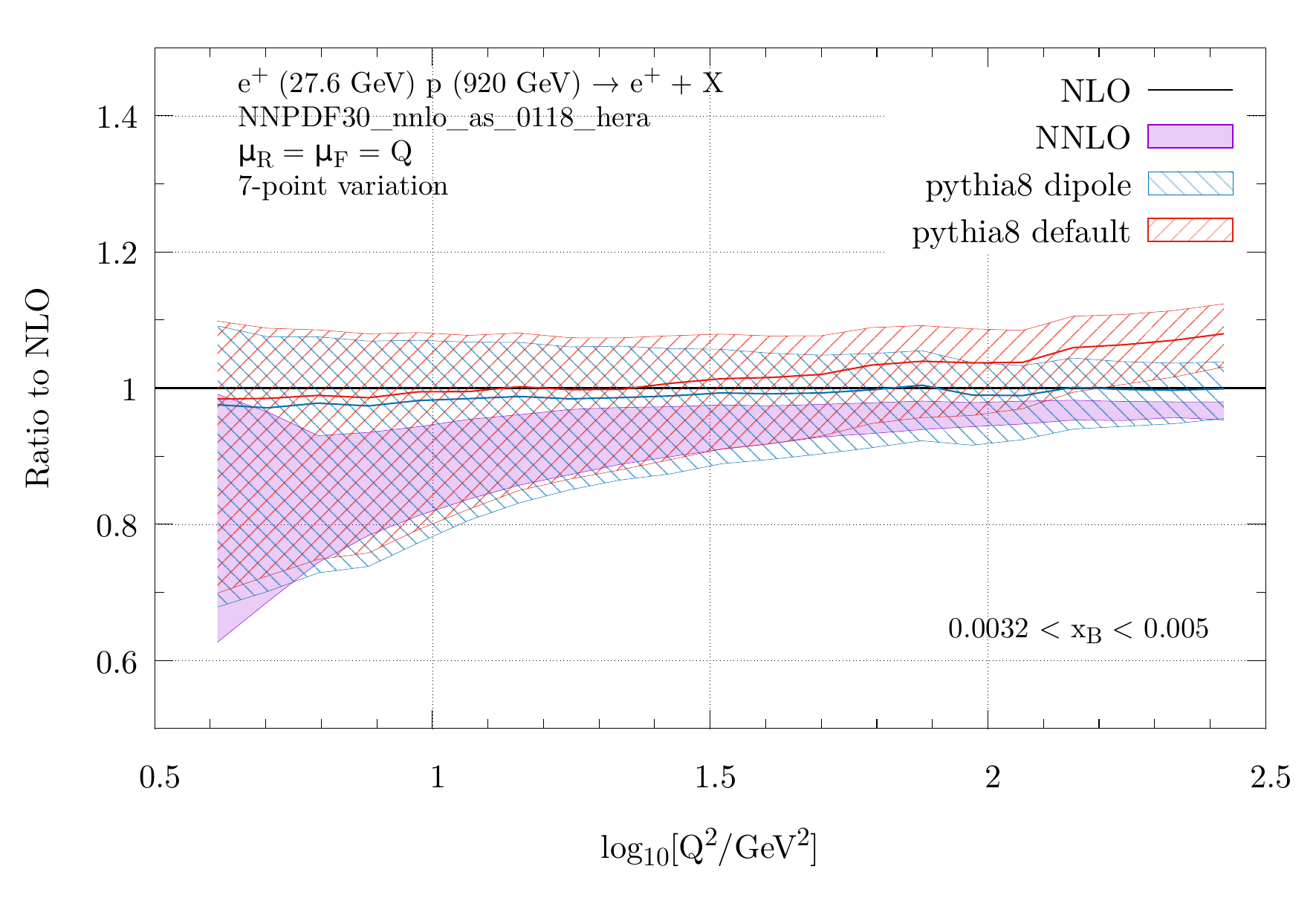}
  \caption{Same as Fig.~\ref{fig:sigmaRratio1}, but for the bin
    $85\,\mathrm{GeV}^2 < Q^2 < 110\,\mathrm{GeV}^2$ as a function of
    $\log{\xdis}$ (left) and for the bin
    $0.0032<\xdis<0.005$ as a function of $\log_{10}{Q^2}$ (right).}
  \label{fig:sigmaRratio2}
\end{figure}
\begin{figure}[tb!]
  \centering\includegraphics[width=0.49\textwidth,page=26]{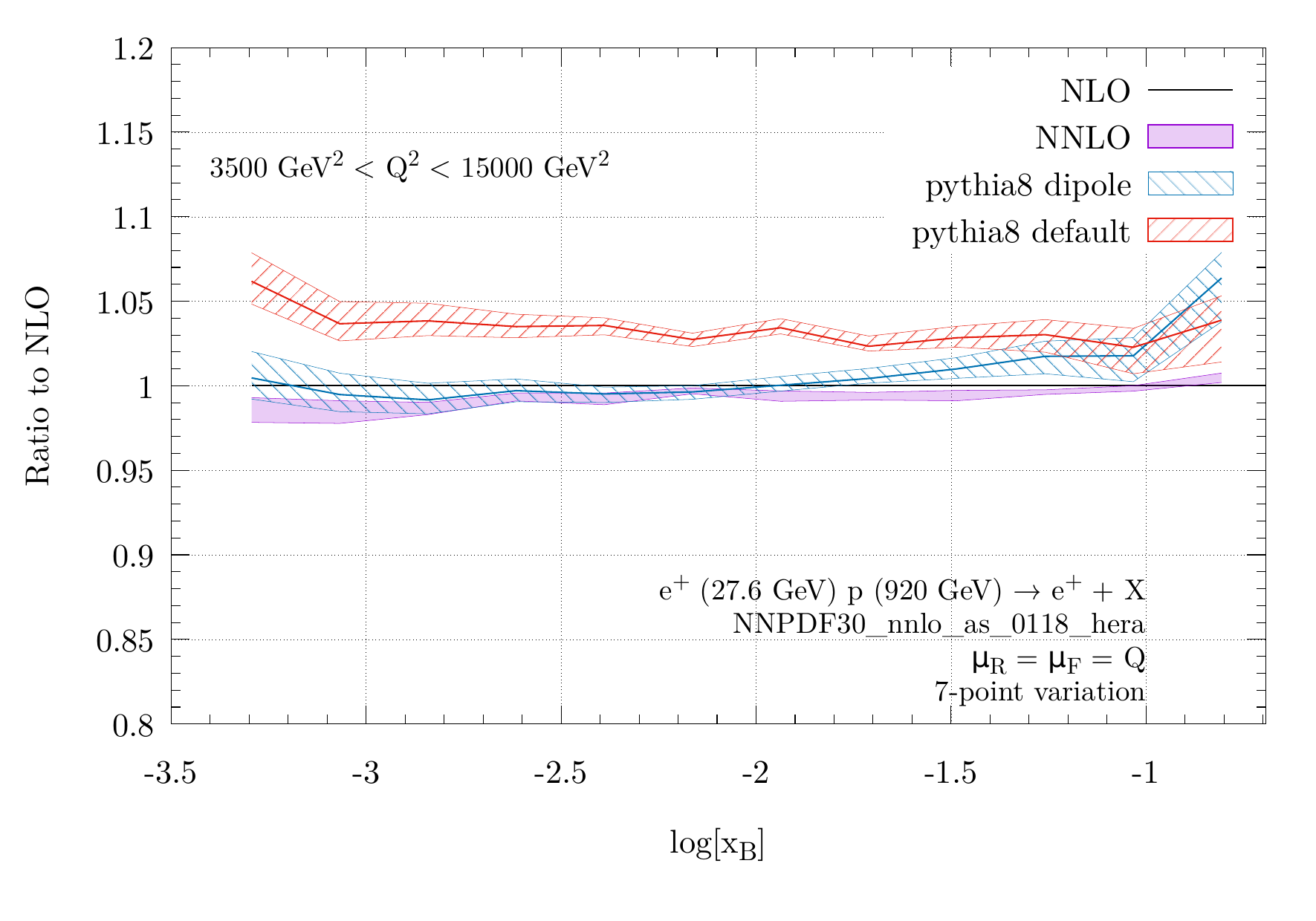}
  \centering\includegraphics[width=0.49\textwidth,page=40]{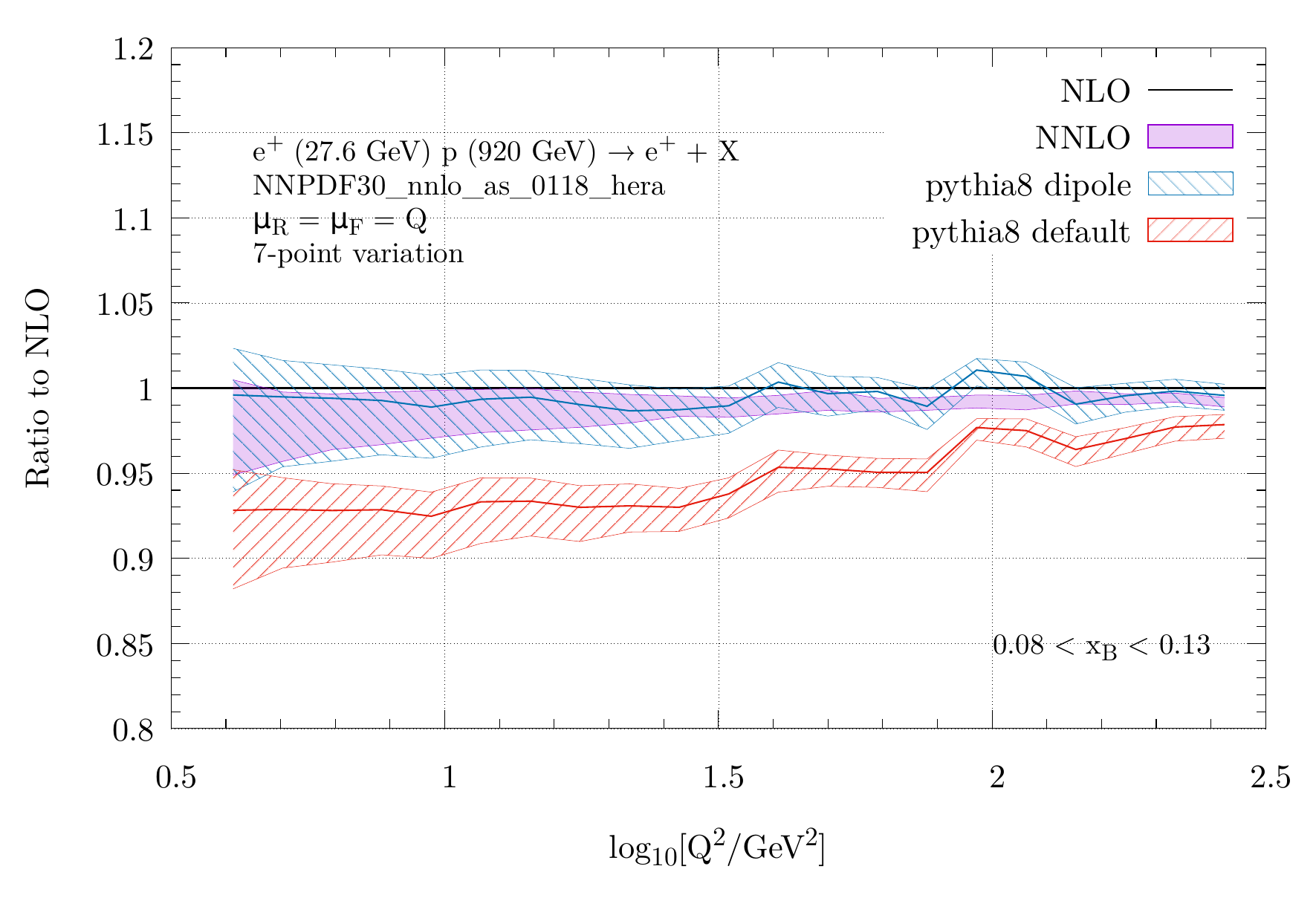}
  \caption{Same as Fig.~\ref{fig:sigmaRratio1}, but for the bin
    $3500\,\mathrm{GeV}^2 < Q^2 < 15000\,\mathrm{GeV}^2$ as a function of
    $\log{\xdis}$ (left) and for the bin
    $0.08<\xdis<0.13$ as a function of $\log_{10}{Q^2}$ (right).}
  \label{fig:sigmaRratio3}
\end{figure}
To make the effect more visible, we select a few representative bins
from the above two figures, and plot the ratios of the various
predictions to the respective fixed-order NLO results in
Figs.~\ref{fig:sigmaRratio1}-\ref{fig:sigmaRratio3}. Here it can be seen very
clearly that for large values of $\xdis$ there is a discrepancy
between the two \PYTHIAE{} showers which is not accounted for by scale
variations or true NNLO corrections. We do not include the scale
variation band for the reference NLO prediction, as it is near
identical to the band obtained with the \PYTHIAE{} dipole shower.

\subsubsection{Impact of alternative momentum mappings}
While the above discussion clearly demonstrates that our \POWHEG{}
implementation achieves NLO accuracy for inclusive quantities, it is
instructive to explore how an implementation using mappings closer to
the standard \POWHEGBOX{} mappings would look like. The main kinematical
difference between hadron-hadron collisions and DIS is that the
incoming lepton momentum in DIS is fixed, whereas the incoming partons
in a hadronic collision are sampled in their energy fractions. In order to simulate DIS it is
therefore necessary that the mappings used by \POWHEG{} preserve the
incoming lepton momentum. The standard FSR mapping in \POWHEG{} already
does this, as it preserves both $x_1$ and $x_2$ (but we stress that
these mappings do not preserve the DIS variables). In contrast, the ISR mappings 
modify both $x_1$ and $x_2$.

It is, however, straightforward to
adapt the ISR mapping such that it only modifies the incoming parton
momentum, but leaves the incoming lepton untouched. 
The minimally modified \POWHEG{} map for ISR can be
straightforwardly obtained by applying an additional longitudinal boost
along the direction of the incoming electron, in order to restore its
original value of the energy, to the kinematic reconstruction detailed
in Sec.~5.1.1 of Ref.~\cite{Frixione:2007vw}. This boost changes the
value of the $x$ fraction associated with the incoming parton, that,
at variance with Eq.~(5.7) of Ref.~\cite{Frixione:2007vw}, now becomes
$x = \frac{\bar{x}}{1-\xi}$, but does not alter the values of
the variable $\xi$, $y$ and $\phi$, which are defined in the partonic
centre-of-mass frame. The only other necessary modifications are then
the expression for upper bound for $\xi$, which now becomes
$\xi<\bar{x}_2$, and for $\kappa_t^{\sss \rm ISR}$, which is now
bounded by $\frac{S-\bar{s}}{4\sqrt{S}}$, being $\sqrt{S}$ the total
hadronic centre-of-mass energy, and $\sqrt{\bar{s}}$ the
underlying-Born centre-of-mass energy, which coincides with the mass
of the recoiling system.
However, with this minimally modified mapping, the
outgoing lepton still takes recoil and hence the DIS variables are not
conserved.

\begin{figure}[t!]
  \centering\includegraphics[width=0.92\textwidth,page=1]{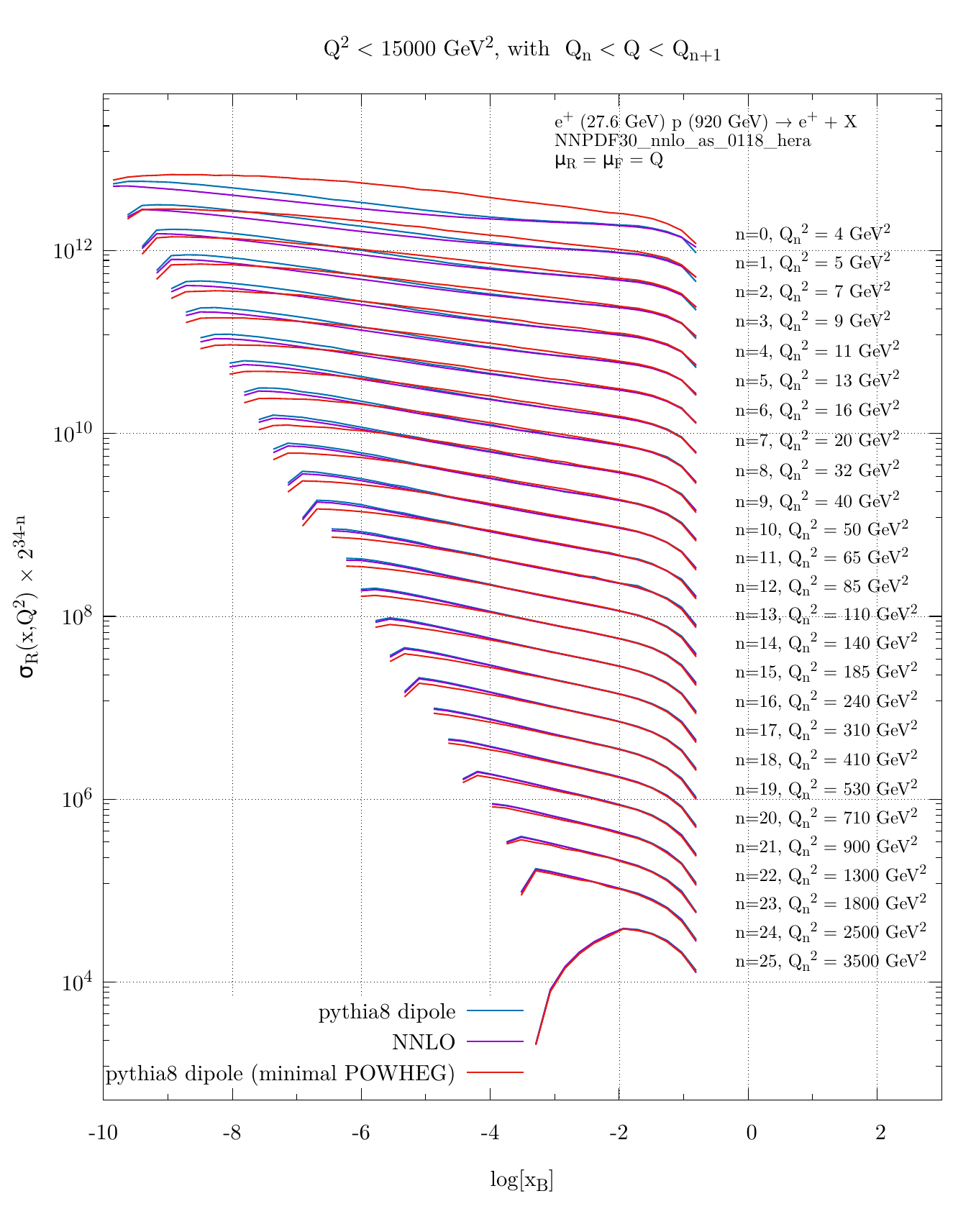}
  \caption{Similar to Fig.~\ref{fig:sigmaRinlogx}, but showing the
    minimally modified momentum mappings that do not preserve DIS
    kinematics in red. We do not show the default \PYTHIAE{} here.  }
  \label{fig:sigmaRinlogx-bad}
\end{figure}
\begin{figure}[tbh!]
  \centering\includegraphics[width=0.92\textwidth,page=2]{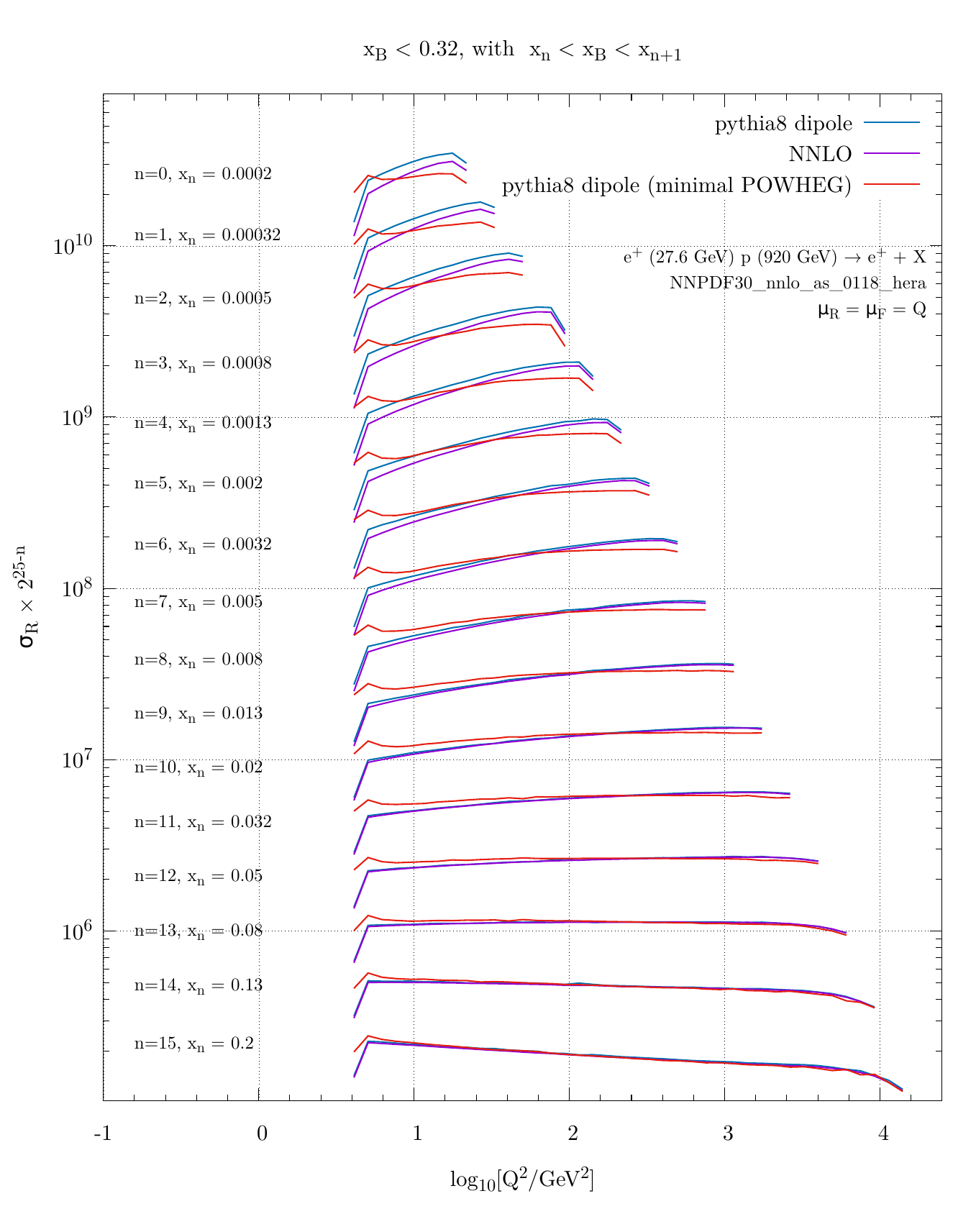}
  \caption{Similar to Fig.~\ref{fig:sigmaRinlogQ}, but showing the minimally
    modified momentum mappings that do not preserve DIS kinematics in red. We do not show the default \PYTHIAE{} here.
  }
  \label{fig:sigmaRinlogQ-bad}
\end{figure}
\begin{figure}[tbh!]
  \centering\includegraphics[width=0.49\textwidth,page=1]{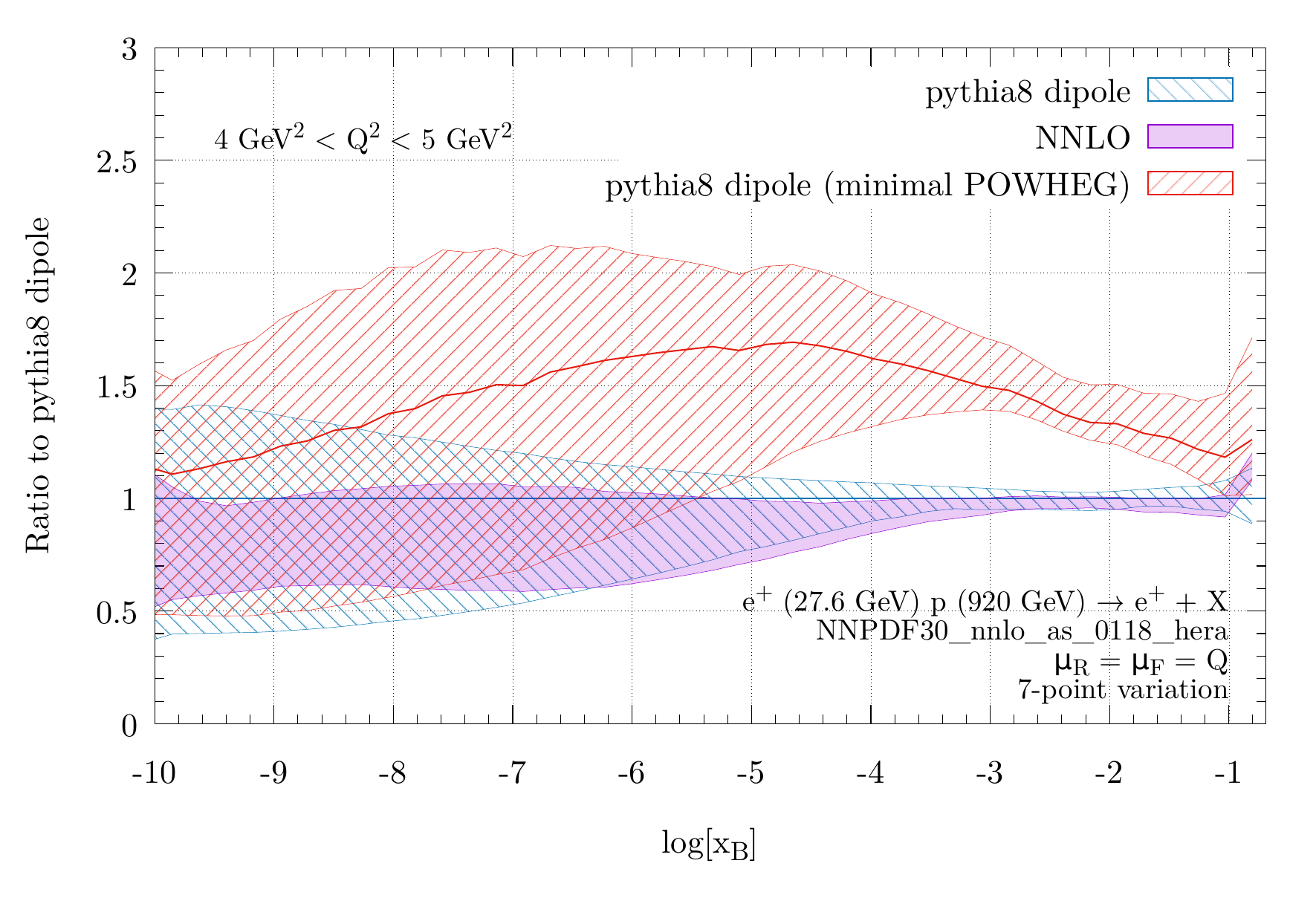}
  \centering\includegraphics[width=0.49\textwidth,page=27]{fig-temp/sigmaR-ratios-bad-powheg.pdf}
  \caption{Similar to Fig.~\ref{fig:sigmaRratio1}, but showing the
    minimally modified momentum mappings that do not preserve DIS
    kinematics in red. We do not show the default \PYTHIAE{} here and
    we normalise to our default ``pythia8 dipole'' result here in
    blue.} 
  \label{fig:sigmaRratio1-bad}
\end{figure}
\begin{figure}[tbh!]
  \centering\includegraphics[width=0.49\textwidth,page=13]{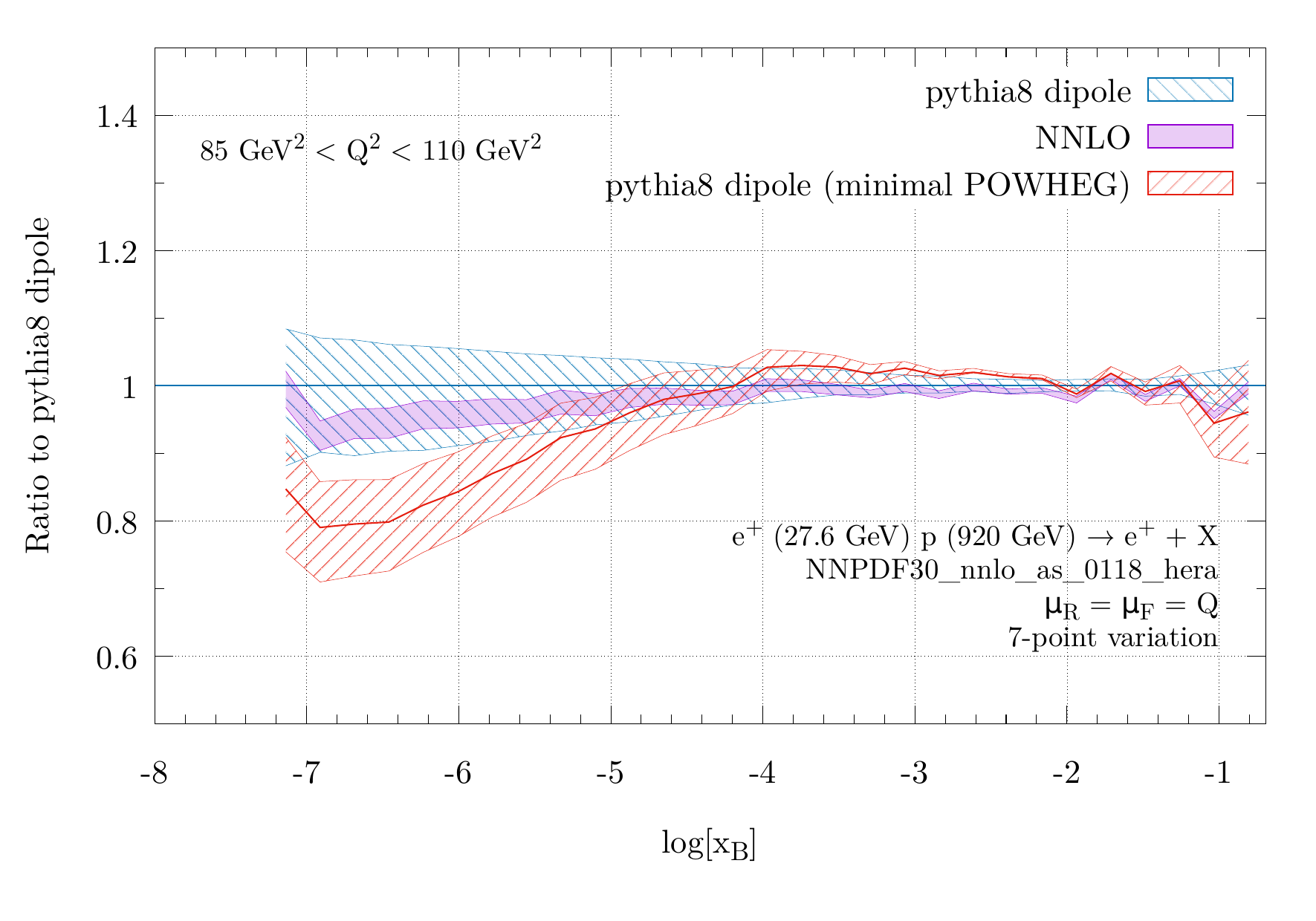}
  \centering\includegraphics[width=0.49\textwidth,page=33]{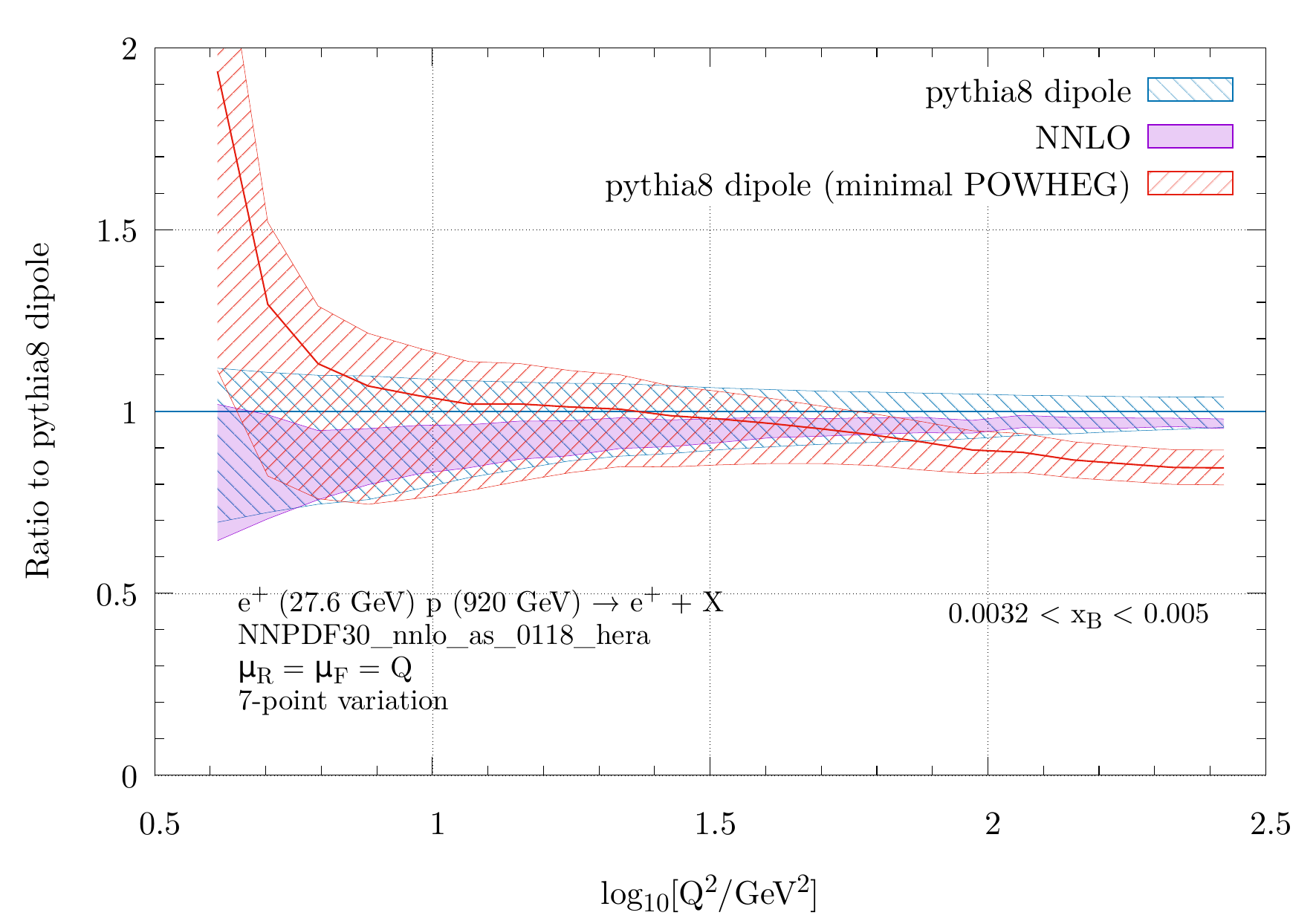}
  \caption{Similar to Fig.~\ref{fig:sigmaRratio2}, but showing the
    minimally modified momentum mappings that do not preserve DIS
    kinematics in red. We do not show the default \PYTHIAE{} here and
    we normalise to our default ``pythia8 dipole'' result here in
    blue.} 
  \label{fig:sigmaRratio2-bad}
\end{figure}
\begin{figure}[tb!]
  \centering\includegraphics[width=0.49\textwidth,page=26]{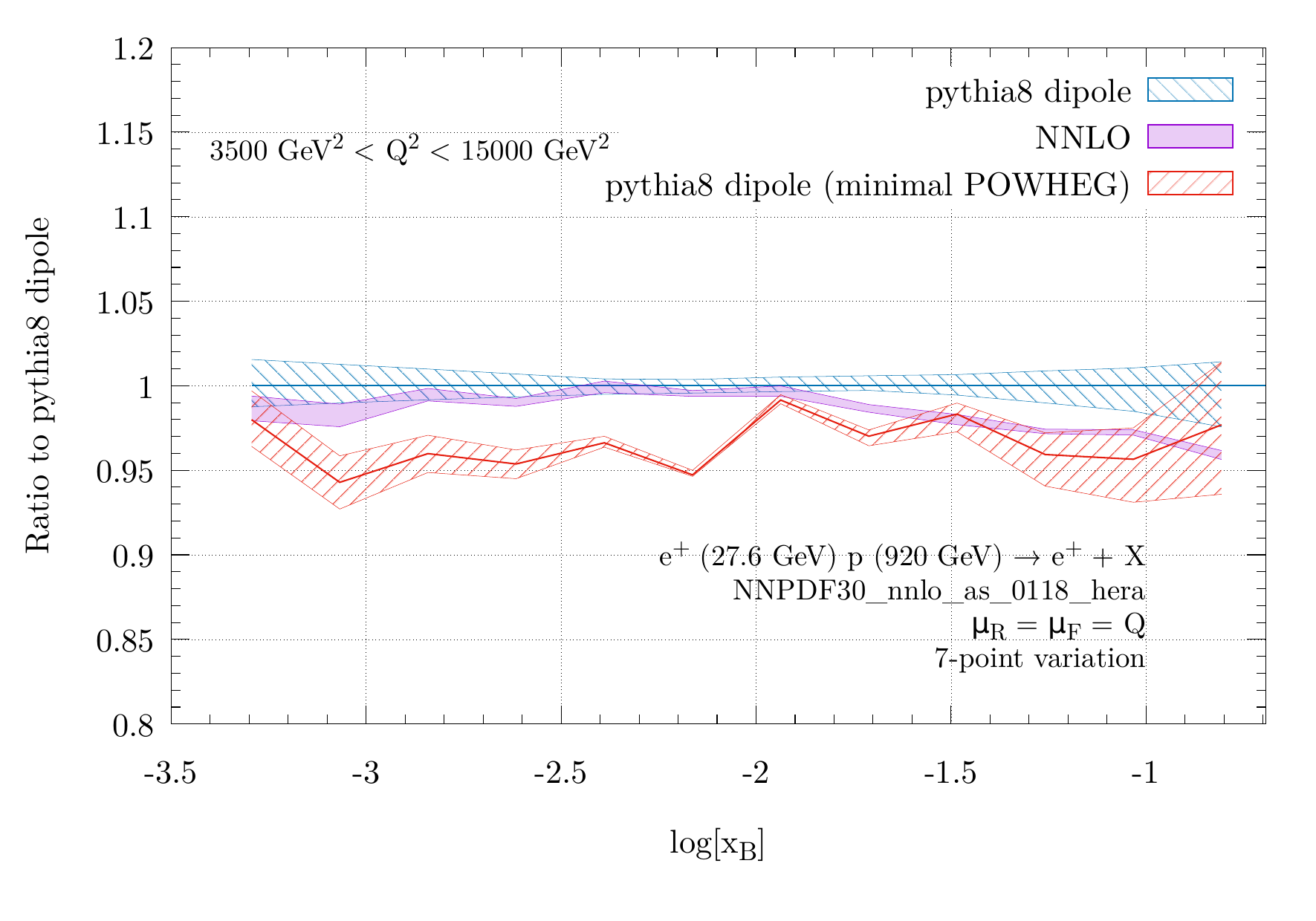}
  \centering\includegraphics[width=0.49\textwidth,page=40]{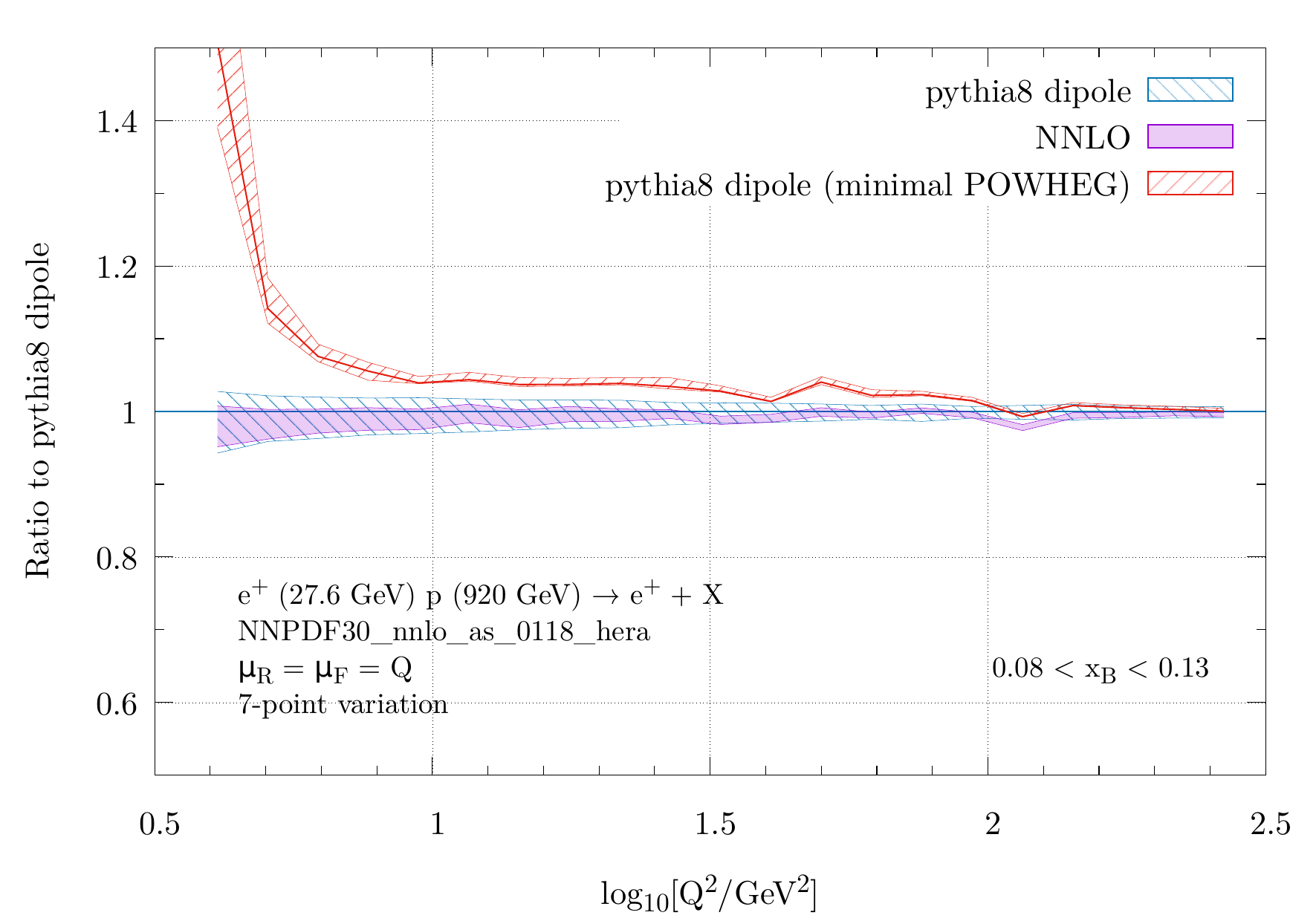}
  \caption{Similar to Fig.~\ref{fig:sigmaRratio3}, but showing the
    minimally modified momentum mappings that do not preserve DIS
    kinematics in red. We do not show the default \PYTHIAE{} here and
    we normalise to our default ``pythia8 dipole'' result here in
    blue.} 
  \label{fig:sigmaRratio3-bad}
\end{figure}

In Figs.~\ref{fig:sigmaRinlogx-bad}-\ref{fig:sigmaRratio3-bad} we show
plots similar to the above, but now we use the minimally modified
momentum mappings (red lines, labelled ``minimal POWHEG'' in the
plots). For reference we show the \POWHEG{}+\PYTHIAE{} dipole
prediction (blue), using our new mappings from the above plots, as
well as NNLO results (purple). As can be seen clearly, the predictions
obtained with these minimally modified mappings exhibit sizable
differences compared to the new mappings for inclusive quantities. In
particular, we observe very large deviations for small $Q$ and $\xdis$. It
can also be seen that the deviations do not approximate the true NNLO
corrections well.

It is interesting to note that LHC processes involving the exchange of
colourless particles in the $t$-channel, like vector boson fusion
(VBF) and single top production, which have been implemented in the
\POWHEGBOX{} in
Refs.~\cite{Nason:2009ai,Jager:2012xk,Alioli:2009je,Frederix:2012dh},
exhibit kinematics that are essentially double-DIS like. For this
reason one would expect that these processes could benefit from a
different momentum mapping. For instance, for VBF, it was observed in
Ref.~\cite{Cacciari:2015jma} that the description of the rapidity
separation between the two hardest jets as predicted with \POWHEG{}
has a very different shape compared to both NLO and NNLO
predictions. It would be interesting to see if this tension can be
resolved with our new mappings. We leave this question for future
work.

\subsection{Exclusive observables and comparison with resummation}
\label{sec:caesar}
In addition to the inclusive cross section considered above,
information on the physics of DIS reactions can be gained from
exclusive observables.
In order to explore the complementarity of fixed-order perturbative
calculations, resummed predictions and NLO+PS simulations, we
performed a comparison of these approaches for two
event shape variables.
Following the Breit-frame%
\footnote{ The Breit frame is defined by $2 \xdis \vec{P} + \vec{q}=0$,
  where $\vec{P}$ denotes the incoming proton momentum and
  $\vec{q}$ the momentum of the virtual boson characterizing the DIS
  topology. In this frame, the exchanged photon has no energy and is anti-aligned to the incoming parton. We follow Appendix 7.11 in Ref.~\cite{Devenish:2004pb} for
  the actual frame transformation.}
definition of event shapes employed in the experimental analyses, we
can distinguish between the remnant and current hemisphere, where
particles in the remnant (current) hemisphere have positive (negative)
pseudo-rapidities when the incoming photon has negative rapidity.
Event shapes are formulated only in terms of particles in the current
hemisphere.
In particular, in this section we consider the thrust distribution
$\tau_{\rm z, Q}$ and broadening $B_{\rm z, E}$, which are defined as
\beq
\label{eq:qthrust}
\tau_{\rm z, Q} = 1 -\frac{ \sum_{h} 2 |\vec{p}_{z,h}|}{Q}\,, 
\eeq
\beq
\label{eq:ebroadening}
B_{\rm z, E} = \frac{\sum_h |\vec{p}_{T,h}|}{2 \sum_h |\vec{p}_{h}|},
\eeq
where the momenta are defined in the Breit frame, the photon
three-momentum determines the $z$-direction, and $h$ denotes all the
hadrons in the current hemisphere.  These observables are
continuously-global and we can obtain resummed predictions at NLL
accuracy, e.g.\ from \texttt{CAESAR}~\cite{Banfi:2004yd}, according to
the practical details in section 3.2 of Ref.~\cite{Banfi:2010xy}. In
particular, the matching of resummation and fixed order is performed
with the mod-R scheme defined in that same reference.

\begin{figure}[tbh!]
  \centering\includegraphics[width=0.49\textwidth]{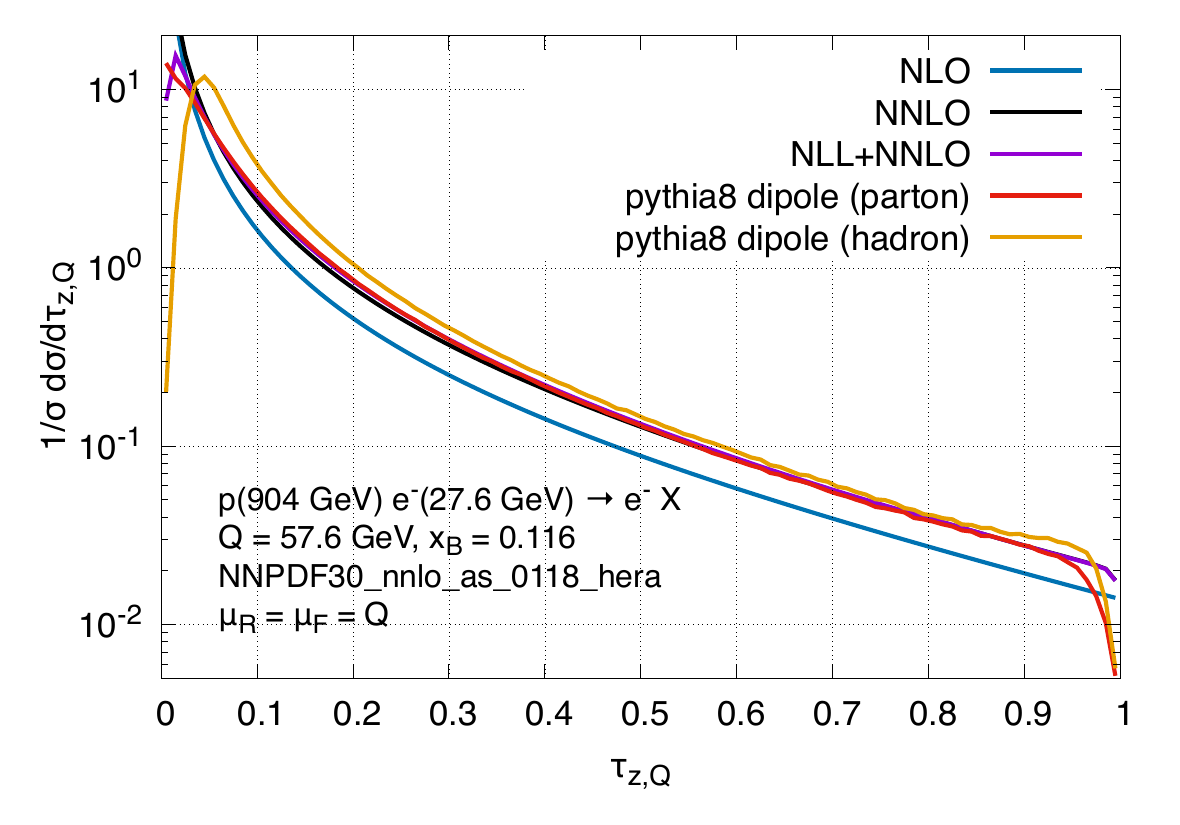}
  \centering\includegraphics[width=0.49\textwidth]{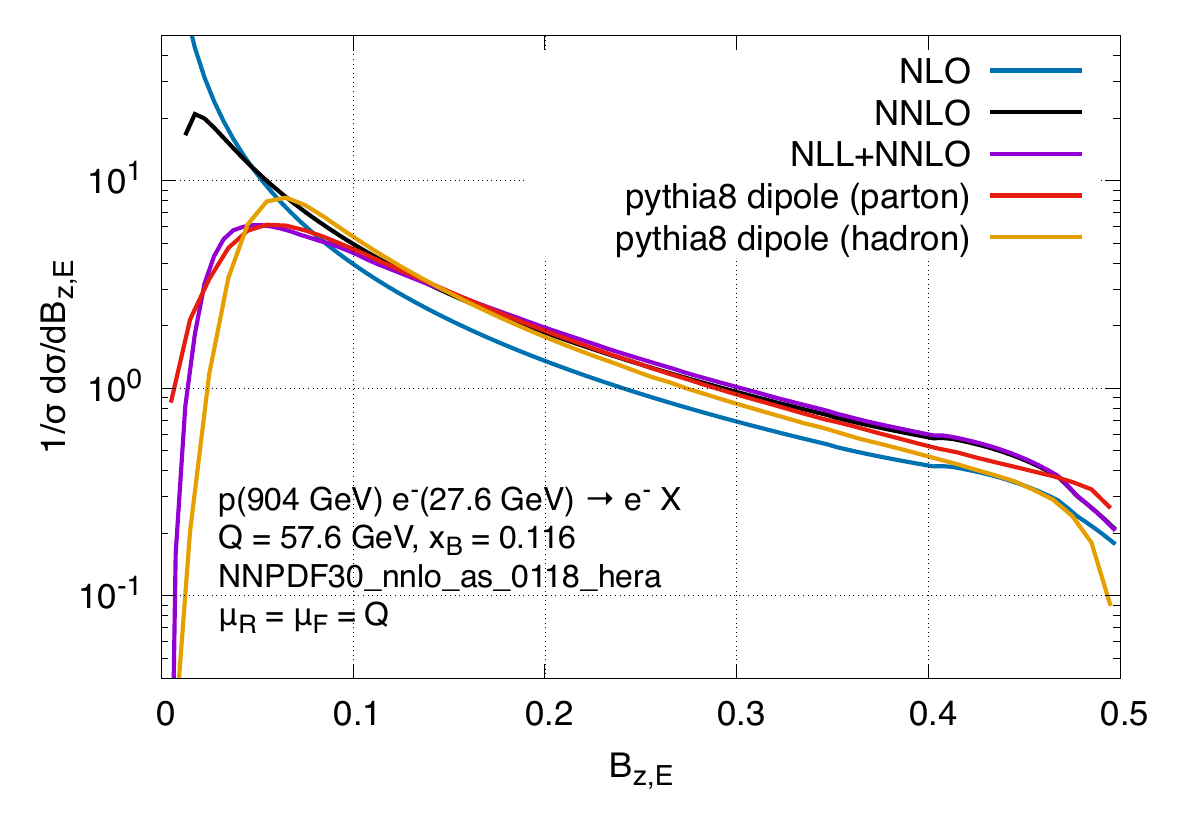}
  \caption{ Thrust distribution normalised with respect to $Q$ (left)
    and broadening (right) for $Q=57.6$~GeV, $\xdis=0.116$ for the
    photon-exchange contribution to $e^- p \rightarrow e^- X$ at NLO
    (blue), NNLO (black), NNLO+NLL (magenta), NLO+\PYTHIAE{} without
    (red) and with non-perturbative effects (orange).
  }
  \label{fig:caesar}
\end{figure}

Fig.~\ref{fig:caesar} 
%
%
%
%
depicts $\tau_{\rm z, Q}$ and $B_{z,E}$ for the photon-exchange
contribution to $e^- p \rightarrow e^- X$ with $E_p = 904.5$~GeV, $E_e =
27.6$~GeV.
We consider  fixed-underlying Born kinematics,
corresponding to $\xdis=0.116$ and $Q=57.6$~GeV.
These settings correspond to the average values of $\xdis$, $Q$ and
$\sqrt s = 316$~GeV reported by the experimental analysis of
Ref.~\cite{H1:2005zsk} for the bin $50\text{~GeV} < Q < 70
\text{~GeV}$.
Notice that this value of $Q$ ensures that mass effects, which are
included in the PS simulation, but not in the NNLO+NLL prediction and
in the \POWHEG{} underlying calculation, are negligible.

Furthermore, an event is included, if the sum over the energies $E_h$ of all hadronic objects $h$ in the current hemisphere exceeds a minimum value $\epsilon_\mathrm{lim}$, 
\beq
\label{eq:elim}
E_{\mathrm{curr}} = \sum_h E_h > \epsilon_\mathrm{lim} = Q/10\,.
\eeq
As discussed in Ref.~\cite{Antonelli:1999bv}, starting from order
$\alpha_s^2$ there can be configurations where the current hemisphere
is populated only by soft large-angle radiation from partons in
the remnant hemisphere.
It is then necessary to introduce a cut on $E_{\mathrm{curr}}$ to
remove sensitivity to these soft emissions in event shapes normalised
with respect to $E_{\mathrm{curr}}$ (or to $\sum_h |\vec{p}_{z,h}|$),
that would be otherwise infrared unsafe.

In the parton-level \PYTHIAE{} curve, we dress \POWHEG{} events with the
QCD \PYTHIAE{} dipole shower. In the
hadron-level curve we also include hadronisation and beam remnants effects.
The fixed-order predictions are obtained with {\tt
  DISENT}~\cite{Catani:1996vz}\footnote{The version of {\tt DISENT}
used here and below includes the bug fix reported in
Refs.~\cite{Borsa:2020ulb,Borsa:2020yxh}.}. Since the event shapes
shown here vanish at Born-level, the predictions shown here are
effectively LO and NLO accurate even though we label the predictions
by their inclusive accuracy, i.e.\ NLO and NNLO.

At NLO, the distributions steeply increase
towards small values.
This behaviour reflects the distinguished kinematics of a LO DIS
configuration with exactly one final-state parton, which at NLO can
only be altered by the emission of a single additional parton.
One more parton can arise at NNLO where we observe a peak at small
values of $\tau_{\rm z, Q}$ and $B_{z,E}$, which is due to the
distributions turning negative (and diverging) due to large negative virtual
corrections.

The divergences of the fixed-order calculations are removed by the resummation of
all-order leading (and next-to-leading) logarithmically enhanced terms.
In the NLO+PS results the parton shower has a similar effect.

We note that, away from the divergence, our \POWHEG{} prediction is much
closer to NNLO than to NLO at the parton level, indicating that higher
order terms induced by the shower and matching approximate well the
true NNLO corrections.

We also notice that, while for the broadening distribution (right
panel), both the NLO+PS (parton) and the NNLO+NLL curves are peaked around the
same value of $B_{\rm z, E}$.
For  $\tau_{z, Q}$ (left panel), the NNLO+NLL Sudakov peak is at a much lower value (i.e.\ $\tau_{z, Q} \lesssim 0.02$) and it is not visible in the parton shower predictions because of the PS cutoff $\mu_{\min} \approx 0.5$~GeV.%

Hadronisation effects are sizable for small values of the event shapes.
In case of the thrust distribution, they correspond to a roughly
constant shift, while for the broadening this shift depends on the
value of $B_{\rm z,E}$.

\subsection{Jet and VBF related observables}
\label{sec:vbf}
\begin{figure}[tb!]
  \centering\includegraphics[width=0.49\textwidth,page=1]{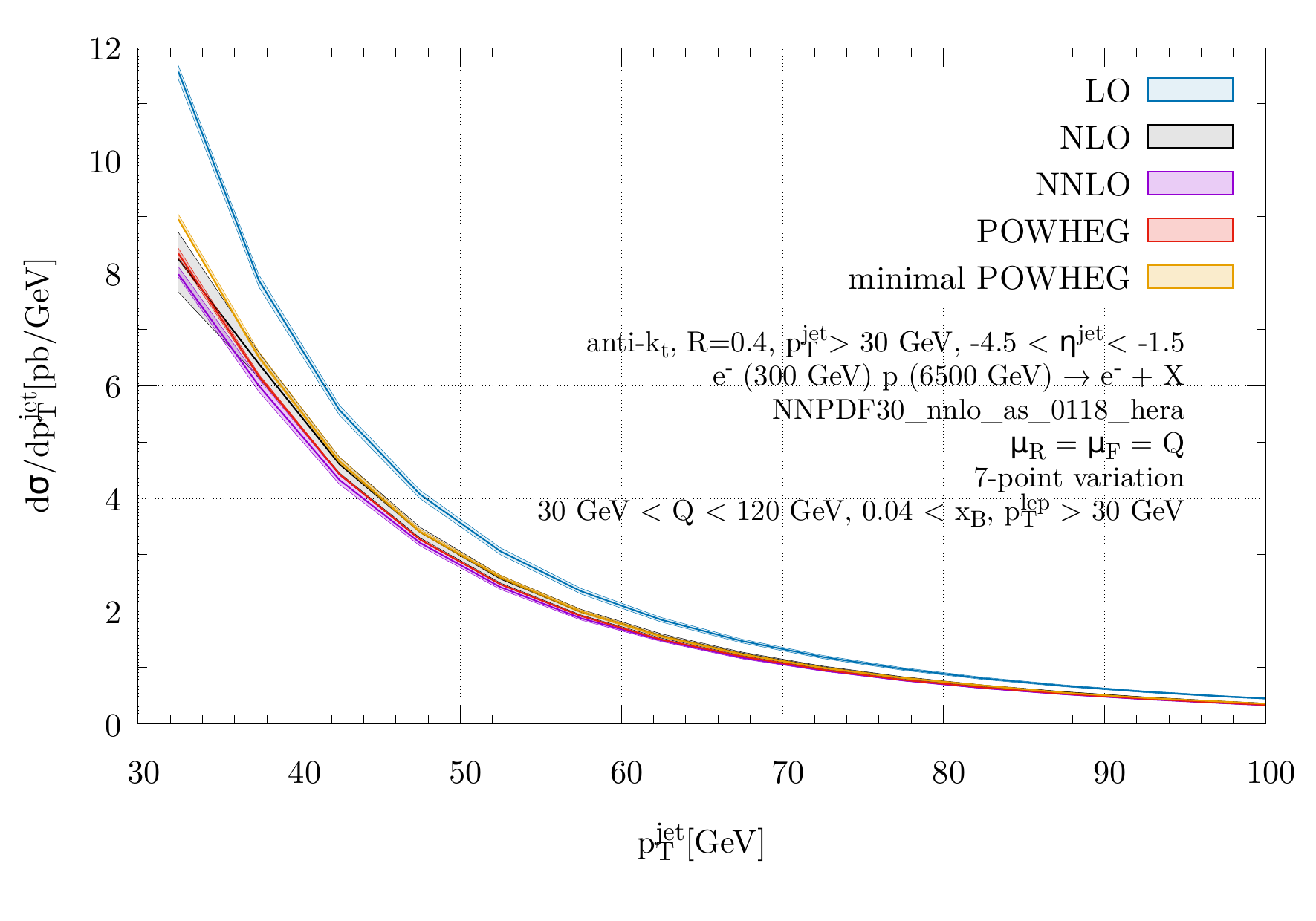}
  \centering\includegraphics[width=0.49\textwidth,page=2]{fig-temp/vbf-plots.pdf}
  \caption{The hardest anti-$k_T$ $R=0.4$ jet in the rapidity window
    $-4.5 < \etaj < -1.5$ for events satisfying the cuts of
    Eq.~\eqref{eq:vbfcuts}. We show LO (blue), NLO (grey), NNLO
    (purple), our new DIS implementation showered with \PYTHIAE{}
    (red) and the minimally modified \POWHEG{} implementation with the
    same shower (orange). On the right we show the ratio to the NLO
    prediction. The bands correspond to a 7-point scale variation
    around the central scale $\mu=Q$. }
  \label{fig:vbf-ptj1}
\end{figure}
\begin{figure}[tb!]
  \centering\includegraphics[width=0.49\textwidth,page=3]{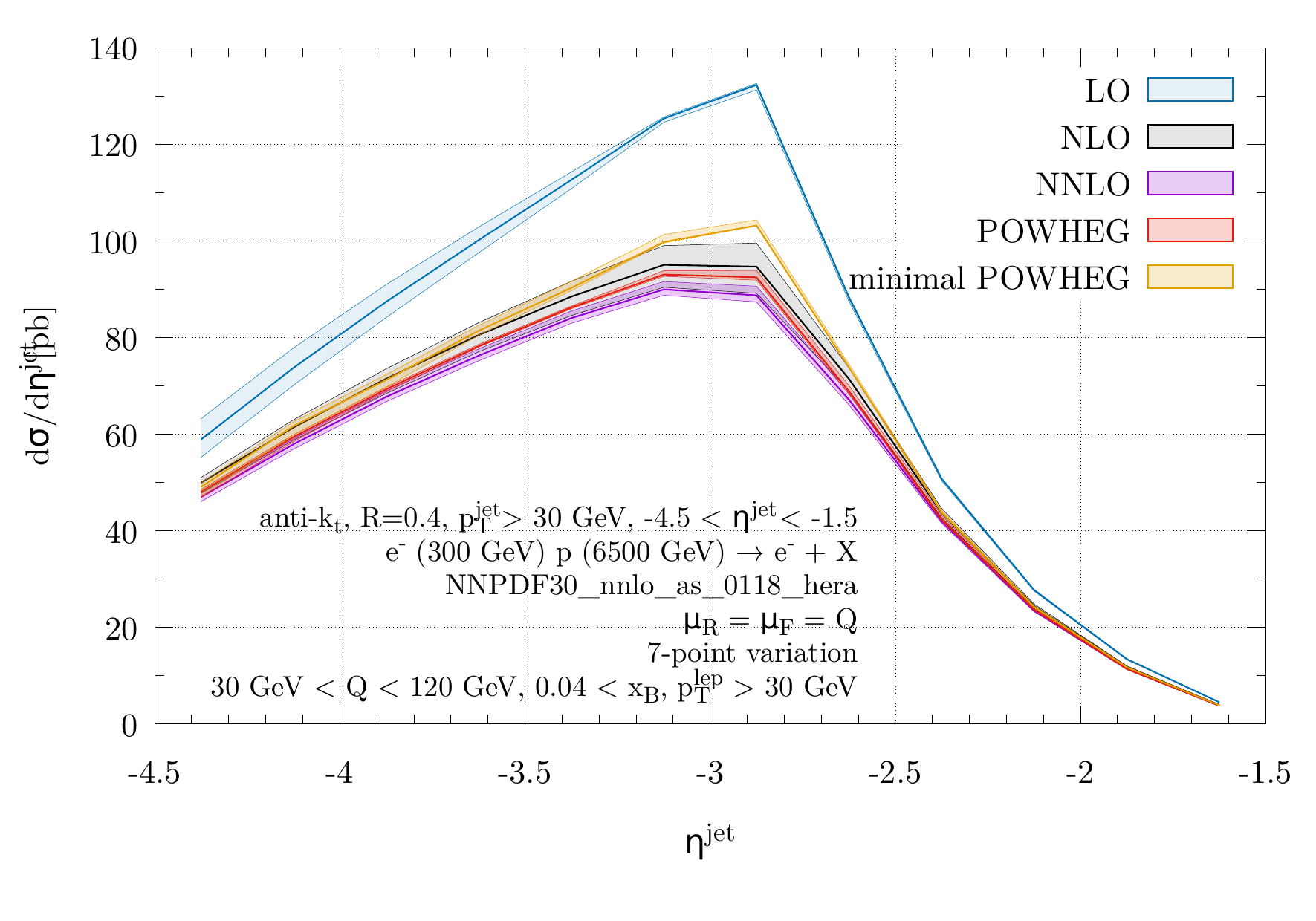}
  \centering\includegraphics[width=0.49\textwidth,page=4]{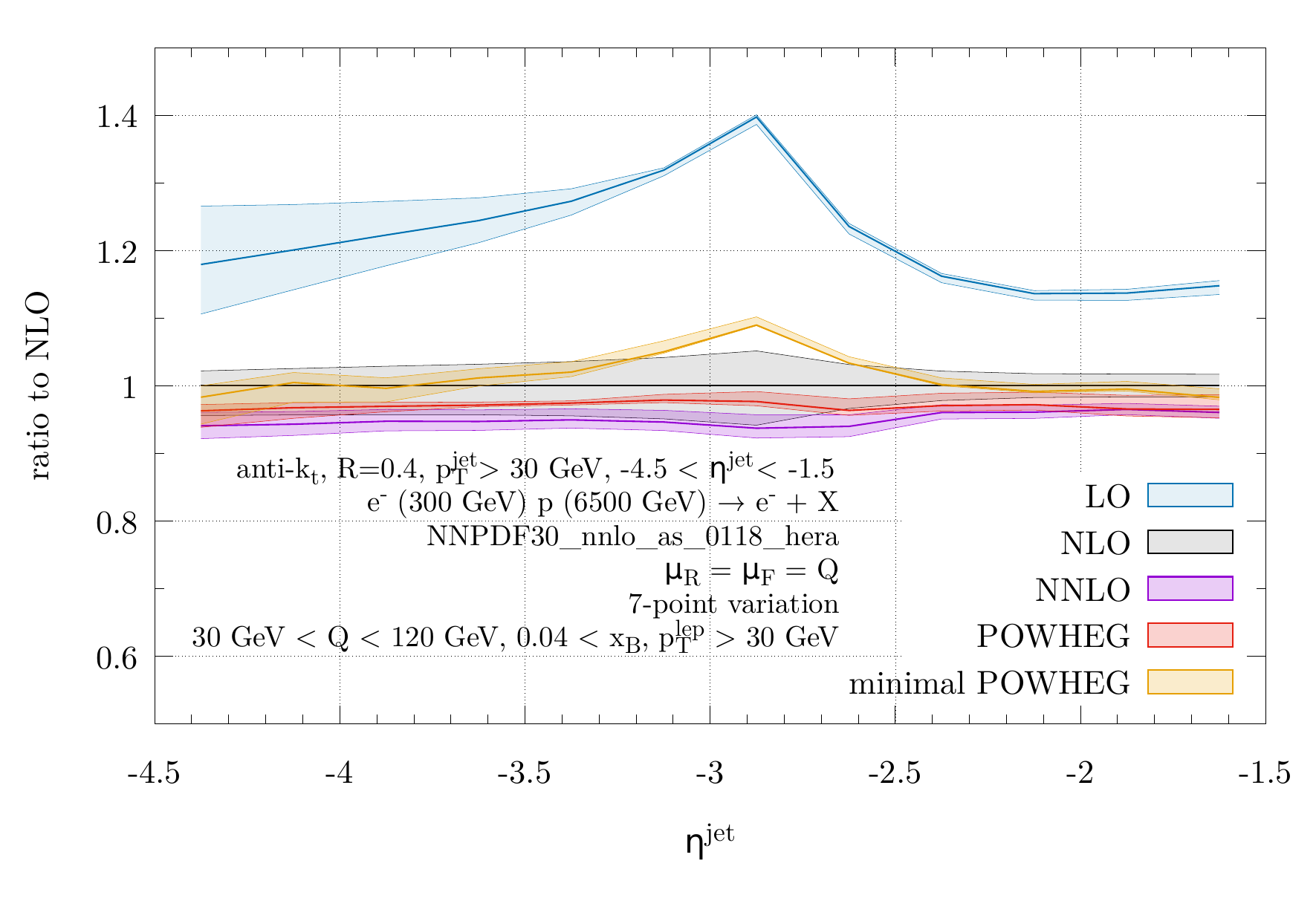}
  \caption{ Same as Fig.~\ref{fig:vbf-ptj1}, but now showing the
    rapidity of the hardest jet satisfying $\ptj>30\,\mathrm{GeV}$.}
  \label{fig:vbf-etaj1}
\end{figure}
Although the full implementation and study of VBF or single top production is
beyond the scope of this paper, it is possible to mimic the kinematics of jets produced
in these processes at the LHC. As discussed above, our \POWHEGBOX{}
implementation reproduces the DIS structure functions exactly at NLO
since the momentum mappings preserve the DIS variables, or
equivalently they leave the lepton momenta untouched. However, if we
instead turn to the kinematics of the DIS jet, we see that it is
clearly modified by the extra radiation. The \POWHEG{} method still
guarantees that we describe the hardest jet with NLO accuracy, but
with higher-order terms in $\as$ induced by the \POWHEG{} Sudakov form
factor (and the subsequent showering).

Typical VBF analyses at the LHC look for two well-separated jets with a
large invariant mass. Therefore, to mimic VBF like topologies, we
produce events in proton-electron collisions, with a proton energy of
$6500\,\mathrm{GeV}$ and an electron energy of
$300\,\mathrm{GeV}$. Additionally we require that
\begin{align}
  30\,\mathrm{GeV} < Q < 120 \,\mathrm{GeV}, \quad 0.04<\xdis, \quad 
  p_{T}^{\mathrm{lep}} > 30\,\mathrm{GeV}.
  \label{eq:vbfcuts}
\end{align}
The cut on $\xdis$ ensures a large invariant mass of the colliding
system, and the range in $Q$ 
is close to the typical momentum transfer in VBF of the order of the
$W$ mass. We also only include photon-mediated DIS to allow for a
comparison with a fixed order code.

We cluster the events in the laboratory frame with the anti-$k_T$
algorithm~\cite{Cacciari:2008gp} with $R=0.4$ using {\tt FastJet}
v3.4.0~\cite{Cacciari:2011ma}. We then look for the jet of hardest
transverse momentum with pseudo-rapidity $-4.5 < \etaj < -1.5$ (the
incoming parton is in the minus $z$-direction). We require this jet to
have $\ptj>30\,\mathrm{GeV}$. We stress that these quantities are
defined in the lab frame, not in the Breit frame, hence they are
non-vanishing already at LO.

In Figs.~\ref{fig:vbf-ptj1}-\ref{fig:vbf-etaj1} we show the transverse
momentum and the rapidity, respectively, of the hardest jet. In
addition to our new \POWHEG{} implementation (shown in red) we also
show fixed-order predictions up to NNLO and events generated with the
minimal \POWHEG{} implementation described in the previous section
(orange). We use the \PYTHIAE{} dipole shower to shower both sets of
\POWHEG{} events. The bands are obtained through the usual 7-point
scale variation around the central scales $\mu_{R/F}=Q$. In
App.~\ref{sec:scale-variations} we show results obtained using two
different central scales, which lead to similar findings to the
  ones presented here.

The fixed order predictions are obtained with the program {\tt
  disorder}~\cite{disorder} which uses {\tt
  DISENT}~\cite{Catani:1996vz} and the projection-to-Born
method~\cite{Cacciari:2015jma} together with the NNLO DIS coefficient
functions~\cite{vanNeerven:1999ca,vanNeerven:2000uj} as implemented in
the DIS structure function branch of {\tt HOPPET}~\cite{Salam:2008qg}.

We note that the minimal \POWHEG{} implementation is typically further
away from the fixed order NLO curve than the new implementation
presented here. The differences between the two \POWHEG{} generators
are, however, compatible in size with the scale uncertainty band,
although the true NNLO corrections (in purple) tend to favour
our new \POWHEG{} implementation. It will be interesting to see if
this pattern persists if one were to use our new mappings in VBF
production or single top.

The size of the scale variation in both \POWHEG{} implementations
deserves some comments. Naively one would expect the scale variation
bands to be commensurate in size with those of the NLO
calculation. However, in \POWHEG{} when one varies the renormalisation
and factorisation scales this only impacts the $\bar{B}$ function. The
scale at which $\as$ is evaluated in the Sudakov form factor remains
the same (the transverse momentum of the emission). The $\bar{B}$
function, introduced in Sec.~\ref{sec:pwgingredients}, is essentially
the NLO inclusive cross section, which has tiny scale variations for
the values of $Q$ probed here. As a consequence the scale variation is
dramatically underestimated -- even more than in the fixed order
prediction. In App.~\ref{sec:damp} we study this effect in more
detail. Although the scale variation band is always underestimated
compared to the fixed order NLO band, including more damping can
ameliorate this effect somewhat.

\section{Phenomenological studies}
\label{sec:pheno}
\begin{figure}[tp]
  \centering
  \includegraphics[width=0.495\textwidth,page=1]{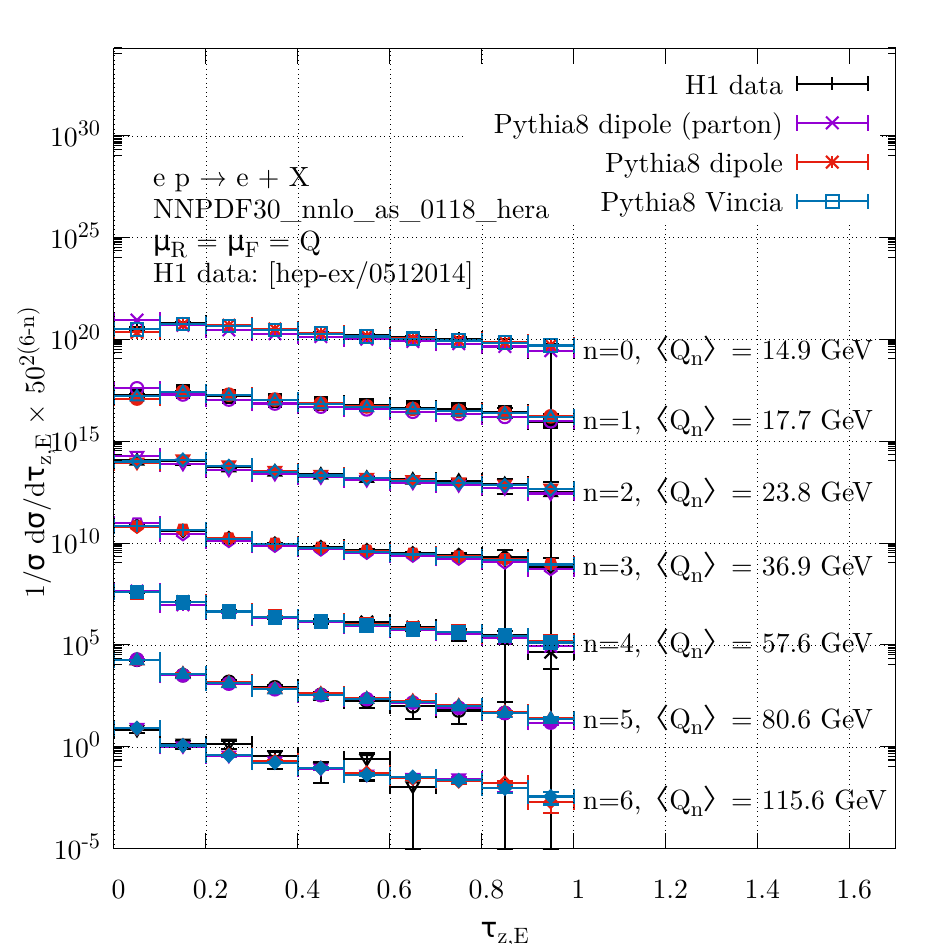}\hfill%
  \includegraphics[width=0.495\textwidth,page=2]{fig-temp/h1_plots_all.pdf} 
    \caption{
     Thrust distribution (left) and broadening (right) for
      different bins in $Q$, at the hadron level for the dipole (red), and {\sc Vincia} (blue) showers,
      and at the parton level (i.e. without hadronisation and beam-remnant
      effects) for the dipole shower (magenta), together with the H1
      data of Ref.~\cite{H1:2005zsk}.
      For a given $Q$-bin~$n$, the average
      value of $Q$ is denoted by $\langle Q_n \rangle$, and the
      corresponding curve is multiplied by a factor of $50^{2(6-n)}$
      for better readability.  }
  \label{fig:thrust+broadening}
\end{figure}
%
%
In this section we present hadron-level predictions for the HERA and EIC experiments.
In particular, we interface our NLO+PS generator with
\PYTHIAE{}~\cite{Bierlich:2022pfr}, which also provides hadronisation
and beam-remnant effects.
The hadron-level predictions are obtained considering both the dipole
shower~\cite{Cabouat:2017rzi}, which we have employed in the previous
section, as well as the \PYTHIAE{} implementation of the default antenna {\sc Vincia}
shower~\cite{Brooks:2020upa}. We do not include QED radiation or
hadron-decay effects.%
\footnote{The implementation interface between our code and the {\sc Vincia} showers largely relies on the works
of Refs.~\cite{Hoche:2021mkv,FerrarioRavasio:2023kjq}.}

\subsection{Comparison to HERA data}
\label{sec:hera}
%
%
We now compare predictions obtained with our new \POWHEGBOX{}
implementation with HERA data analyzed by the H1~Collaboration in
Ref.~\cite{H1:2005zsk} corresponding to an integrated luminosity of
$\lint = 106$~pb$^{-1}$. Following their study, we consider collisions
of electrons or positrons of energy $E_e=27.6$~GeV and protons of
energy $E_p=820$~GeV or $E_p=920$~GeV resulting in centre-of-mass
energies $\sqrt{s}$ of 301~GeV and 319~GeV, respectively. Electron and
positron samples are generated independently and combined in the
results discussed below corresponding to the composition of the data
sample of Ref.~\cite{H1:2005zsk}, i.e.\ $e^+p$: $\sqrt{s}=301$~GeV,
$\lint = 30~$pb$^{-1}$; $e^-p$: $\sqrt{s}=319$~GeV, $\lint =
14~$pb$^{-1}$; $e^+p$: $\sqrt{s}=319$~GeV, $\lint = 62~$pb$^{-1}$.
%
%
\begin{figure}[tp]
  \centering
  \includegraphics[width=0.33\textwidth,page=1]{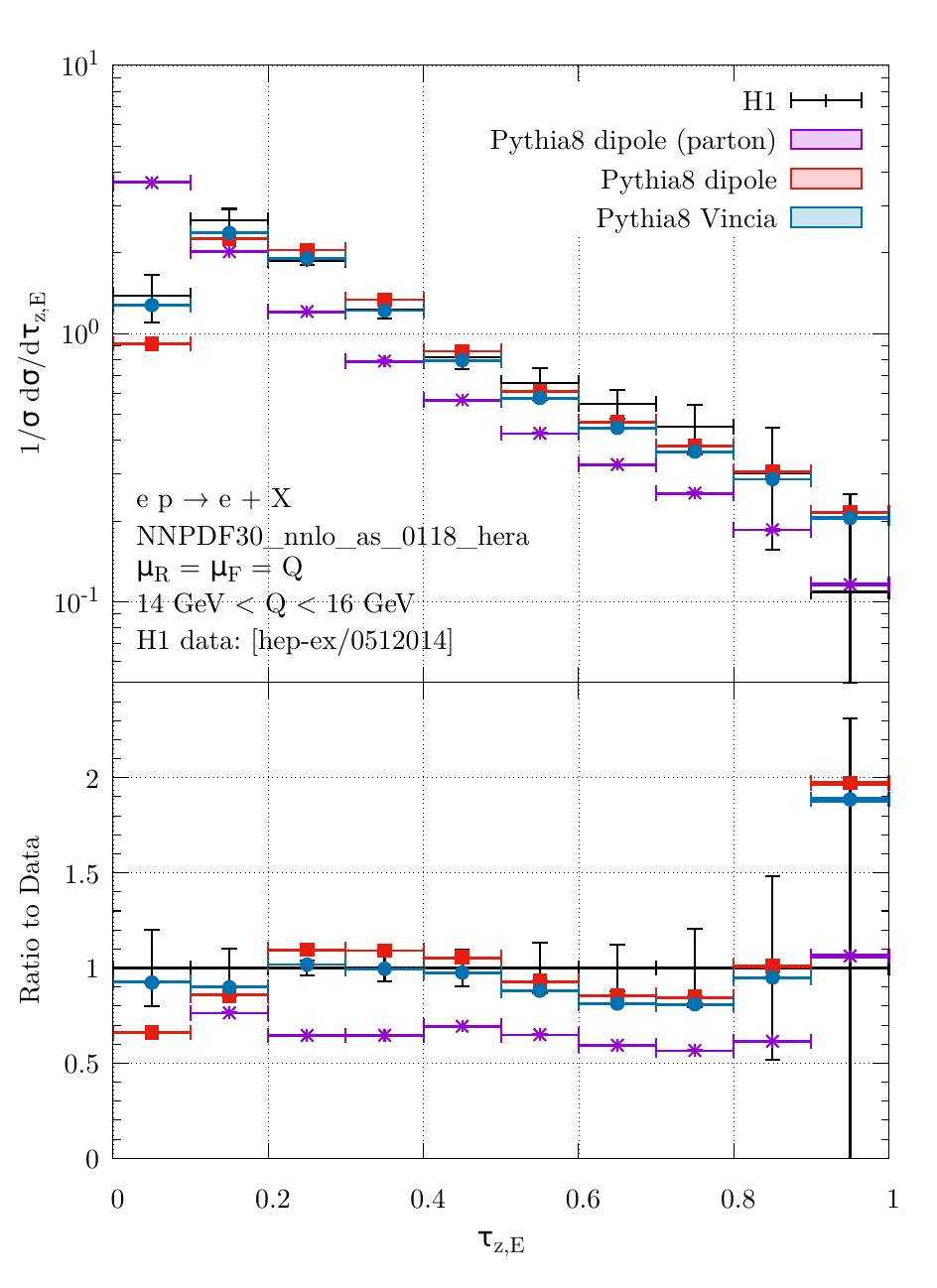}%
  \includegraphics[width=0.33\textwidth,page=5]{fig-temp/h1_plots.pdf}%
  \includegraphics[width=0.33\textwidth,page=9]{fig-temp/h1_plots.pdf}
  \caption{
Thrust distribution at the hadron level for the dipole (red), and {\sc Vincia} (blue) showers,
      and at the parton level (i.e. without hadronisation and beam-remnant
      effects) for the dipole shower (magenta), together with the H1
      data of \cite{H1:2005zsk},
    for the bins $14~\text{GeV}< Q< 16$~GeV (left),
    $30~\text{GeV}< Q< 50$~GeV (middle), and $70~\text{GeV}< Q<
    100$~GeV (right).  The bands represent the 7-point scale variation
    of $\mu_R$ and $\mu_F$ by a factor of two around the central value
    $Q$ for the \POWHEG results{}. The lower panels show the ratio of the
   predictions to data.}
  \label{fig:tau-band}
\end{figure}
We use the {\tt NNPDF30\_nnlo\_as\_0118\_hera} set already mentioned in Sec.~\ref{sec:validation}. Only events in the range 
\bea
\label{eq:qyrange} 
&14~\text{GeV} < Q < 200~\text{GeV}\,,& \nonumber\\ 
&0.1< \ydis < 0.7\,, &
\eea
are taken into account. 
Furthermore, following the H1 analysis, we accept events where the energy in the current hemisphere exceeds a minimum value $\epsilon_\mathrm{lim}$, according to Eq.~\eqref{eq:elim}. 
As discussed in Sec.~\ref{sec:caesar}, this cutoff ensures the
collinear and infrared safety of event shapes which are normalised
with respect to $E_{\mathrm{curr}}$ (or $\sum_h |\vec{p}_{z,h}|$), by
removing events with no hard, but only arbitrarily soft partons in the
current hemisphere.

Event shape variables constitute a class of quantities particularly
suited to probe the interplay of the hard scattering and the
hadronisation mechanism governing DIS processes (see
e.g.\ Ref.~\cite{Dasgupta:2003iq}). 
Following Ref.~\cite{H1:2005zsk}, we introduce the thrust variable $\tau_{\rm z, E}$ is defined by
\beq
\label{eq:thrust}
\tau_{\rm z, E} = 1 -T_{\rm z, E} \quad
\text{with} \quad 
T_{\rm z, E} = \frac{\sum_h |\vec{p}_{z,h}|}{\sum_h |\vec{p}_{h}|}\,,
\eeq where the summations run over all hadronic objects $h$ in the
current hemisphere, $\vec{p}_{h}$ denotes the Breit-frame
three-momentum of parton $h$ and $\vec{p}_{z,h}$ its component along
the $z$~axis, which is chosen along the direction of the virtual
boson.\footnote{Note that this definition of thrust differs
from the one for $\tau_{\rm z, Q}$ of \refeq{eq:qthrust} used in the
previous section. } The thrust variable is a measure of the momentum
components of the hadronic system parallel to the $z$~axis in the
Breit frame.
For the broadening we use the definition of Eq.~\eqref{eq:ebroadening}.

\begin{figure}[tp]
  \centering
  \includegraphics[width=0.33\textwidth,page=2]{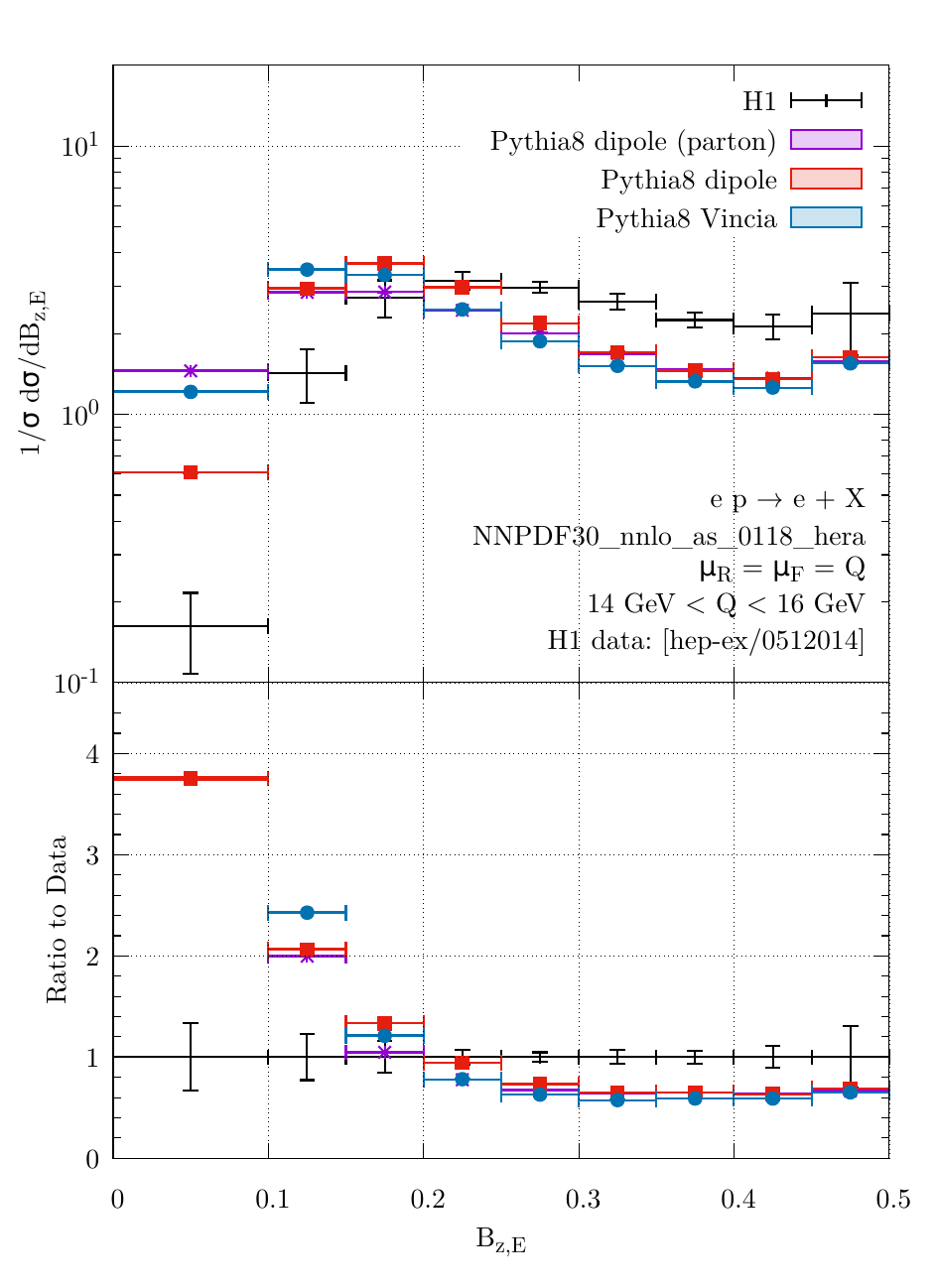}%
  \includegraphics[width=0.33\textwidth,page=6]{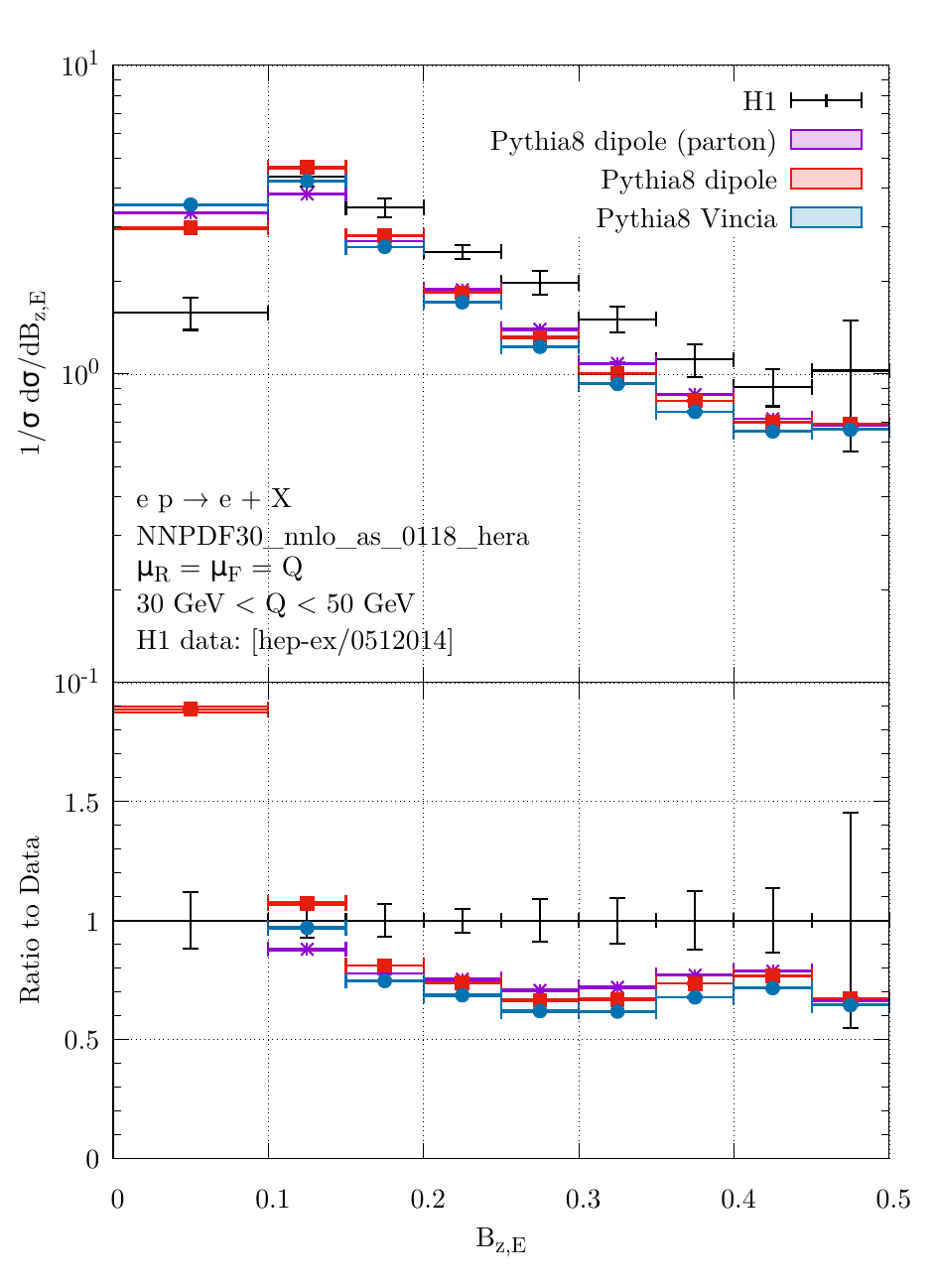}%
  \includegraphics[width=0.33\textwidth,page=10]{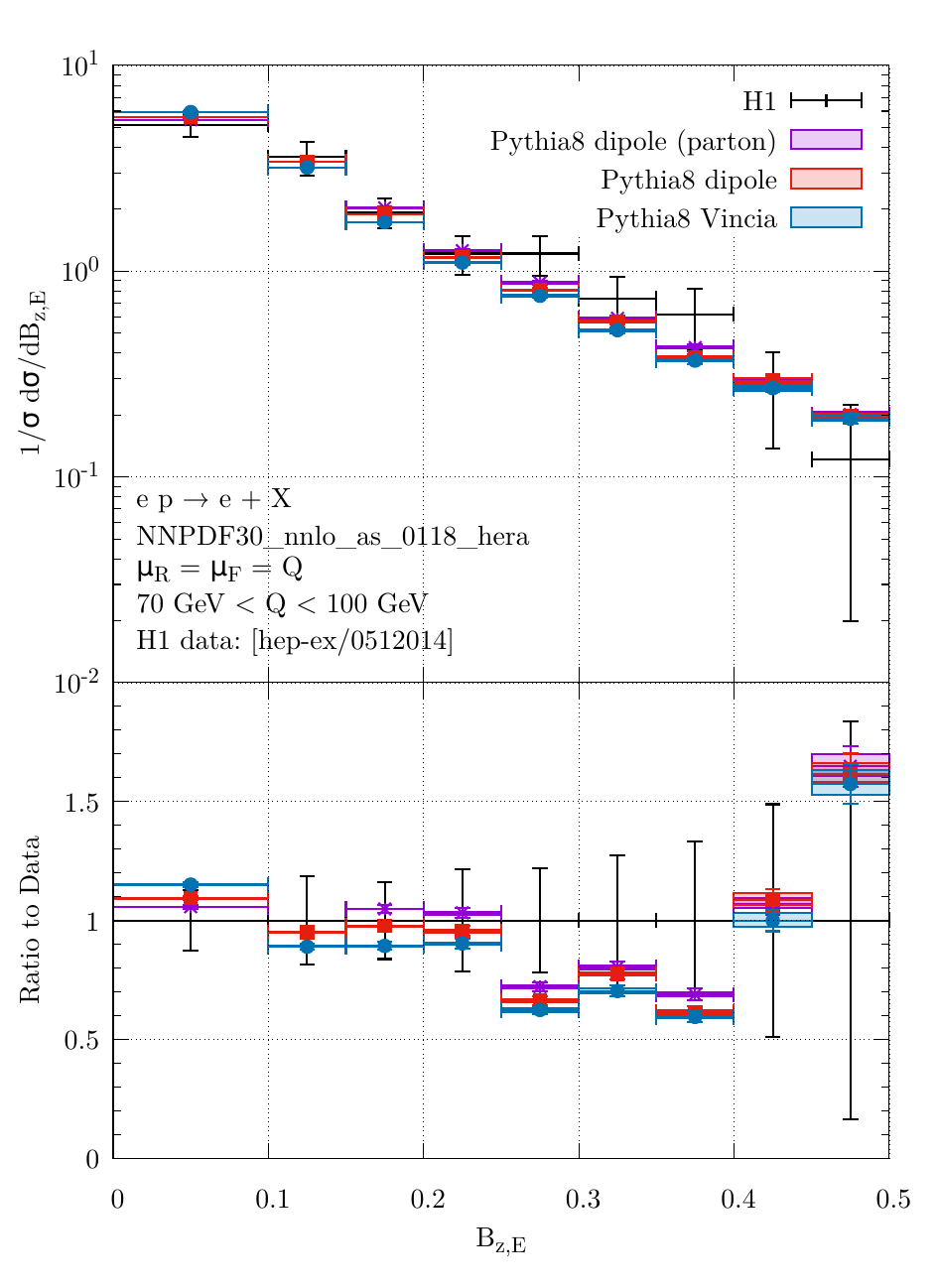}
  \caption{
    Same as Fig.~\ref{fig:tau-band}, but for the broadening distribution.
    }
  \label{fig:broadening-band}
\end{figure}

%
In Fig.~\ref{fig:thrust+broadening} 
we show $\tau_{\rm z, E}$ and $B_{\rm z, E}$, respectively, for
various $Q$~bins at the same time, together with the H1 data of
\cite{H1:2005zsk}.
To assess the impact of soft-physics effects, we also produce
parton-level predictions in which hadronisation and beam-remnant
effects are not included.  We do so only for the dipole shower, as the
modelling of these effects is the same for {\sc Vincia} predictions. 
In Figs.~\ref{fig:tau-band}~and~\ref{fig:broadening-band} we consider
the same event shapes for selected ranges of $Q$ together with the
7-point variation of $\mu_R$ and $\mu_F$.
%
%
%
We select the lowest $Q$-bin, $14~\text{GeV}< Q< 16$~GeV, which is
dominated by photon exchange contributions, one with intermediate
values of $Q$, and one including the value where $Q$ coincides with
the mass of the $Z$~boson.

For $\tau_{\rm z, E}$ we find good
agreement of our hadron-level predictions with H1 data. Especially at low values of $Q$
hadronisation effects are crucial for a reasonable description of
data. As expected, at higher values of $Q$ the impact of these effects becomes less
relevant. While agreement between predictions and data is generally
worse for the broadening, a similar trend as in the thrust
distribution can be observed, with hadronisation effects being
particularly important at low values of $Q$ and differences between
the dipole and the antenna showers being small throughout.  Scale
uncertainties are generally small for both distributions, but largest
towards their respective upper ends, which reflects the relevance of
higher-order corrections in these kinematic regions. The smallness of
the scale variation in these plots can mostly be attributed to the
fact that the plots show normalised distributions.

\begin{figure}[tp]
  \centering
  \includegraphics[width=0.495\textwidth,page=3]{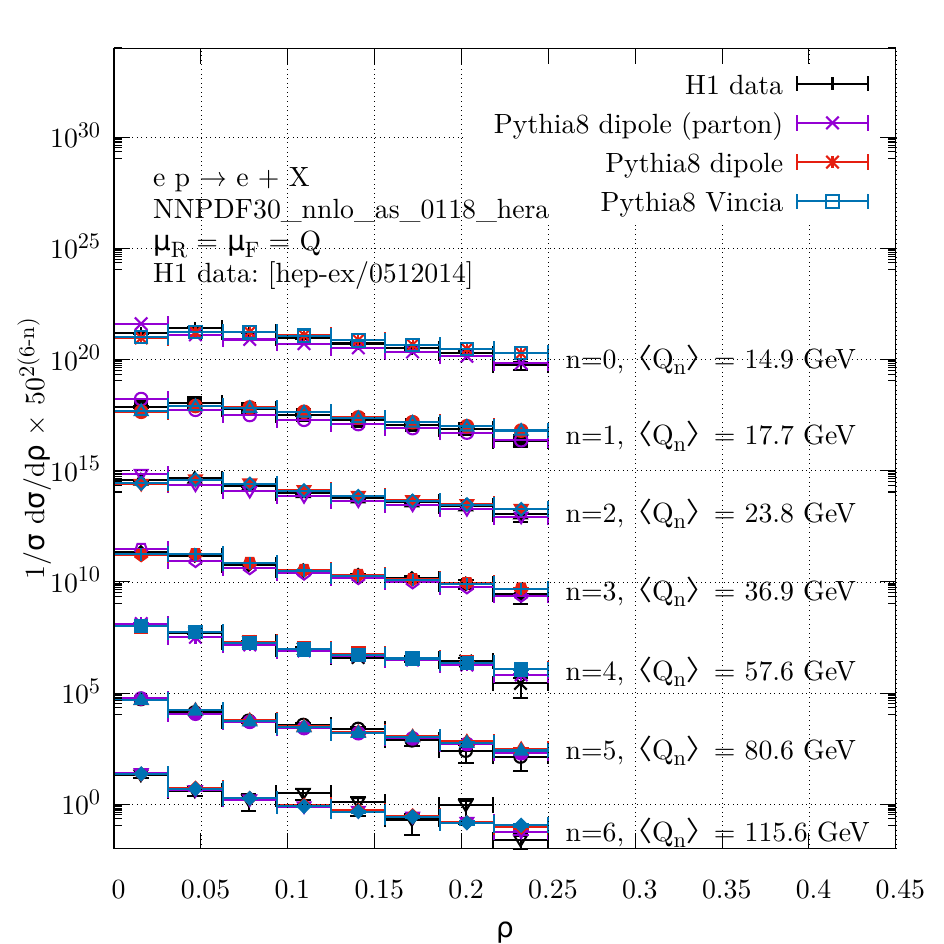}\hfill%
  \includegraphics[width=0.495\textwidth,page=4]{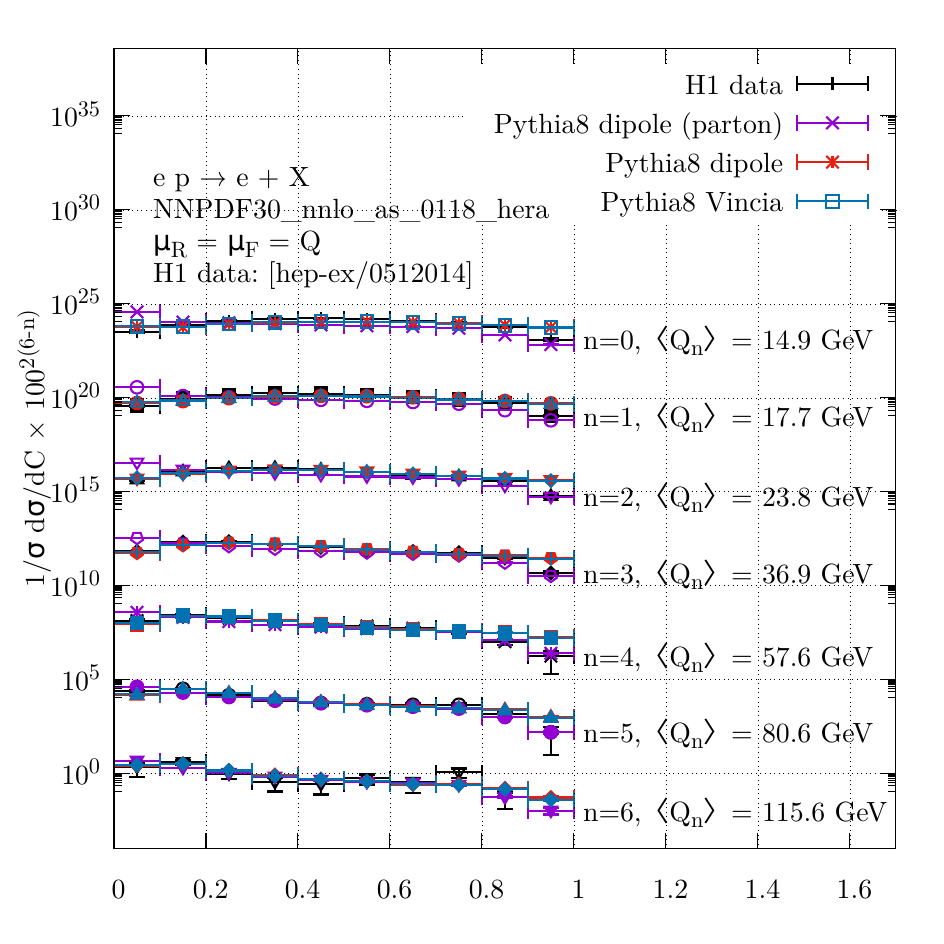} 
  \caption{
      Same as Fig.~\ref{fig:thrust+broadening}, but for the squared jet mass (left) and the
      $C$-parameter (right).
    }
  \label{fig:rho+cpr}
\end{figure}

Similarly to the event shapes of
Eqs.~(\ref{eq:thrust})~and~(\ref{eq:ebroadening}), the squared jet
mass $\rho$ and the $C$-parameter are defined as \beq
\label{eq:rho}
\rho = \frac{\left(\sum_h E_h\right)^2 -\left(\sum_h
  \vec{p}_{h}\right)^2}{\left(2\sum_h |\vec{p}_{h}|\right)^2}\,, \eeq
and \beq
\label{eq:c-param}
C = \frac{3}{2}\frac{\sum_{h,h'} |\vec{p}_{h}|
  |\vec{p}_{h'}|\cos^2\theta_{hh'}}{\left(\sum_h
  |\vec{p}_{h}|\right)^2}\,, \eeq where $h$ and $h'$ are two different
hadronic objects in the current hemisphere separated by an angle
$\theta_{hh'}$.
In Fig~\ref{fig:rho+cpr} 
%
%
%
we show $\rho$ and $C$ for various $Q$~bins at the same time. 
%
%
\begin{figure}[tp]
  \centering
  \includegraphics[width=0.33\textwidth,page=3]{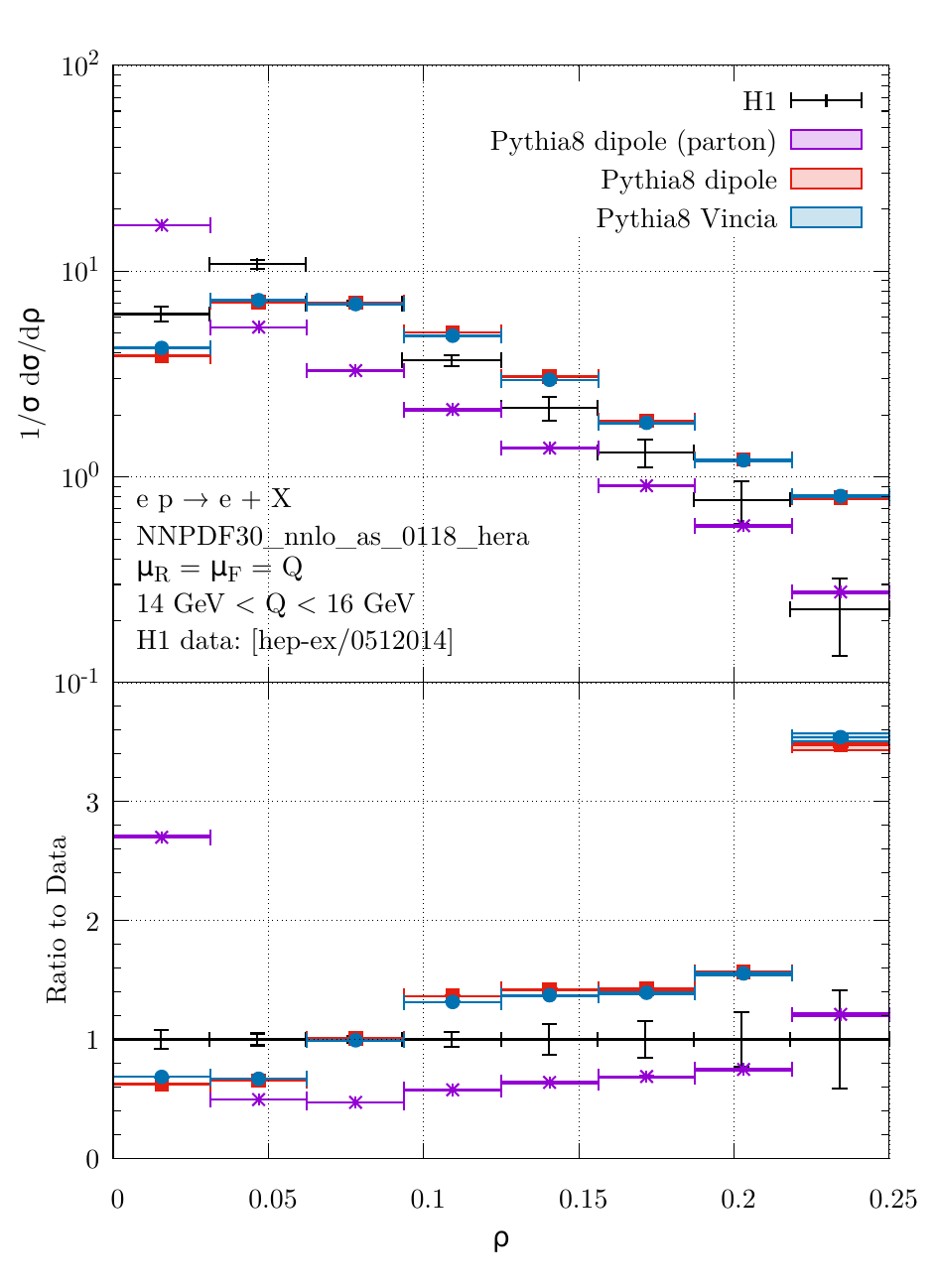}%
  \includegraphics[width=0.33\textwidth,page=7]{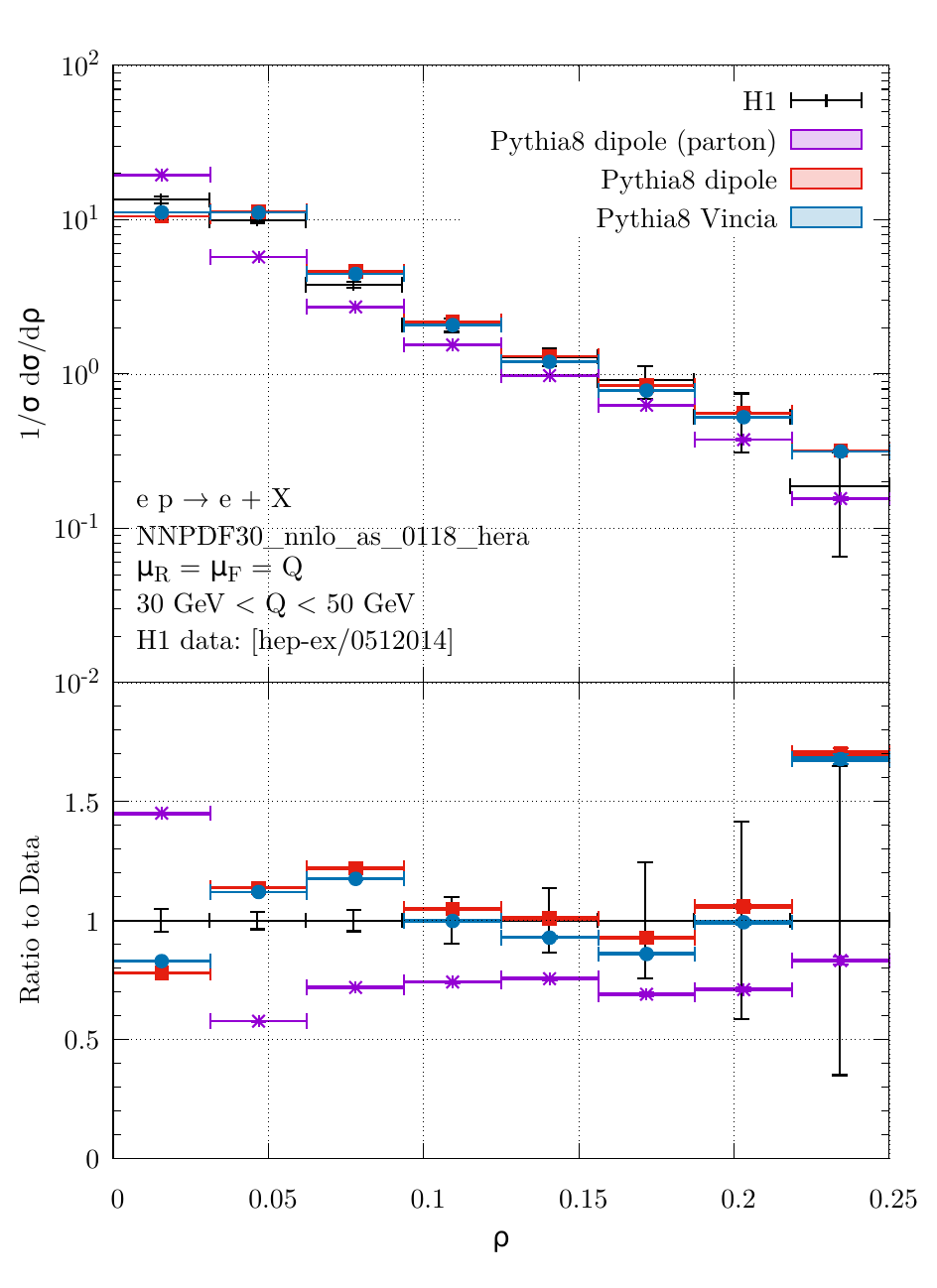}%
  \includegraphics[width=0.33\textwidth,page=11]{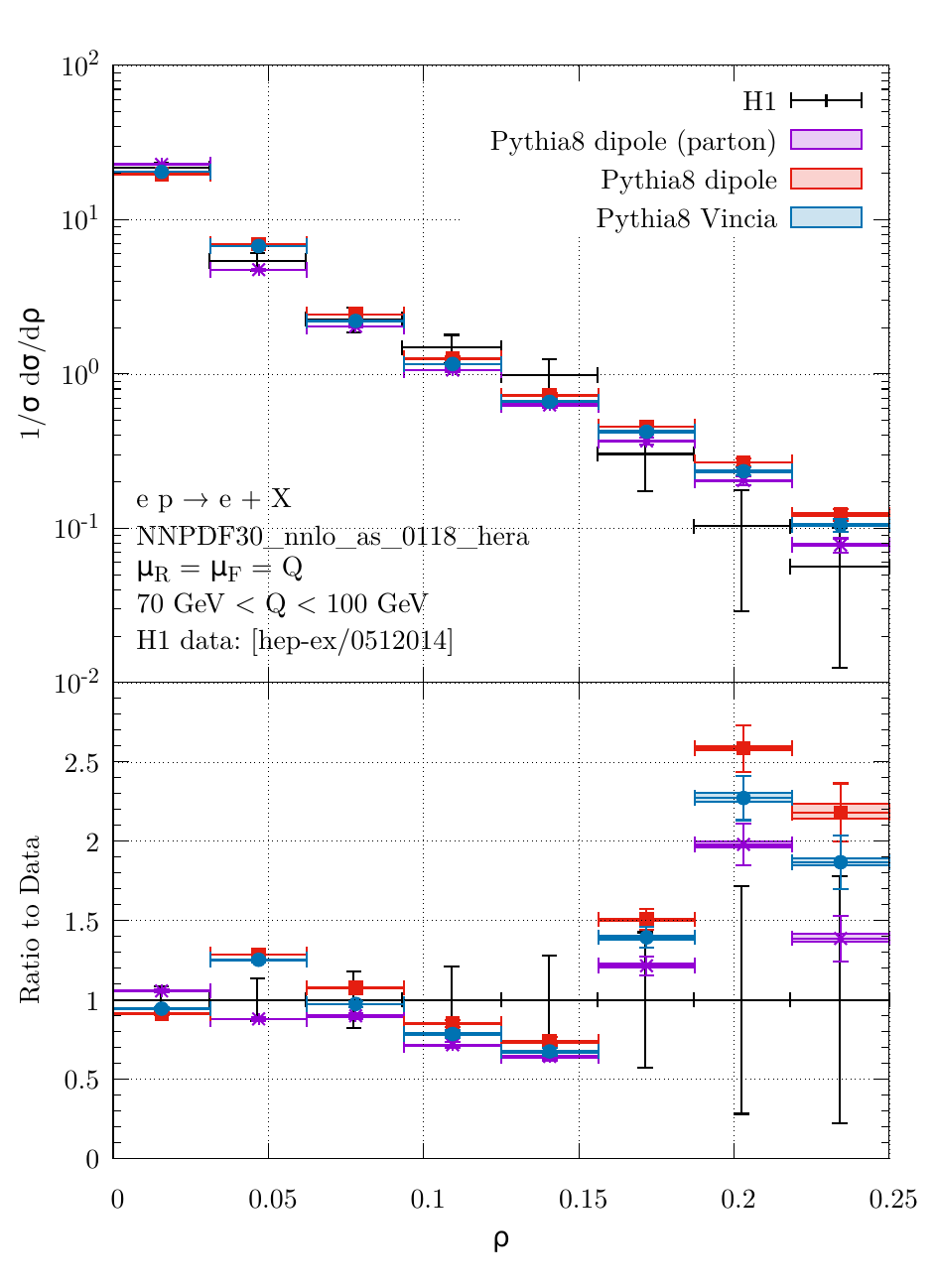}
  \caption{
    Same as Fig.~\ref{fig:tau-band}, but for the squared jet mass.
  }
  \label{fig:rho-band}
\end{figure}
%
%
\begin{figure}[tp]
  \centering
  \includegraphics[width=0.33\textwidth,page=4]{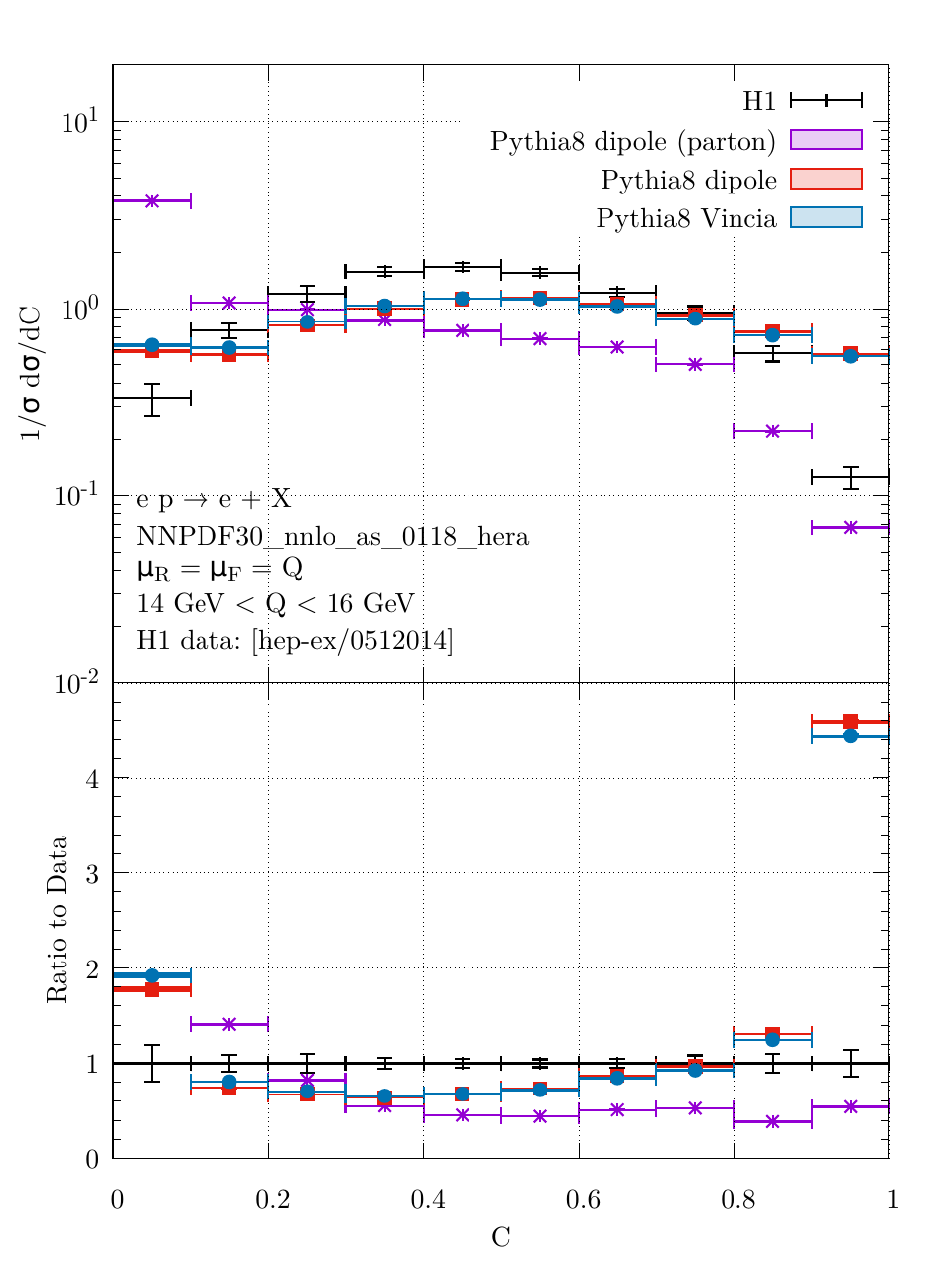}%
  \includegraphics[width=0.33\textwidth,page=8]{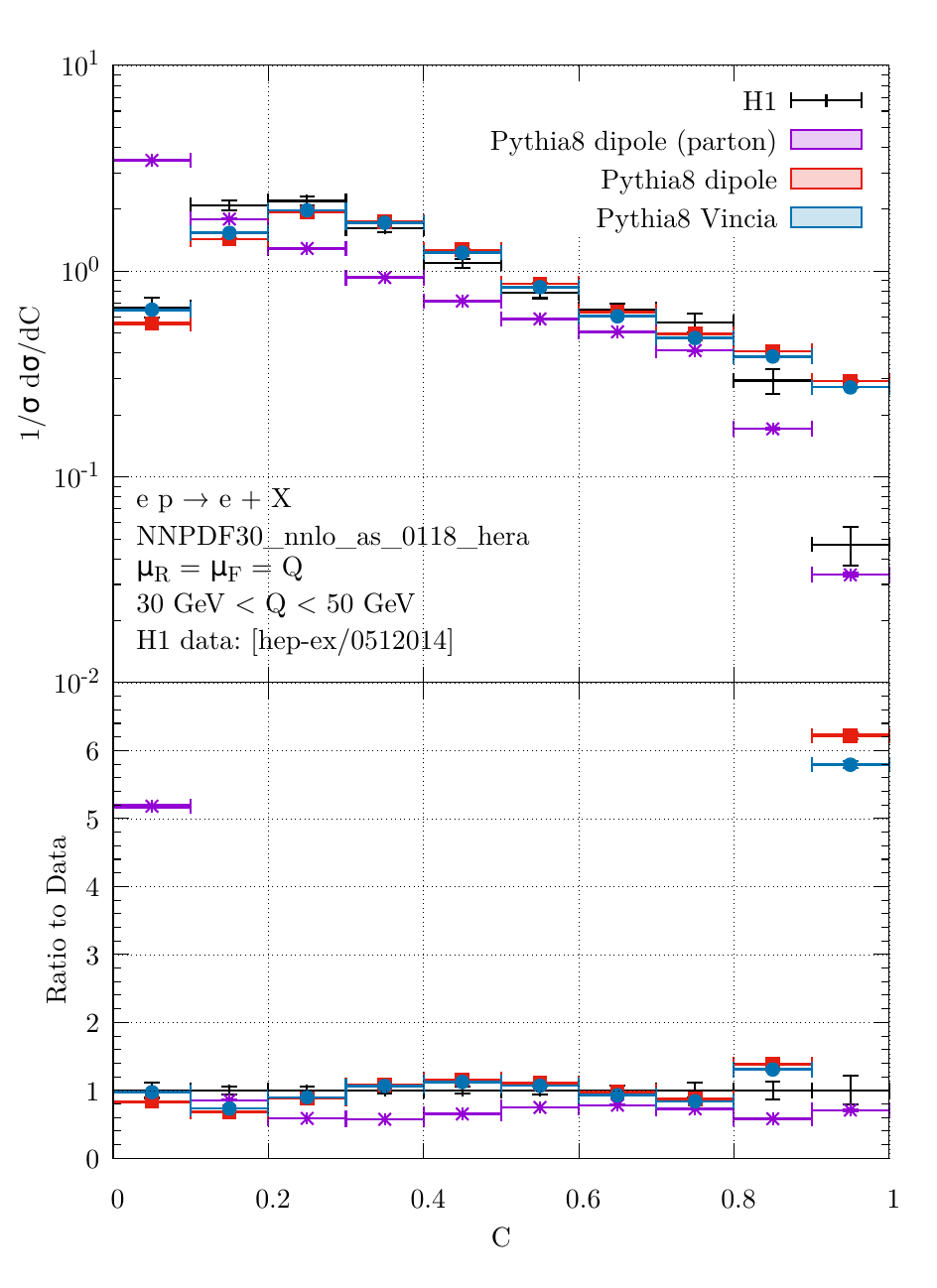}%
  \includegraphics[width=0.33\textwidth,page=12]{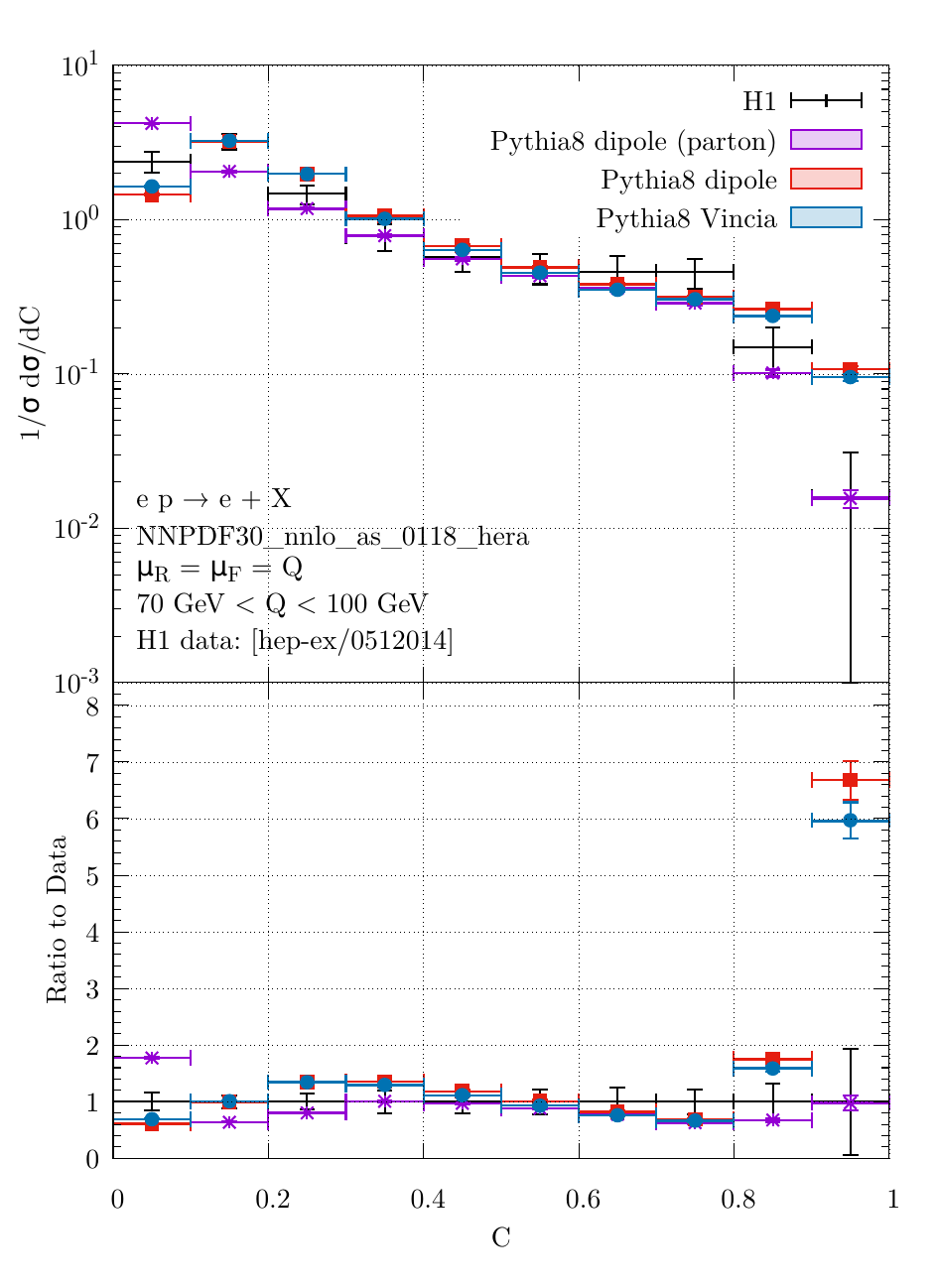}
  \caption{
        Same as Fig.~\ref{fig:tau-band}, but for the $C$-parameter distribution.
  }
    \label{fig:cpr-band}
\end{figure}
%
%
Figs.~\ref{fig:rho-band}~and~\ref{fig:cpr-band} depict the same
distributions for selected bins in $Q$ together with scale uncertainty
bands. 
For $\rho$ and $C$ we observe a similar pattern as in the case of
thrust and broadening. Taking hadronisation effects into account is
crucial for a reasonable agreement between simulation and data. Even
after the inclusion of hadronisation effects, at low values of $Q$ our
predictions for both distributions deviate from data. Better agreement
is found at intermediate values of $Q$. At large values of $Q$ for
most bins predictions agree with data considering the large
statistical uncertainties of the latter.

We note that a large impact of non-perturbative effects has also been
reported in~\cite{Knobbe:2023ehi} for the so-called 1-jettiness
distribution which is closely related to the thrust distribution.

\subsection{Predictions for the EIC}
\label{sec:EIC}
After employing our new \POWHEGBOX{} implementation for the
description of H1 legacy results, we turn to DIS at the future EIC. We
consider electron-proton collisions with $E_e = 18$~GeV,
$E_p=275$~GeV, both in the neutral current (NC) and charged current
(CC) modes with the incoming lepton either remaining intact or being
converted into a neutrino.

Following Ref.~\cite{Borsa:2022cap}, the DIS kinematics are restricted by 
\bea
\label{eq:eic-qy}
&25~\text{GeV}^2 < Q^2 < 1000~\text{GeV}^2\,,& \nonumber\\ 
&0.04< \ydis < 0.95\,.  &
\eea
In contrast to the settings used in the HERA analysis of
Sec.~\ref{sec:hera}, for our EIC predictions we employ the {\tt
  PDF4LHC15\_nlo\_100\_pdfas} parton distribution
set~\cite{Butterworth:2015oua} to account for LHC constraints on the
proton structure.
Jets are reconstructed in the laboratory frame with the anti-$k_T$
algorithm~\cite{Cacciari:2008gp}  using an $R$-parameter of $R=0.8$ and
restrictions on transverse momentum and pseudorapidity,\footnote{At variance with Ref.~\cite{Borsa:2022cap} we use the standard $E$-scheme recombination, rather than the $E_T$ one.} \beq
\label{eq:eic-jets}
\ptj > 5~\text{GeV}\,,\quad
|\etaj| < 3\,. 
\eeq

\begin{figure}[tb!]
  \centering
  \includegraphics[width=0.44\textwidth,page=1]{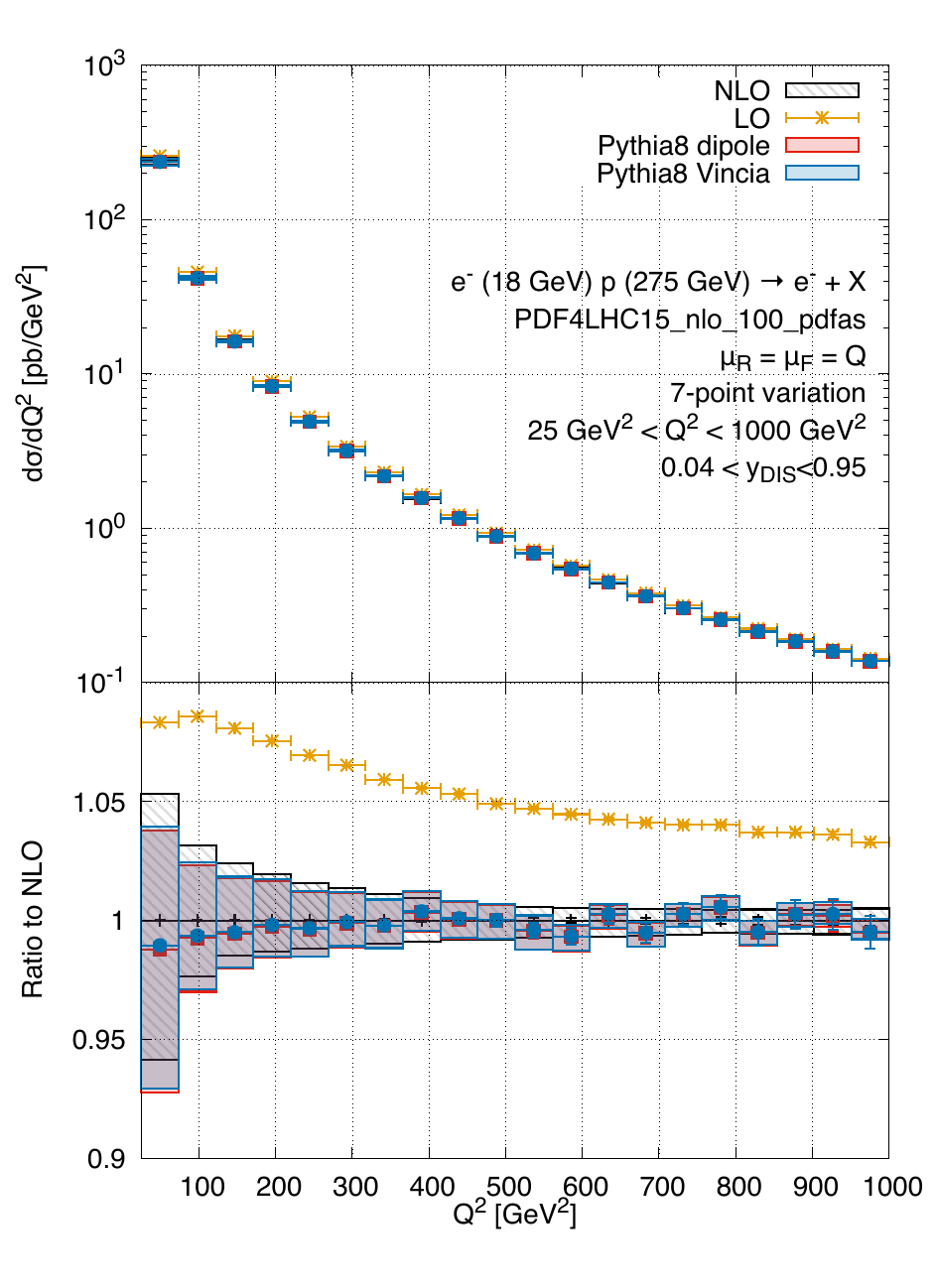}
  \includegraphics[width=0.44\textwidth,page=2]{fig-temp/eic_NC_plots.pdf}
  \caption{Distributions of $Q^2$ (left) and $\xB$ (right) for NC DIS
    at the EIC with $\sqrt{s}=140$~GeV and within the cuts of
    Eq.~(\ref{eq:eic-qy}) at LO (orange), NLO (magenta), and
    NLO+PS results, obtained  with dipole shower (red) or {\sc Vincia} 
    (blue) \PYTHIAE{} showers. Hadronisation and beam remnant effects are included in the NLO+PS simulations.
    Error bars indicate statistical uncertainties, bands are
    obtained by a 7-point scale variation of $\mu_R$ and $\mu_F$ by a
    factor of two around the central value $Q$.  The lower panels
    display the ratios to the respective NLO results. 
}
  \label{fig:eic-nc-qx}
\end{figure}

Fig.~\ref{fig:eic-nc-qx} 
%
%
%
%
displays the $Q^2$ and $\xB$~distributions of the NC cross section
within the cuts of Eq.~(\ref{eq:eic-qy}) at LO, NLO, and NLO+PS
accuracy (with the inclusion of hadronisation and beam remnant
effects) for two different shower versions.  The NLO corrections
change the LO results in a non-uniform way, slightly shifting the
$Q^2$~distribution to larger values. Also the shape of the
$\xB$~distribution is modified by NLO corrections with a
tendency to smaller $\xB$~values at LO.
Both the {\sc Vincia} and the dipole showers preserve the lepton
kinematics, hence the NLO+PS results agree with the NLO
result.\footnote{The 1\% difference at very small $Q^2$ values is
induced by the reshuffling procedure that \POWHEG{} applies to
introduce heavy-quark mass effects in the momenta that are written in
the LHE files, and is not related to the parton shower or
non-perturbative effects.
}
For the NLO+PS results, we also performed a 7-point scale variation,
modifying the renormalisation and factorisation scales independently
by factors of two around their central value $Q$.

\begin{figure}[tb!]
  \centering
  \includegraphics[width=0.42\textwidth,page=3]{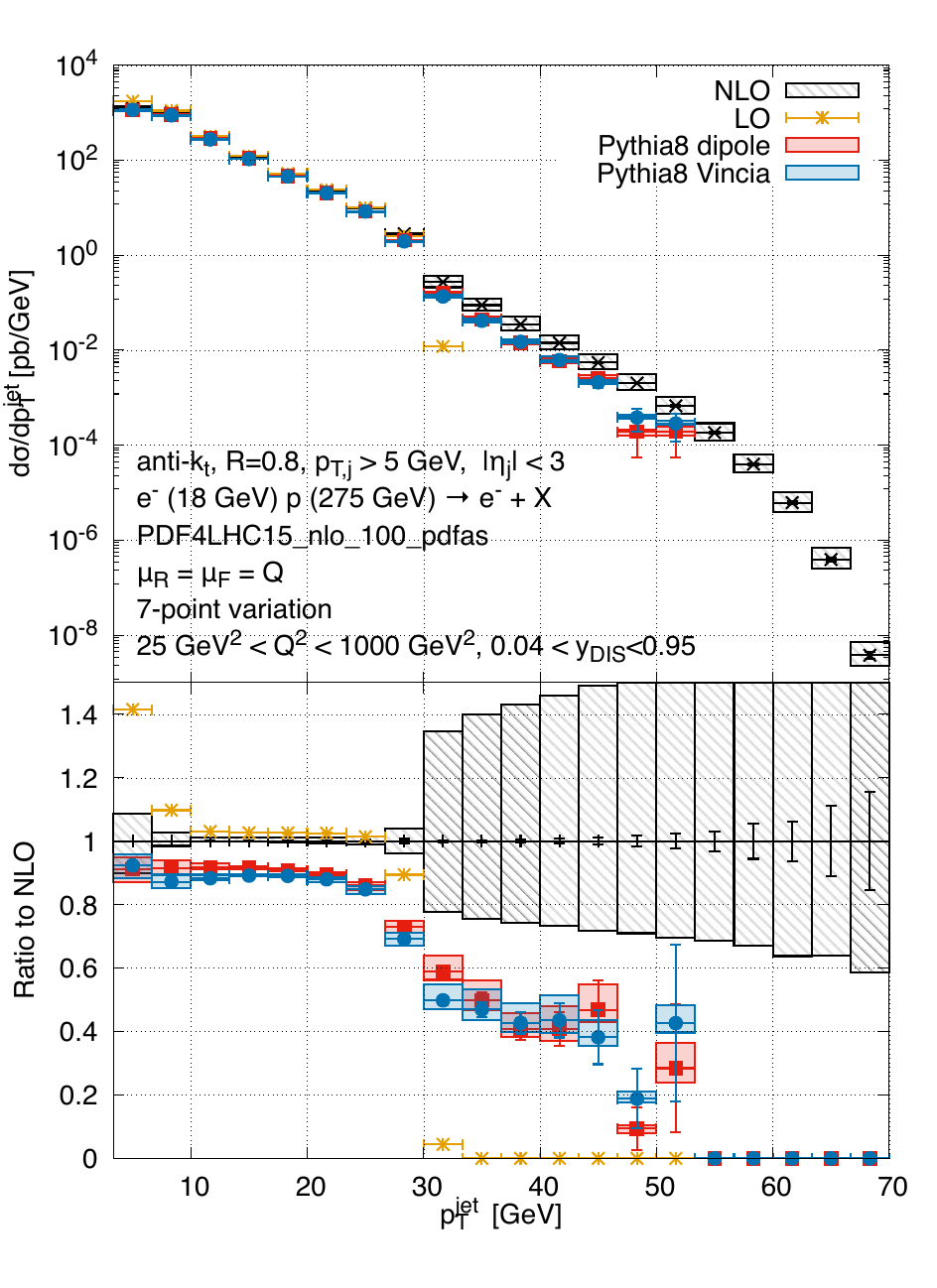}
  \includegraphics[width=0.42\textwidth,page=5]{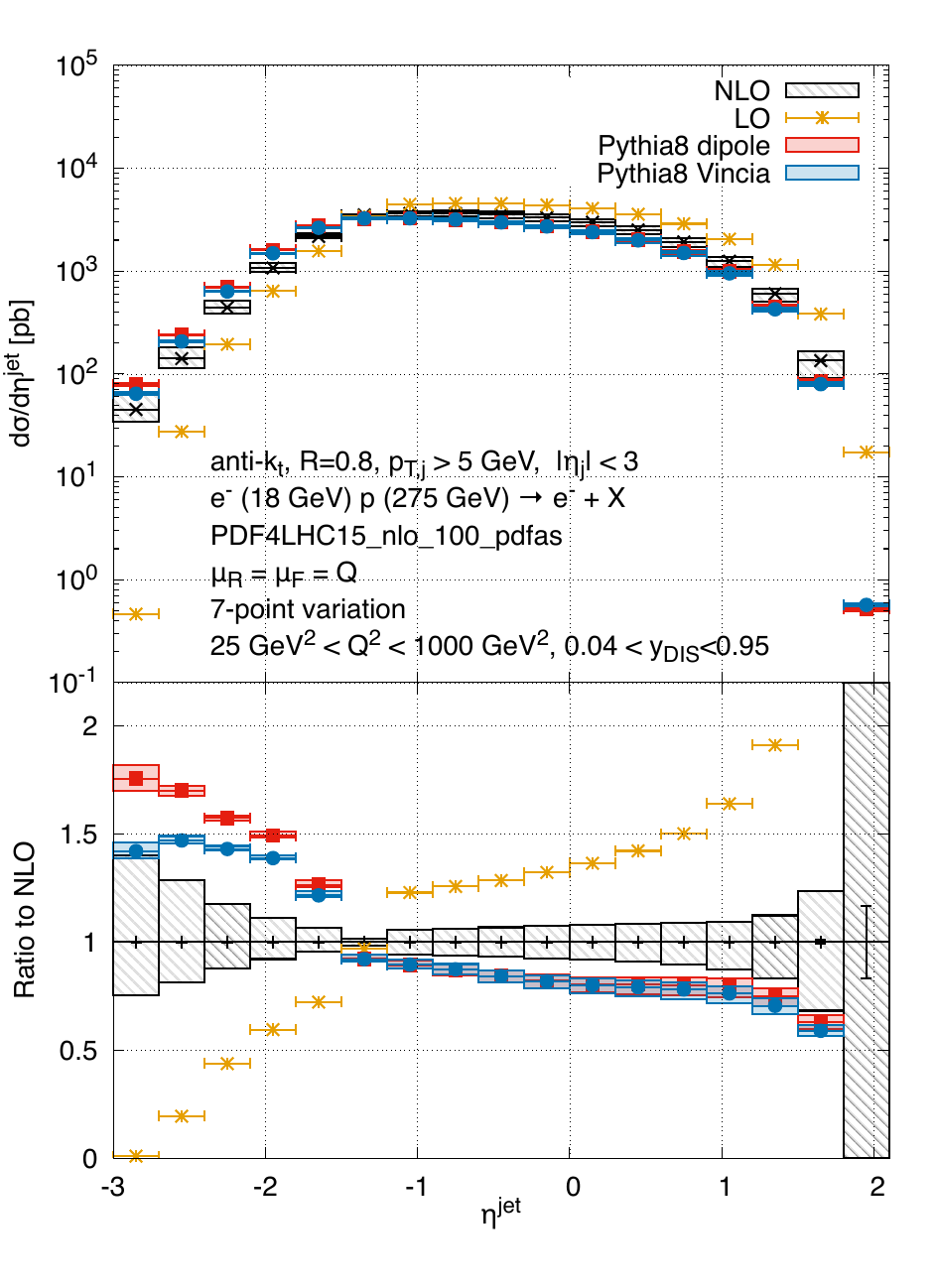}
  \caption{Distributions of $\ptj$ (left) and $\etaj$ (right) for NC
    DIS at the EIC with $\sqrt{s}=140$~GeV and within the cuts of
    Eqs.~(\ref{eq:eic-jets})--(\ref{eq:eic-qy}).}
  \label{fig:eic-nc-jet}
\end{figure}

In Fig.~\ref{fig:eic-nc-jet} 
%
%
%
we display the transverse-momentum and pseudorapidity distributions of the hardest jet reconstructed with cuts of Eqs.~(\ref{eq:eic-jets})--(\ref{eq:eic-qy}).
We remind the reader that these quantities are defined in the lab frame, hence they are non-vanishing already at LO.  
At LO, the accessible range of transverse momentum is limited by the upper limit on $Q^2$ of 1000~GeV$^2$, $\ptj < Q \sim 32$ GeV. 
Beyond LO, the transverse momentum available for the hadronic system can instead be distributed among various final-state partons resulting in non-vanishing contributions to the respective cross sections beyond this threshold. Here NLO predictions are effectively leading order accurate, which is reflected by the larger scale uncertainly bands. 
We observe that the NLO corrections considerably reduce the $\ptj$~distribution at low values, and the parton shower slightly enhances that effect. The shape of the pseudorapidity distribution is modified by NLO corrections in an asymmetric way with largest effects at high values of $|\etaj|$. In this range an additional, though smaller shape distortion is caused by the parton shower. 
The impact of the shower is large in kinematic regions that are not accessible at LO, but require the presence of additional radiation, such as the large transverse-momentum region, or for very negative values of $\etaj$. In particular, we find that the dipole and the {\sc Vincia} shower agree remarkably well with each other, except for $\etaj \lesssim -2$, where differences between the two shower models reach 10-15$\%$.
Scale uncertainties are generally smaller than differences between fixed-order and NLO+PS results.

\begin{figure}[tp]
  \centering
  \includegraphics[width=0.42\textwidth,page=1]{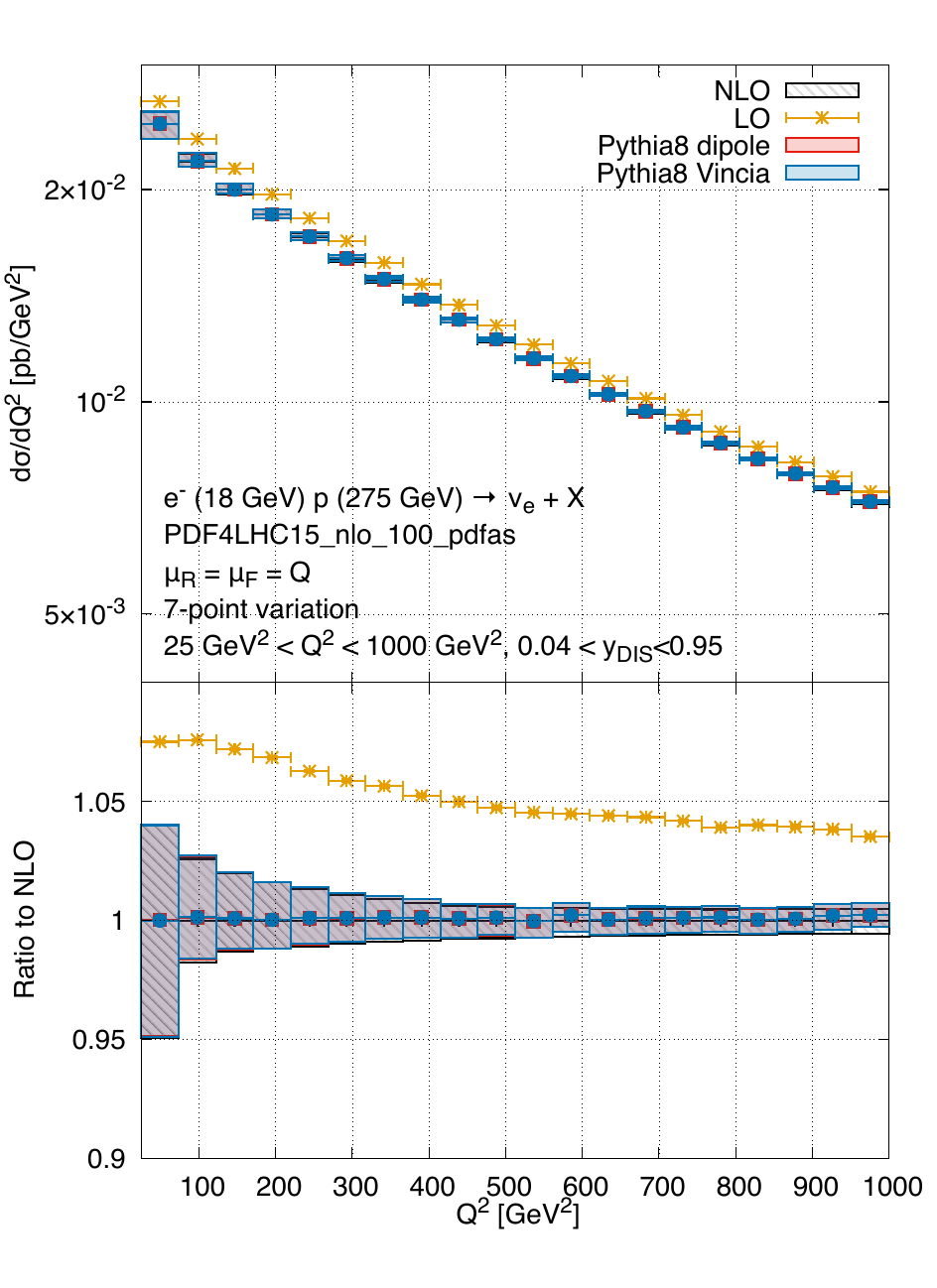}
  \includegraphics[width=0.42\textwidth,page=2]{fig-temp/eic_CC_plots.pdf}
  \caption{Same as Fig.~\ref{fig:eic-nc-qx}, but for the charged current channel. }
  \label{fig:eic-cc-qx}
\end{figure}

We now consider the CC case.
The main difference between the NC and
the CC processes is that the former can proceed via the exchange of a
virtual photon or a $Z$~boson between the scattering electron and
proton, and so is divergent for small $Q^2$ or $\ptj$ values, while the
CC cross section is entirely due to weak boson exchange contributions,
which leads to a finite cross section also for vanishing $Q^2$.
Nonetheless, in the most characteristic distributions, radiative
corrections display similar features as in the NC case.
The $Q^2$~and $\xB$~distributions depicted in Fig.~\ref{fig:eic-cc-qx} 
%
%
exhibit negative NLO corrections of  
about 5\% to 10\% over the entire range of $Q^2$. At small values of $\xB$ the NLO corrections are small and negative, while they become positive beyond $\xB\approx 0.3$ and reach values of almost 25\% at large $\xB$.  Parton-shower effects are small in each case.   

\begin{figure}[tp]
  \centering
  \includegraphics[width=0.42\textwidth,page=3]{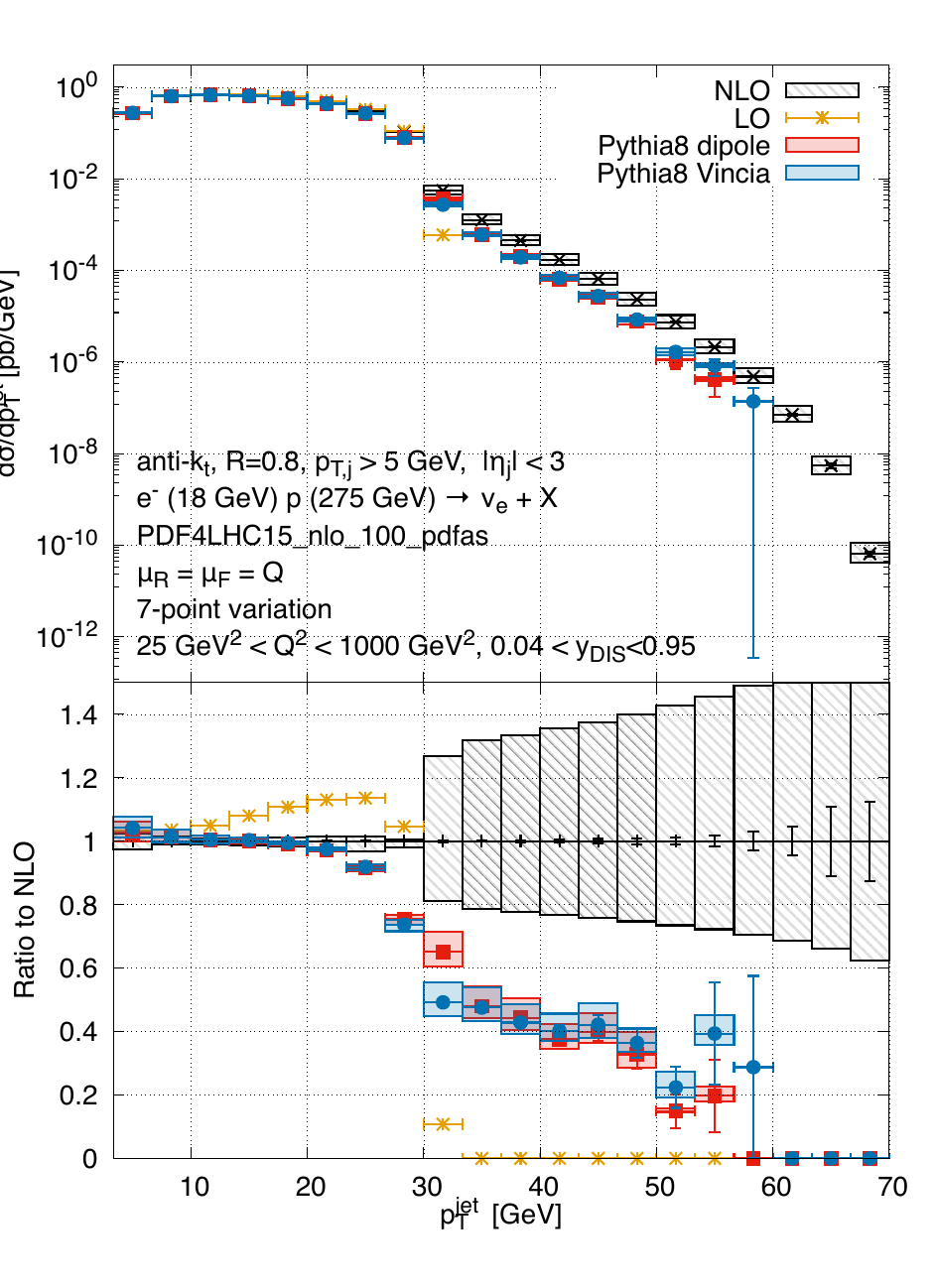}
  \includegraphics[width=0.42\textwidth,page=5]{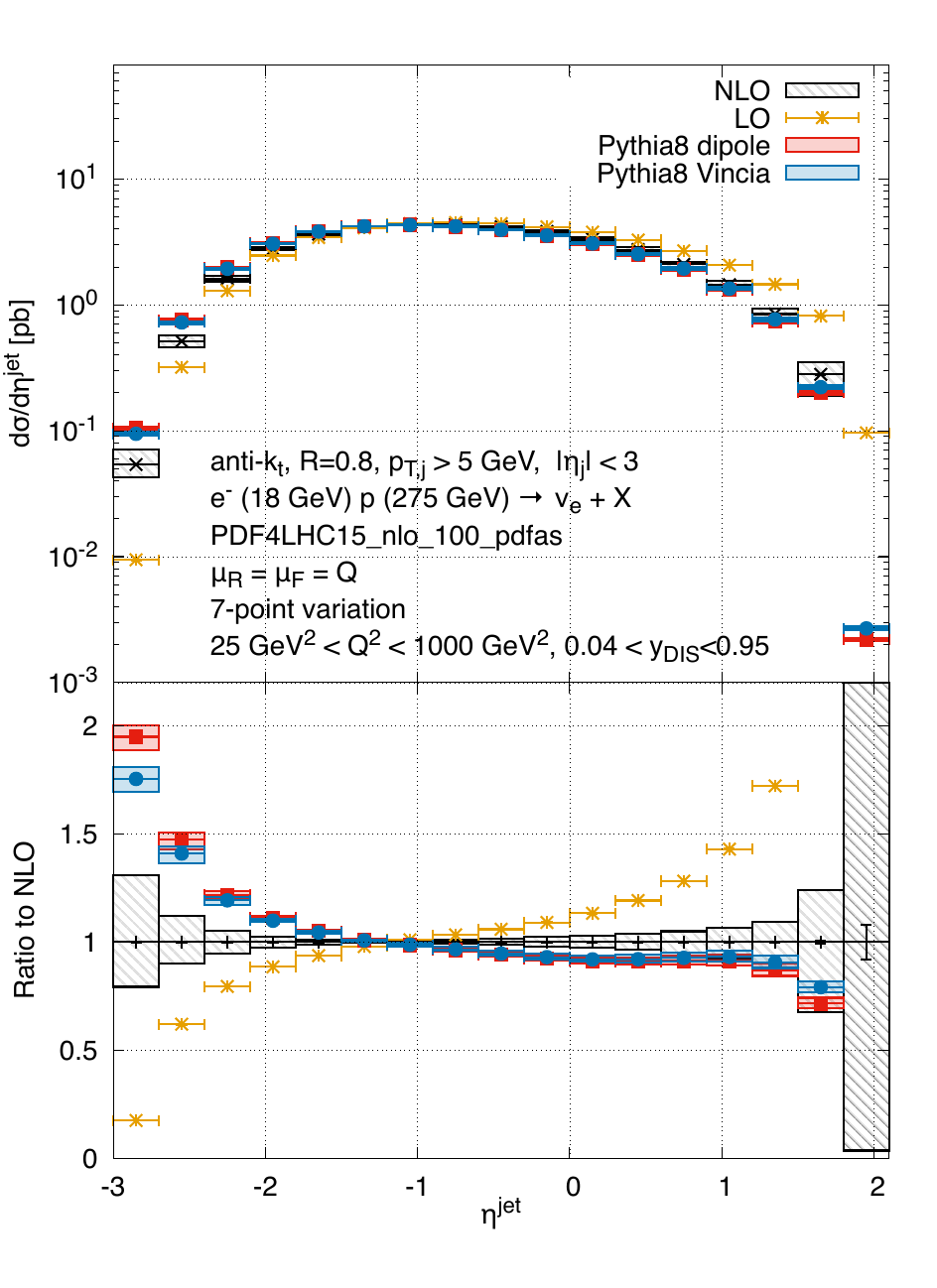}
  \caption{Same as Fig.~\ref{fig:eic-nc-jet}, but for the charged current channel. }
  \label{fig:eic-cc-jet}
\end{figure}

The transverse-momentum and pseudorapidity distributions illustrated in Fig.~\ref{fig:eic-cc-jet} 
%
%
turn out to be less sensitive to NLO corrections at low $\ptj$ in the
CC than in the NC case, but receive small negative NLO corrections at
intermediate transverse momenta. The different behaviour of this jet
distribution at low $\ptj$ can be traced back to the presence of
photon-exchange contributions in the NC case.
The pseudorapidity distribution, which is most sensitive to
perturbative corrections at large values of $|\etaj|$ where the cross
section itself is small, exhibits a similar behaviour as in the NC
case.

\section{Summary and conclusions}
\label{sec:summary}
In this paper we have presented the first implementation of an NLO+PS event generator for DIS in the \POWHEGBOX{}. 
The code will be made publically available there. While the
\POWHEGBOX{} allows for almost automated generation of hadron-hadron
collisions, the kinematics of the DIS process required us to address a
number of problems.

In particular we had to modify the FKS momentum mappings that are used
in the \POWHEGBOX{}, in order to preserve both the incoming and
outgoing lepton momenta. The standard ISR map in the \POWHEGBOX{} is
such that it modifies the kinematics of both the incoming
legs. Although the FSR map does not modify the incoming energy
fractions, it does not preserve the DIS variables, $\xdis$, $\ydis$,
and $Q$. We have shown that if one makes only minimal modifications to
these mappings, such that the ISR map conserves the incoming lepton
momentum, but the FSR maps is untouched, the resulting \POWHEG{}
generator substantially modifies even very inclusive distributions,
even outside the NLO scale uncertainty. On the other hand, with our
new and significantly different mappings, which do preserve DIS
kinematics, NLO accuracy is numerically retained.
Since the momentum mapping introduced preserves the DIS variables, it
is possible to be fully differential in $\xB$ and $Q^2$ or to consider
specific ranges in $\xB$ and $Q^2$.

We have presented several phenomenological studies. Firstly, we
compared our results to event-shape distributions measured by
H1. Overall, we observe a reasonable agreement, but for certain event
shapes, there are discrepancies between the shapes of our theoretical
predictions and the data. This is not unexpected as event shapes are
described only at LO+PS in our generator, starting at ${\cal
  O}(\alpha_s)$.
In the future, we plan to extend the description of DIS processes in
\POWHEG{} to include DIS + one jet. This extension would allow us to
achieve NLO accuracy for both inclusive and one-jet quantities within
the \MINLO{} framework. By doing so, we aim to improve the precision
and reliability of our predictions for a broader range of observables
in DIS and in particular to achieve NLO accuracy both for inclusive
and one-jet quantities.

We then considered a possible future setup at the future EIC. We find
that NLO corrections can be important and must be included to have an
accurate description of this process. This is the case both for the
$Q^2$ and $x_B$ dependence of the inclusive cross section, as well
as for the transverse momentum and rapidity distribution of the
leading jet, where NLO corrections give rise to sizable shape
difference compared to LO and parton shower effects are also
important.

Although the present study focused on DIS, our work has implications
also for LHC processes which involve the exchange of colourless
particles in the $t$-channel like VBF and single top production. We
leave it for future work to investigate the impact of the new momentum
mappings in these processes.
Very recently a family of NLL-accurate parton showers for DIS and VBF was
presented in Ref.~\cite{vanBeekveld:2023lfu}. It will be also interesting
in the future to investigate the matching of these showers to our
\POWHEG{} generator. Our code can be downloaded from the following SVN repository: \\ \begin{center} {\tt svn://powhegbox.mib.infn.it/trunk/User-Processes-RES/DIS} \end{center}

\begin{acknowledgments}
AB and FR would like to thank the University of Oxford and the Rudolf
Peierls Center for Theoretical Physics for hospitality while part of
this work was carried out. FR would also like to thank the Theory
Department at CERN for hospitality. Part of this work was carried out
while AK and SFR were supported by the European Research Council (ERC)
under the European Union’s Horizon 2020 research and innovation
programme (grant agreement No.\ 788223, PanScales).  The work of BJ
and FR was supported by the German Research Foundation (DFG) through
the Research Unit FOR 2926. They furthermore acknowledge support by
the state of Baden-W\"urttemberg through bwHPC and the DFG through
grant no.~INST 39/963-1 FUGG.
The work of AB has been supported by the Science Technology and
Facilities Council (STFC) under grant number ST/T00102X/1. The
fixed-order calculations matched to NLL resummations were performed
using the Cambridge Service for Data Driven Discovery (CSD3), part of
which is operated by the University of Cambridge Research Computing on
behalf of the STFC DiRAC HPC Facility (\url{www.dirac.ac.uk}). The
DiRAC component of CSD3 was funded by BEIS capital funding via STFC
capital grants ST/P002307/1 and ST/R002452/1 and STFC operations grant
ST/R00689X/1. DiRAC is part of the UK National e-Infrastructure. SFR
and AK acknowledge the use of computing resources made available by
CERN.

\end{acknowledgments}

\appendix
\section{Phase-space parameterisation}
\label{app:ps}
We consider the LO process $\ell(\bar{k}_i), q(\bar{p}_i) \rightarrow \ell(\bar{k}_f), q(\bar{p}_f)$. In the centre-of-mass frame, the momenta of the particles can be explicitly written as
\begin{align}
  \bar{k}_i =& \frac{\sqrt{\bar{s}}}{2}\left(1,0,0,+1\right)\\
  \bar{p}_i =& \frac{\sqrt{\bar{s}}}{2}\left(1,0,0,-1\right)\\
  \bar{k}_f =& \frac{\sqrt{\bar{s}}}{2}\left(1,+\sin\bar\phi\sin\bar\theta,+\cos\bar\phi\sin\bar\theta,+\cos\bar\theta\right)\\
  \bar{p}_f =& \frac{\sqrt{\bar{s}}}{2}\left(1,-\sin\bar\phi\sin\bar\theta,-\cos\bar\phi\sin\bar\theta,-\cos\bar\theta\right),
\end{align}
where $\bar{s}=( \bar{k}_i + \bar{p}_i)^2$ is the partonic centre-of-mass energy. 
Introducing $\ydis=(1+\cos\bar\theta)/2$ and $\xdis = \bar{s}/S$, being $S=(P+\bar k_i)^2$ and $P$ the incoming proton momentum, it is easy to see that the LO phase space can be written as
\begin{equation}
  \int \dd\bar{\Phi}_2 = \int \dd x \frac{\dd^3 \vb k_f}{2k_f^0 (2\pi)^3} \frac{\dd^3 \vb p_f}{2p_f^0 (2\pi)^3} (2\pi)^4 \delta^{(4)}\qty(k_i + p_i - k_f - p_f )=
  \int \frac{\dd x \dd \ydis \dd \bar\phi}{16\pi^2}.
  \label{eq:phi2}
\end{equation}
If we consider the emission of an extra parton with momentum $p_r$, the phase space $\dd\Phi_3$ is given by 
\begin{align}
  \int
  \dd\Phi_3 = \int
  \dd x\, \dd\phi_3 = \int
  \dd x \frac{\dd^3 \vb k_f}{2k_f^0 (2\pi)^3} \frac{\dd^3 \vb p_f}{2p_f^0 (2\pi)^3} \frac{\dd^3 \vb p_r}{2p_r^0 (2\pi)^3} (2\pi)^4 \delta^{(4)}\qty(k_i + p_i - k_f - p_f - p_r),
	\label{eq:ps3}
\end{align}
where $x$ is the longitudinal momentum fraction of the incoming parton and
$\dd\phi_3$ is the final state three particle phase space.
We use $p$ and $k$ to denote the recoiled momenta, while $\bar{p}$ and $\bar{k}$
are employed for the underlying Born kinematics. 
Bold-face notation is used for three-momenta. 

Like for the LO case, in the partonic centre-of-mass frame we can write for the incoming partons
\begin{align}
\label{eq:kp-cms}
	k_i &= \frac{\sqrt{s}}{2} \qty(1,0,0,+1)\,,\\
	p_i &= \frac{\sqrt{s}}{2} \qty(1,0,0,-1)\,,
\end{align}
where $s=(k_i+p_i)^2= x S$. 
The explicit parametrisation of the three final-state particles will differ in case of ISR and FSR.


\subsection{Phase-space parameterisation for initial-state radiation}
In order to evaluate the phase-space of Eq.~(\ref{eq:ps3}) for the
case of ISR we parameterisation the momentum of the radiated parton in
terms of the FKS variables~\cite{Frixione:1995ms} $\xi, y$ and $\phi$,
as
\begin{align}
  p_r &= \xi \frac{\sqrt{s}}{2}\qty(1,\sqrt{1-y^2} \cos\phi, \sqrt{1-y^2} \sin\phi, y)\,,
  \label{eq:pr}
\end{align}
while the momentum of the final-state lepton reads
\begin{align}
  k_f &= \xi_k \frac{\sqrt{s}}{2}\qty(1,\sqrt{1-y_k^2} \cos\bar \phi, \sqrt{1-y_k^2} \sin\bar \phi, y_k)\,.   
\label{eq:kf}
\end{align}
Momentum conservation implies that the momentum of the other final-state parton is $p_f = p_i+k_i-p_r-k_f$.
In terms of these variables one has $\dd^3\vb p_r = dp_r^0 (p_r^0)^2
d\phi dy$ and $\dd^3\vb k_f = dk_f^0 (k_f^0)^2
d\bar \phi d\ck$.
After performing the integration over $\dd^3 \vb p_f$ 
one obtains for the phase-space integral of Eq.~(\ref{eq:ps3})
\begin{align}
  \int\dd\Phi_3 &= \frac{1}{8 (2\pi)^5} \int \dd x \; \dd k_f^0 \; \dd \bar \phi \; \dd \ck \; \dd p_r^0 \; \dd \phi \; \dd y \; \frac{k_f^0 p_r^0}{p_f^0} \delta\qty(k_i^0 + p_i^0 - k_f^0 - p_f^0 - p_r^0),
  \label{eq:ps3a}
\end{align}
where  the final-state parton's energy is fixed to $p_f^0 = \sqrt{\qty(-\vb k_f - \vb p_r)^2}$.

In terms of the DIS variables $\ydis$, introduced in Eq.~\eqref{eq:dis-variables}, and $Q^2=-(k_i-k_f)^2$  one has 
\beq
	\ck = \frac{s (1- \ydis) -Q^2}{s (1 - \ydis) + Q^2}\,,\quad 
	k_f^0 = \xi_k \frac{\sqrt{s}}{2} = \frac{s (1-\ydis) + Q^2}{2 \sqrt{s}}\,,
\label{eq:xikdef}
\eeq
and, thus, 	
	\beq	
	\dd k_f^0 \; \dd \ck = 
	 \dd \ydis \; \dd Q^2 \; \frac{1}{2k_f^0}\,.
         \eeq
The phase-space integral of Eq.~(\ref{eq:ps3a}) then becomes 
\begin{align}
  \int\dd\Phi_3
  &= \frac{1}{16 (2\pi)^5} \int \dd x \; \dd \ydis \; \dd \bar \phi \; \dd \lambda \; \dd \xi \; \dd \phi \; \dd y \;
  \frac{s\, \ydis\, p_r^0}{p_f^0} \delta\qty(k_i^0 + p_i^0 - k_f^0 - p_f^0 - p_r^0)\,,   
\label{eq:phi3-lambda}	
\end{align}
where $\lambda \equiv Q^2/(2k_i q) = Q^2/(s \ydis) = x_B/x$ and where 
$p_r^0$ and $p_f^0$ are given by 
\begin{align}
  p_r^0  =  \xi \frac{\sqrt{s}}{2}\,, \qquad 
  p_f^0  =  \frac{\sqrt{s}}{2} \sqrt{\xi^2 +\xi_k^2 + 2 \xi \xi_k y y_k + 2 \xi \xi_k\sqrt{1-y^2}\sqrt{1-y_k^2} \cos\left(\phi-\bar \phi\right) }\,,
\label{eq:pr0pf0def}
\end{align}  
with $\xi_k$ defined in Eq.~\eqref{eq:xikdef}. 
The integration bounds in Eq.~\eqref{eq:phi3-lambda} are given by
\begin{align}
  & 0 < x,\lambda,\xi,\ydis <1\,,\quad 0<\phi,\bar \phi < 2 \pi\,, \quad -1 < y < 1 \,.  
 \end{align} 
Furthermore it is clear that the integrand depends on $ \Delta \phi=\phi-\bar \phi$, but not $\phi,\bar \phi$ individually.

Writing explicitly the 
argument of the delta function 
in terms of the remaining integration variables one has, 
\begin{align}
\label{eq:arg-delta}
&  \frac{\sqrt{s}}{2} \bigg[\left(\ydis(1-\lambda)+1-\xi \right) 
  -\left(4 \xi  \sqrt{\lambda  \qty(1-y^2) \qty(1-\ydis) \ydis} \cos \Delta \phi \right. \nn\\
   & \left. -2 \xi  y \qty(\ydis(1+\lambda)-1)+\left((\lambda -1) \ydis+1\right){}^2+\xi ^2\right)^{1/2}\bigg]=0.
\end{align}
In order to perform the integration over $\lambda$ in Eq.~(\ref{eq:phi3-lambda}) with the help of the delta function, is it useful to write its argument as 
\begin{align}
	&\delta\qty(k_i^0 + p_i^0 - k_f^0 - p_f^0 - p_r^0) = \frac{ 1}{D} \qty(\delta(\lambda - \lambda_+) + \delta(\lambda-\lambda_-)). 
\end{align}
For the two zeros of the argument we find 
\begin{align}
  \lambda_\pm &= \frac{\pm 2 \xi  \cos \Delta \phi \sqrt{A}+\xi ^2 \left(1-y^2\right) (1-\ydis) \cos \left(2 \Delta \phi\right)+2 (1-\xi ) (2 \ydis -\xi(1+y))}{\ydis (\xi(1+y)-2)^2},
  \label{eq:lambdasol}
\end{align}
with  
\begin{align}
	D&= \left| \frac{s \left(\xi  \cos \Delta \phi \sqrt{\lambda  \left(1-y^2\right) (1-\ydis) \ydis}+\lambda  \ydis (1-\xi  y-(1-\lambda ) \ydis)\right)}{4 p_f^0 \lambda }+\frac{\sqrt{s} \ydis}{2}\right|\,,
\label{eq:D}
        \\
A&=\left(1-y^2\right)^2 (1-\ydis)^2 \left[\xi ^2  \cos ^2\Delta \phi+\frac{(2-\xi  (1+y)) (2 \ydis-\xi  ((1-y) \ydis+y+1))}{\left(1-y^2\right) (1-\ydis)}\right]\,,
\label{eq:A}
\end{align}
where $p_f^0$ is given in Eq.~\eqref{eq:pr0pf0def}. 

These solutions have been obtained by reshuffling and squaring
Eq.~(\ref{eq:arg-delta}) twice. 
Therefore, one needs to verify whether the solutions
in Eq.~\eqref{eq:lambdasol} satisfy the original equation,
Eq.~(\ref{eq:arg-delta}). This restricts the range of physically
allowed values that $\lambda_\pm$ can assume.
Furthermore, one needs to make sure that the
arguments of the square-roots in Eqs.~\eqref{eq:lambdasol} and
\eqref{eq:D} are positive. 
The argument of the delta function, given in Eq.~(\ref{eq:arg-delta}), 
results in the condition 
\begin{align}
\label{eq:arg-delta1}
(1-\lambda ) \ydis+1-\xi&=\sqrt{B+4 \xi  \sqrt{\lambda  \left(1-y^2\right) \left(1-\ydis\right) \ydis} \cos \left(\Delta \phi\right)}\,,
\end{align}
with 
\begin{align}
B &= -2 \xi  y \left(\lambda  \ydis+\ydis-1\right)+\left((\lambda -1) \ydis+1\right){}^2+\xi ^2\,.
\end{align}
For the allowed range of the variables $\lambda,\ydis,\xi$, the left-hand side
of this equation is always positive. The right-hand side of
the equation is clearly also always positive, therefore both sides can
be squared without generating spurious solutions, resulting in
\begin{align}
\xi  \ydis (\lambda +\lambda  y+y-1)+2 (1-\lambda) \ydis-\xi  (y+1)=2 \xi  \sqrt{\lambda  \left(1-y^2\right) \left(1-\ydis\right) \ydis} \cos \Delta \phi .
\label{eq:lambdaRestriction}
\end{align}
In this equation, the sign of the right-hand side is determined by the
sign of the factor $\cos \Delta \phi$. Therefore only
solutions for $\lambda_\pm$ are allowed, where the left-hand side has
the same sign as $ \cos \Delta \phi$. The two solutions have the
property that $\lambda_+$ ($\lambda_-$) becomes equal to $\lambda_-$
($\lambda_+$) when the sign of $\cos \Delta \phi$ is changed.
By inserting $\lambda_\pm$ in the l.h.s.\ of the above equation one
finds that in the case of $\cos\Delta\phi > 0$ only the
$\lambda_-$ solution gives the correct sign up to a maximum value of
$\xi$ equal to $\xi_{0} = \frac{2 \ydis}{1-y \ydis+\ydis+y}$.
On the other hand, in the case of $\cos\Delta \phi < 0$
the $\lambda_-$ solution is always a correct solution and the
$\lambda_+$ solution is only correct if $\xi > \xi_0$.

Rewriting $x=\xdis/\lambda$, for the phase-space integral of Eq.~(\ref{eq:phi3-lambda}) can be rewritten as
\begin{align}
	\int\dd\Phi_3 &= \frac{1}{16 (2\pi)^5} \int \dd \xdis{} \; \dd \ydis \; \dd \bar \phi \; \dd \lambda \; \dd \xi \; \dd \phi \; \dd y \; \left[\delta(\lambda - \lambda_+) + \delta(\lambda-\lambda_-)\right]\nn\\
	&\qquad \times  \frac{\bar s\ydis \xi}{\lambda \left| \lambda  \ydis (\xi(1+y)-2)-\xi  \cos \left(\Delta \phi\right) \sqrt{\lambda  \left(1-y^2\right) (1-\ydis) \ydis}\right|}\,.
\end{align}
This three-particle phase-space integral can be expressed in terms of the two-particle phase-space integral in Eq.~\eqref{eq:phi2}, 
and the radiation variables $\xi, \phi, y$ as 
\begin{align}
	\int \dd \Phi_3 &= \frac{1}{32 \pi^3}\int \dd \Phi_2 \; \dd \lambda \; \dd \xi \; \dd \phi \; \dd y \; 
	\left[\delta(\lambda - \lambda_+) + \delta(\lambda-\lambda_-)\right]\\
	&\qquad \times  \frac{\bar s\ydis \xi}{\lambda \left| \lambda  \ydis (\xi(1+y)-2)-\xi  \cos \left(\Delta \phi\right) \sqrt{\lambda  \left(1-y^2\right) (1-\ydis) \ydis}\right|}\,. 
\end{align}
In the above equation, the $\lambda$ integral is now constrained to the 
range $\bar x < \lambda < 1$, for each $\lambda_\pm$ solution, and the
$\xi$ integral is constrained as described above.
One can observe that in the collinear and soft limits only $\lambda_-$
is a valid solution. Schematically, the $\xi$ integration can then be written as 
\begin{align}
	\int \dd \xi f(\xi) \qty(\delta(\lambda - \lambda_+) + \delta(\lambda-\lambda_-)) &=: \int \dd \xi \qty(f_+(\xi) + f_-(\xi))\\
	&= \int_0^{\ximax} \dd \xi f_-(\xi) + \int_{\xi_0}^{\ximax} \dd \xi f_+(\xi),
\end{align}
where $\xi_0$ is given above and $\ximax$ depends on the value $\xi_1$ where
$\lambda_+=\lambda_-$ and $A$ in Eq.~\eqref{eq:A} vanishes. Explicitly, one has 
\begin{align}
  \xi_{1} = \frac{ 4 \ydis}{1+y+2 \ydis +\sqrt{C}}\,,
\end{align}
with
\begin{align}
  C = 1+2y(1-2\ydis) - 4  \cos \Delta \phi^2 \ydis (1-\ydis)+y^2\left(1-4 \sin \Delta \phi^2(1-\ydis)\right)\,, 
\end{align}
and
\begin{align}
\xi_{\rm max} = \max\left(\xi_0,\xi_1 \theta(-\cos\Delta \phi )\right)\,. 
\end{align}  
Therefore, in the case where $\lambda_+$ is not a valid solution $\xi_0 = \ximax$.

The integral can then be written as 
\begin{align}
	\int \dd \xi \qty(f_+(\xi) + f_-(\xi)) &= \int_0^{\ximax'} \dd \xi \qty(f_-(\xi) \Theta\qty(\ximax-\xi) + f_+(2\ximax-\xi) \Theta\qty(\xi-\ximax)),
\end{align}
with $\ximax' = 2 \ximax - \xi_0$.
Lastly, one can make the transformation to $\xitilde=\xi/\ximax'$,
to obtain 
\begin{align}
  \int \dd \xi \qty(f_+(\xi) + f_-(\xi))
  = \int_0^1 \dd & \xitilde \ximax' \left(f_-(\xitilde \ximax') \Theta(\ximax-\xitilde\ximax') \right. \nonumber \\ 
&  \left. + f_+(\ximax'(1-\xitilde)+\xi_0) \Theta(\xitilde \ximax' - \ximax)\right)\,. 
\end{align}

\subsection{Phase-space parameterisation for  final-state radiation}
Here, we use the same notation as in the previous section and work in the
centre-of-mass frame. We write the phase space as
\begin{align}
\int\dd\Phi_3 = \int\dd x \dd\phi_3 = \int \dd x \frac{\dd^3 \vb k_f}{2k_f^0 (2\pi)^3} \frac{\dd^3 \vb p_f}{2p_f^0 (2\pi)^3} \frac{\dd^3 \vb p_r}{2p_r^0 (2\pi)^3} (2\pi)^4 \delta^{(4)}\qty(k_i + p_i - k_f - p_f - p_r).
\label{eq:FSRPS}
\end{align}
We introduce $k$, the sum of the momenta of the two outgoing partons,  
\begin{align}
	k = p_f + p_r\,.
\end{align}

We parameterise $k$ as
\begin{align}
	k = (k^0,\uk \sqrt{1-\ck^2} \cos \bar \phi,\uk\sqrt{1-\ck^2} \sin \bar \phi,\uk \ck),\quad {\rm with}\quad \uk = |\vb k|,
\end{align}
where $k^0 = p_f^0 + p_r^0$. If we use $k$ instead of $p_f$, the phase space becomes
\begin{align}
  \int\dd\Phi_3 
  = \frac{1}{256 \pi^5} \int \dd x \; \dd^3 \vb k_f \; \dd \uk \; \dd \ck \; \dd \bar \phi \; \dd^3 \vb p_r \frac{\uk^2}{k_f^0 p_f^0 p_r^0} \delta^{(4)}\qty(k_i + p_i - k_f - k).
\end{align}
Next, we need rotate $k$ along the $z$-axis, such that  
\begin{align}
	k^{\sss \rm (R)} = (k^0,0,0,\uk) = (p_f^0 + p_r^0,0,0,\uk).
\end{align}
 By doing that rotation we are constricting the two integration variables $\ck$ and $\bar \phi$. Hence, we have to transform them into new angles. To that end, we choose the corresponding angles of the incoming parton.
The rotation matrix $R$ can be written explicitly as
\begin{align}
	R= \left(
	\begin{array}{cccc}
	1 & 0 & 0 & 0 \\
	0 & \ck \cos ^2\bar \phi+\sin ^2\bar \phi & (\ck-1) \sin \bar \phi \cos \bar \phi & -\sqrt{1-\ck^2} \cos \bar \phi \\
	0 & (\ck-1) \sin \bar \phi \cos \bar \phi & \ck \sin ^2\bar \phi+\cos ^2\bar \phi & -\sqrt{1-\ck^2} \sin \bar \phi \\
	0 & \sqrt{1-\ck^2} \cos \bar \phi & \sqrt{1-\ck^2} \sin \bar \phi & \ck \\
	\end{array}
	\right)\,.
	\label{eq:Rotation}
\end{align}
In the rotated frame one has 
\begin{align}
	k^{\sss \rm (R)} &= (k^0,0,0,\uk),\\
	p_i^{\sss \rm (R)} &= \frac{\sqrt{s}}{2} \qty(1,\sqrt{1-\ck^2} \cos \bar \phi, \sqrt{1-\ck^2} \sin \bar \phi, -\ck),\\
	k_i^{\sss \rm (R)} &= \frac{\sqrt{s}}{2} \qty(1,-\sqrt{1-\ck^2} \cos \bar \phi, -\sqrt{1-\ck^2} \sin \bar \phi, \ck),\\
	k_f^{\sss \rm (R)} &= k_f^0 \qty(1,\sqrt{1-c_f^2} \cos \phi_f, \sqrt{1-c_f^2} \sin \phi_f, c_f),\\
	p_r^{\sss \rm (R)} &= \frac{\sqrt{s}}{2} \xi \qty(1,\sqrt{1-c_\psi^2} \cos \phi_r, \sqrt{1-c_\psi^2} \sin \phi_r, c_\psi).
\end{align}
Note that $c_\psi$ and $\phi_r$ are not the FKS variables. One can
perform a change of variables $\ck \rightarrow - c_p$ and $\bar \phi
\rightarrow \phi_p$ to obtain the usual angles of the incoming parton
in the rotated frame.

We can integrate over $\vb k_f$ to obtain $\vb k_f = -\vb k$ and $k_f^0 = \uk$.
Thereby, the momentum of the outgoing lepton in the rotated frame is simply
\begin{align}
	k_fc^{\rm \sss (R)} = \uk (1,0,0,-1).
\end{align}
The phase space now becomes
\begin{align}
\begin{split}
	\int\dd\Phi_3 &= \frac{1}{256 \pi^5} \int \dd x \; \dd \uk \; \dd c_p \; \dd \phi_p \; \dd^3 \vb p_r^{\rm \sss (R)} \frac{\uk}{p_r^0\sqrt{\qty(p_r^0)^2 + \uk^2 - 2 c_\psi p_r^0 \uk}}\\
	&\qquad \times \delta\qty(\sqrt{s} - \uk - \sqrt{\qty(p_r^0)^2 + \uk^2 - 2 c_\psi p_r^0 \uk} - p_r^0)\,, 
\end{split}
\end{align}
where we used that $p_f^0 = \sqrt{\qty(p_r^0)^2 + \uk^2 - 2 c_\psi p_r^0 \uk}$.

We now computed the DIS variables in the rotated frame
\begin{align}
	& q^{\rm \sss (R)} = k_i^{\rm \sss (R)} - k_f^{\rm \sss (R)}
	= \qty(\frac{\sqrt{s}}{2} - \uk, -\frac{\sqrt{s}}{2}\sqrt{1-c_p^2} \cos \phi_p, -\frac{\sqrt{s}}{2}\sqrt{1-c_p^2} \sin \phi_p, -\frac{\sqrt{s}\ck}{2} + \uk),\nonumber\\
	& \ydis = \frac{p_i q}{p_i k_i}
	= 1-\frac{\uk (1+c_p)}{\sqrt{s}},\nonumber \\
	& Q^2 = -q^2 = \sqrt{s} \uk (1-c_p).
\end{align}
In order to express the phase space as a function of the underlying Born one, we 
 replace $\uk$ and $c_p$ by DIS variables using 
\begin{align}
	\uk &= \frac{s(1-\ydis)+Q^2}{2\sqrt{s}},\\
	c_p &= \frac{s(1-\ydis)-Q^2}{s(1-\ydis)+Q^2},\\
	\dd \uk \; \dd c_p &= \dd \ydis \; \dd Q^2 \; \frac{\sqrt{s}}{s(1-\ydis)+Q^2}.
\end{align}
We also replace the $Q^2$ integral introducing 
$\lambda = Q^2/(s\ydis)$ 
\begin{align}
	\dd Q^2 &= \dd \lambda \; s\, \ydis.
\end{align}
Additionally, we transform $\vb p_r^{\rm \sss (R)}$ into spherical coordinates as indicated above. The phase space becomes
\begin{align}
	\int\dd\Phi_3 &=\frac{1}{1024 \pi^5} \int \dd x \; \dd \ydis \; \dd \lambda \; \dd \phi_p \; \dd \xi \; \dd c_\psi \; \dd \phi_r \frac{s^{3/2} \xi \ydis}{\sqrt{F}} \\
&\qquad \times \delta\qty(\frac{1}{2} \sqrt{s} \left(-\sqrt{F}-\xi +(1-\lambda ) \ydis+1\right))\,, 
 \end{align}
with
\begin{align}
F = -2 \xi  c_{\psi }-2 (\lambda -1) \ydis \left(\xi  c_{\psi }-1\right)+\xi ^2+(\lambda -1)^2 \ydis^2+1\,. 
 \end{align}

Next we remove the last delta distribution by integrating over $\lambda$. The root of the argument of the delta distribution is
\begin{align}
  \lambda_0 = \frac{\xi  \left(1-c_{\psi }\right)+\ydis \left(\xi (1+ c_{\psi }) -2\right)}{\ydis \left(\xi (1+ c_{\psi }) -2\right)}
  \label{eq:lambda0}
\end{align}
and we get the additional Jacobian factor of
\begin{align}
	\frac{2 \left(\xi ^2 \left(c_{\psi }+1\right)-2 \xi  \left(c_{\psi }+1\right)+2\right)}{\sqrt{s} \ydis \left(\xi  \left(c_{\psi }+1\right)-2\right){}^2}\,. 
\end{align}
Therefore, the integration over $\lambda$ yields
\begin{align}
	\int \dd \Phi_3 &= \frac{1}{512 \pi^5} \int \dd x \; \dd \ydis \; \dd \phi_p \; \dd \xi \; \dd c_\psi \; \dd \phi_r \frac{s \xi }{2-\xi  \left(c_{\psi }+1\right)} .
\end{align}
The last step is to transform $c_\psi$ and $\phi_r$ into the FKS
variables $y$ and $\phi$. Here, $y$ is the cosine of the angle between the
emitter and the radiation, while $\phi$ denotes the azimuthal angle of
$p_r$ around $k$, where $k_i$, i.e.\ the $z$-axis of the usual centre-of-mass frame, 
serves as origin for the angle. We can transform the
momenta back into the usual centre-of-mass frame using $R^{-1}$ of
\eqref{eq:Rotation}. For the FKS variables we get 
\begin{align}
	y &= 1-\frac{2 \left(1-c_{\psi }\right)}{2-(2-\xi ) \xi  \left(c_{\psi }+1\right)},\\
	\phi &= (\phi_r - \phi_p - \pi) \mod 2\pi, 
\end{align}
which leads to 
\begin{align}
	\dd c_\psi &= \dd y \frac{\left(2-(2-\xi ) \xi  \left(c_{\psi }+1\right)\right){}^2}{4 (1-\xi )^2},\qquad 
	\dd \phi_r = \dd \phi.
\end{align}
Therefore the phase space becomes 
\begin{align}
	\int \dd \Phi_3 &= \frac{1}{256 \pi^5} \int \dd x \; \dd \ydis \; \dd \phi_p \; \dd \xi \; \dd y \; \dd \phi \frac{(1-\xi ) \xi  s}{(2-\xi  (1-y)) (2-(2-\xi ) \xi  (1-y))}.
\end{align}
Finally, we change variable to from $x$ to $\xdis{} = \lambda_0 x $ and factor out the Born phase space to get
\begin{align}
	\int \dd \Phi_3 &= \frac{1}{16 \pi^3} \int \dd \Phi_2 \;  \dd \xi \; \dd y \; \dd \phi \frac{(1-\xi ) \xi  \bar s}{\lambda _0^2 (2-\xi  (1-y)) (2-(2-\xi ) \xi  (1-y))}.
\end{align}

\section{Generation of radiation}
\label{app:genrad}

In this appendix we describe how we modify the default implementation
of the event generation for radiation, in order to be able
to handle DIS.
In practice, for every singular region $\alpha$ (associated with a given underlying Born $f_b$), we want to find a function $U_{\alpha}(\xi, y)$ such that
\begin{align}
  \frac{R_{\alpha}(\bar{\Phi}_b, \Phi_{\rm \sss rad}^{(\alpha)})}{B_{f_b}(\bar{\Phi}_b)} \dd \Phi_{\rm \sss rad}^{(\alpha)} \leq U_{\alpha}(\xi, y) \dd \xi \dd y \frac{\dd \phi}{2\pi}.
\end{align}
We also need to introduce a dimensioned variable
$\kappa_t^{\alpha}(\xi, y)$ that approaches the transverse momentum of
the emission in the soft-collinear limit, so that we can integrate
analytically
\begin{equation}
\Delta^{(U)}_{\alpha}(k_{\rm \sss T}) = \exp\left(-\int  U_{\alpha}(\xi, y) \dd \xi \dd y \Theta(\kappa_t^{\alpha}(\xi,
y)>k_{\rm \sss T})\right).
  \end{equation}
In practice, we generate a random number $r$, and we determine $k_{\rm \sss T}$ by solving $\Delta^{(U)}_{\alpha}(k_{\rm \sss T})=r$.
We then generate $\xi$ uniformly in $U_{\alpha}(\xi, y(\xi, k_{\rm \sss T})$, while $\phi$ is generated uniformly between $0$ and $2\pi$.
The emission is then accepted with probability
\begin{equation}
  \frac{R_{\alpha}(\bar{\Phi}_b, \Phi_{\rm \sss rad}^{(\alpha)})}{B_{f_b}(\bar{\Phi}_b) U_{\alpha}(\xi,y)}.
\end{equation}
To handle DIS we need two upper bound functions, one for FSR and one for ISR.

\subsection{Generation of final-state radiation}
\label{app:genradfsr}

\subsubsection{The standard \POWHEGBOX{} implementation}
\label{app:genradfsrstandard}

For final state radiation (FSR) in general the cross section can have
a logarithmic divergence in the soft ($\xi \to 0$) or collinear ($y
\to 1$) limit, therefore it is convenient to parameterise the upper
bound for the generation of radiation as follows (see App.~C of
Ref.~\cite{Alioli:2010xd})
\begin{equation}
U(\xi, y) d \xi d y = \tilde{N}\frac{\alpha_s(\kappa_t^2)}{\xi (1-y)} d \xi d y\,,
\end{equation}
where $\kappa_t$ can be seen as the \POWHEG{} evolution variable and reads
\begin{equation}
\kappa_{t}^2 = \frac{s}{2}\xi^2 (1-y).
\label{eq:kappat}
\end{equation}
Notice that for FSR, $s$ does not change between Born and real contributions.
\POWHEG{} chooses convenient values for $\bar{b}_0$ and $\bar{\Lambda}$ such that
\begin{equation}
\alpha_s(\kappa_t^2) \leq \frac{1}{\bar{b}_0 \log \frac{\kappa_t^2}{\bar{\Lambda}^2}}.
\end{equation}
The emitted parton can carry at most an energy fraction 
\begin{equation}
\xi \leq \xi_{\max} =\frac{s-M^2_{\textbf{rec}}}{s},
\end{equation}
where $M_{\text{rec}}$ is the mass of the recoiling system, which coincides with the final-state lepton in our case. 
Since $y \geq -1 $, Eq.~\eqref{eq:kappat} implies that $\kappa_t^2 \leq \xi^2 s$.
So we have (notice that we absorbed some constants in $N$)
\begin{align}
\mathcal{S}(\kappa_t^2) =& \int U(\xi, y) \,d \xi\, d y \,d \phi =  2\pi N \int_0^{\xi_{\max}} \frac{d\xi}{\xi} \int_{0}^{\xi^2 s} \frac{d t}{t} \frac{1}{\bar{b}_0  \log \frac{t}{\bar{\Lambda}^2}} \Theta(t> \kappa_t^2) \nonumber \\
=& 2\pi N\Theta(\kappa_t^2 < \xi_{\max}^2 s) \int_{\kappa_t^2}^{\xi_{\max}^2 s} \frac{d t}{t} \frac{1}{\bar{b}_0  \log \frac{t}{\bar{\Lambda}^2}} \int_{\sqrt{\frac{t}{s}}}^{\xi_{\max}} \frac{d\xi}{\xi} \nonumber \\
=&\frac{\pi N}{\bar{b}_0} \Theta(\kappa_t^2 < \xi_{\max}^2 s)
\left\{
\log \frac{\xi^2_{\max} s}{\bar{\Lambda}^2} \log \left[\frac{\log(\xi^2_{\max} s/\bar{\Lambda}^2)}{\log(\kappa_t^2/\bar{\Lambda}^2)} - \log \frac{\xi^2_{\max} s}{\kappa_t^2}\right]
\right\}\nonumber \\
=& {\frac{\pi N}{\bar{b}_0} \Theta(\kappa_t^2 < \kappa_{t,\max}^2)
\left\{
\log \frac{\kappa_{t,\max}^2}{\bar{\Lambda}^2} \log \left[\frac{\log(\kappa_{t,\max}^2/\bar{\Lambda}^2)}{\log(\kappa_t^2/\bar{\Lambda}^2)} - \log \frac{\kappa_{t,\max}^2}{\kappa_t^2}\right]
\right\}},
\end{align}
with $\kappa_{t,\max}^2 =\xi_{\max}^2 \bar{s}$
To generate $\kappa_t^2$, one extracts a random number $r$, and solves numerically
\begin{equation}
r = \exp^{-\mathcal{S}(\kappa_t^2)}. 
\end{equation}
Since $d\xi U \propto d \log \xi$, one then generates uniformly $\log
\xi$ between $\frac{1}{2}\log \frac{k_t^2}{s}$ and $\log \xi_{\max} =
\log(\kappa_{t}^2) $ and one gets $y$ from
Eq.~\eqref{eq:kappat}. Finally the variable $\phi$ is generated
uniformly.

Next one builds the radiation phase space $\Phi_{n+1}(\bar{\Phi}_n,
\xi, y, \phi)$, and accepts the generated point with probability
equal to the ratio between the real over Born cross section and the upper bound.
If the point is rejected, $\kappa_{t,\max}^2$ is set to the last
generated value.

We note that in the \POWHEGBOX{} there are also alternative
implementations of the upper bound. The one presented here, which
corresponds to setting \texttt{rad\_iupperfsr 1}, is the one we start
from.  Indeed in our case, since the recoiling system is given only by
the final-state lepton, we have $M_{\text{rec}}=0$, and the other
upper bound options do not work in this case.

\subsubsection{The DIS case}
\label{app:genradfsrdis}
In our DIS phase space, the centre-of-mass energy of the underlying Born is $\lambda$-times smaller than the one of the real contribution, with $\lambda$ given by
\begin{equation}
\lambda = \frac{\bar{x}}{x} =1-  \frac{\xi  (1-\xi)(1-y)}{y_{\text{DIS}} (2-\xi  (1-y))},
\end{equation}
where $\bar{x}=\xdis$ is at the Born level, $x$ is the incoming parton energy fraction after the emission. 

Since $\lambda \to 1$ both in the soft or in the collinear limit, one 
can still use as ordering variable
\begin{equation}
\kappa_{t}^2 = \frac{\bar{s}}{2}\xi^2 (1-y),
\label{eq:kappatfsr}
\end{equation}
which now involves explicitly the underlying Born centre-of-mass
energy. 
Neglecting the mass of the recoiling lepton, $M_{\rm rec}=0$ implies
that in this case $\xi_{\max}=1$. One then proceeds as before
generating a radiation phase-space point and accepting or rejecting it
using the standard hit and miss technique.


\subsection{Generation of initial-state radiation}
\label{app:genradisrdis}

\subsubsection{The standard \POWHEGBOX{} implementation}
\label{app:genradisrstandard}

For initial state, the standard \POWHEG{} code handles together the
$+$ and $-$ collinear regions, thus, in the default setup
(corresponding to \texttt{rad\_iupperisr=1}) one uses an upper bound
of the form
\begin{equation}
U(\xi, y) d \xi d y = \tilde{N}\frac{\alpha_s(\kappa_t^2)}{\xi (1-y^2)} d \xi d y\,. 
\end{equation}
The ordering variable is defined as
 \begin{equation}
\kappa_{t}^2 = \frac{s}{4}\xi^2 (1-y^2) = \frac{\bar{s}}{4(1-\xi)}\xi^2 (1-y^2), 
 \end{equation}
which depends on $(1-y^2)$ to account for both singularities. The
normalisation is such that in the limit $y\to1$, this expression agrees
with the FSR case, Eq.~\eqref{eq:fsr-kappa}.

Conversely to the FSR case, in order to handle ISR in DIS we need to
change the definition of $\kappa_t(\xi,y)$, and hence the generation
of the radiation variables.

\subsubsection{Implementation of ISR for DIS}
In this case, however, since there is only a singularity associated
with $y\to 1$\footnote{Actually in the code it is $y=-1$, but in analogy with FSR we here use $1$.},  it is more
appropriate to use as upper-bound 
\begin{equation}
U(\xi, y) d \xi d y = \tilde{N}\frac{\alpha_s(\kappa_t^2)}{\xi (1-y)} d \xi d y\,, 
\end{equation}
which is more similar to the FSR case, and  as ordering variable
\begin{equation}
\kappa_t^2 = \frac{\xi^2}{2-\xi(1+y)} \bar{s}(1-y).
\label{eq:fsr-kappa}
\end{equation}
This choice satisfies the appropriate limits because for $\xi \to 0$
we have $\kappa_t^2 \to \frac{\bar{s}\xi^2(1-y)}{2}$, and for $y\to 1$
$\kappa_t^2 \to
\frac{\bar{s}\xi^2(1-y)}{2(1-\xi)}\approx\frac{\bar{s}\xi^2(1-y)}{2}$.\footnote{Note
  that we have discarded the option $\kappa_t^2 =
  \frac{\xi^2}{2(1-\xi)} \bar{s}(1-y)$, because $\xi$ can go up to 1
  for non-singular configurations. This indeed only happens when, in
  the event frame, the radiated parton becomes anti-parallel to the
  final-state lepton and the emitter becomes parallel to the
  final-state lepton, hereby taking a substantial recoil.}  We can
replace the $y$ integration with a $\kappa_t^2$ one using
\begin{equation}
y=\frac{\xi ^2 \bar{s}+(\xi -2) \kappa _t^2}{\xi  \left(\kappa _t^2-\xi  \bar{s}\right)}\,. 
\label{eq:y-isr}
\end{equation}
The requirement $-1\leq y \leq 1$, leads to
\begin{equation}
\sqrt{\frac{\kappa_t^2}{\bar{s}}}<\xi < 1,
\end{equation}
which means that our ordering variable is bounded by
\begin{equation}
\kappa_{t,\max}^2 = \bar{s},
\end{equation}
like for final-state radiation.
It is also easy to see that this is the upper-bound since $\kappa_t^2$ in Eq.~\eqref{eq:fsr-kappa} increases with $\xi$ and for $\xi=1$ one obtains $\kappa_{t,\max}^2 = \bar{s}$.
We have then 
\begin{align}
  \mathcal{S}(\kappa_t^2) & = N \int \alpha_s(t) \frac{dy}{1-y} \frac{d\xi}{\xi} d \phi\nonumber \\
  & = 2\pi N \int_{\kappa_t^2}^{\kappa_{t,\max}^2} \alpha_s(t)\frac{dt}{t} \int_{\sqrt{\frac{t}{\bar{s}}}}^1\frac{d\xi}{\xi -\frac{t}{\bar{s}}}\nonumber \\
  & = 2\pi N \int_{\kappa_t^2}^{\kappa_{t,\max}^2} \alpha_s(t)\frac{dt}{t} \log \left(1+\sqrt{\frac{\bar{s}}{t}}\right)\,.
\end{align}
As before we define 
\begin{equation}
V(t) = 2\pi N \alpha_s(t)\log \left(1+\sqrt{\frac{\bar{s}}{t}}\right)\,,
\end{equation}
and define an upper bound using 
\begin{align}
\log \left(1+\sqrt{\frac{\bar{s}}{t}}\right) \leq \frac{1}{2}\log\left(\frac{4\bar{s}}{t}\right)\,,
\end{align}
which follows from the fact that $t$ is always smaller than $\bar{s}$. Thus we have
\begin{align}
\bar{\mathcal{S}}(\kappa_t^2) =& \frac{\pi N}{b_0} \int_{\log(\kappa_t^2/\bar{\Lambda}^2)}^{\log(\kappa_{t,\max}^2/\bar{\Lambda}^2)} \frac{d\ell}{\ell} \left[{\log \left(\frac{4\bar{s}}{\bar{\Lambda}^2}\right)}-\ell \right] \nonumber \\
=& \frac{\pi N}{b_0}\left\{ \log \left(\frac{4\bar{s}}{\bar{\Lambda}^2}\right) \log \frac{\log(\kappa_{t,\max}^2/\bar{\Lambda}^2) }{ \log(\kappa_{t}^2/\bar{\Lambda}^2)}
-\log \frac{\kappa_{t,\max}^2}{\kappa_t^2}
\right\}\,. 
\end{align}
At this point, $\kappa_t^2$ is then sampled uniformly in
$\exp(-\bar{\mathcal{S}}(\kappa_t^2))$.  Then, before generating
$\xi$, one accepts $\kappa_t$ with probability
$V(\kappa_t^2)/\bar{V}(\kappa_t^2)$.
Next, one needs to
generate $\xi=1-x$ in the range $\kappa_t/\sqrt{\bar{s}}<\xi<1 $ with probability proportional to $1/\left(\xi - \kappa_t^2/\bar s\right)$. 
One then computes $y$ using Eq.~\eqref{eq:y-isr}. 
Finally one needs to check if the resulting variables $x_{1,2}$, which
only depend on the Born variable $\bar{x}_{1,2}$ and $\xi$ and $y$,
are smaller than 1. If this is not the case, one restarts the
generation setting the starting scale equal to $\kappa_t^2$, till one
obtains values of $x_{1,2}<1$. At this point $\phi$ is chosen
randomly.

As a last step one builds the radiation phase space and accepts the point
with probability equal to the ratio between the real over Born cross
section and the upper bound. Notice that when doing this, one needs to
compute the real matrix element for both branch cuts and sum them.  
In the singular regions, only the negative (``$-$'') branch cut is possible, however far away from this limit both are possible. We then choose the negative branch cut with probability
\begin{equation}
\frac{R_{\alpha}(\Phi_{n+1}^{(-)}(\bar{\Phi}_n, \xi, y, \phi))}{R_{\alpha}(\Phi_{n+1}^{(-)}(\bar{\Phi}_n, \xi, y, \phi))+R_{\alpha}(\Phi_{n+1}^{(+)}(\bar{\Phi}_n, \xi, y, \phi))},
  \end{equation}
where the label $(\pm)$ denotes which branch cut is used, and the positive cut otherwise.

\subsubsection{Alternative implementation of ISR for DIS}
We also implemented a different way to generate ISR in DIS, which we
use as a check of our default treatment of ISR.
In the alternative treatment, we use the standard ISR \POWHEGBOX{}
upper bounding function
\begin{align}
	U(\xi, y) = N \frac{\alpha_s(\kappa_t^2)}{\xi (1-y^2)}.
\end{align}
As ordering variable $\kappa_t^2$ we choose a quantity that is equal to the transverse momentum of the radiation $k_T^2$ used in \POWHEG{} in the soft and collinear limit respectively
\begin{align}
	\kappa_t^2 = \frac{\bar s \xi^2 (1-y^2)}{4 (1-\xi y^2)}.
\end{align}
For any given Born configuration there exists a combination of radiation variables $(y, \phi)$ such that $\xi_\text{max} = 1$. Therefore the maximum value of $\kappa_t$ is given by 
\begin{align}
	\kappa_{\rm t, max}^2 = \frac{\bar s}{4}.
\end{align}
For convenience we introduce 
\begin{align}
	r = \frac{\kappa_t^2}{\bar s}= \frac{\xi^2 (1-y^2)}{4 (1-\xi y^2)},
	\label{eq:yaISRrDef}
\end{align}
which satisfies 
\begin{align}
	r\le r_{\max} = \frac{1}{4}.
\end{align}
One can invert Eq.~\eqref{eq:yaISRrDef} to get
\begin{align}
	y_\pm = \pm \sqrt{\frac{\xi ^2-4 r}{\xi  (\xi -4 r)}}.
	\label{eq:yaISRyDef}
\end{align}
Now, $p_T^2$ is to be generated in
\begin{align}
	\Delta^{(U)}(p_T) = \exp \qty[- \int U(\xi,y)\; \theta(\kappa_T - p_T)\; \dd \xi\; \dd y\; \dd \phi].
\end{align}
From Eq.~\eqref{eq:yaISRyDef} we get that, in order to have $-1 \le y
\le 1$, $\xi$  has to satisfy $2\sqrt{r}\le \xi \le 1$. Since the
integrand in symmetric in $y$, we can consider only the positive range
and multiply by a factor two.  We follow then similar steps as before.
After changing integration variable from $y$ to $r$, we have
\begin{align}
	- \log \Delta^{(U)}(p_T) &= N \int_{\frac{p_T^2}{\bar s}}^{1/4} \frac{\dd r}{r} \int_{2\sqrt{r}}^1 \dd\xi\int_0^{2\pi}\dd\phi \; \alpha_s(r \bar s)\frac{1-y_+^2 \xi}{y_+(1-\xi)\xi}\\
	&= 2 \pi N \int_{\frac{p_T^2}{\bar s}}^{{1/4}} \frac{\dd r}{r} \int_{2\sqrt{r}}^1 \dd \xi\;\sqrt{\frac{\xi }{(\xi -4 r) \left(\xi ^2-4 r\right)}}\nonumber 
        \\
	&= 4 \pi N \int_{\frac{p_T^2}{\bar s}}^{{1/4}}  \frac{\dd r}{r} \frac{\alpha(r \bar s)}{\sqrt{1-2 \sqrt{r}} } \Bigg\{ 2 \sqrt{r} K\left(2+\frac{1}{\sqrt{r}-\frac{1}{2}}\right)-2 \sqrt{r} F\left(\frac{\pi}{4}|2+\frac{1}{\sqrt{r}-\frac{1}{2}}\right)\nonumber\\	
	& \qquad -\left(2 \sqrt{r}+1\right) \Bigg[\Pi\left(\frac{2}{1-2 \sqrt{r}}|2+\frac{1}{\sqrt{r}-\frac{1}{2}}\right)\nonumber\\
	& \qquad -\Pi \left(\frac{2}{1-2 \sqrt{r}};\frac{\pi}{4}|2+\frac{1}{\sqrt{r}-\frac{1}{2}}\right)\Bigg]\Bigg\}\label{eq:yaISRbigINTEGRAND}\, ,
\end{align}
where $F\qty(\phi|m):=\int_0^\phi(1-m\sin^2\theta)^{-1/2} \dd \theta$ is the incomplete elliptic integral of the first kind, $K\qty(k) = F(\pi/2|m)$, and $\Pi(n;\phi|m):= \int_0^\phi (1-n\sin^2\theta)^{-1}(1-m\sin^2\theta)^{-1/2}$ is the incomplete elliptic integral of the third kind with $\Pi(n|m) = \Pi(n;\pi/2|m)$.
In the next step, we use that
\begin{align}
   \int_{2\sqrt{r}}^1
  \dd \xi\;\sqrt{\frac{\xi }{(\xi -4 r) \left(\xi ^2-4 r\right)}} < \frac{1}{2} \log\left(\frac{4}{r}\right)\,, 
\label{eq:xiint-ISR}
\end{align}  
to introduce an upper bound $  \bar \Delta^{(U)}(p_T)$ 
\begin{align}
	- \log \Delta^{(U)}(p_T) \le - \log \bar\Delta^{(U)}(p_T) &= \pi N \int_{\frac{p_T^2}{\bar s}}^{{1/4}}  \frac{\dd r}{r} \frac{\log \left(\frac{4}{r}\right)}{b_0 \log \left(\frac{r \bar s}{\Lambda ^2}\right)} \label{eq:yaISRsmallINTEGRAND}\nonumber\\
	&=\frac{\pi  N}{b_0} \left(\log \left(\frac{4 \bar s}{\Lambda ^2}\right) \log \left(\frac{\log \left(\frac{\bar s}{4\Lambda ^2}\right)}{\log \left(\frac{p_T^2}{\Lambda ^2}\right)}\right)+\log \left(\frac{
		4p_T^2}{\bar s}\right)\right).
\end{align}
In order to generate $p_T^2$, the equation $ \log x_{p_T} = \log
\bar\Delta^{(U)}(p_T)$ is solved numerically for $p_T$, where $x_{p_T}$
is a random number generated uniformly between 0 and 1.
To obtain the $p_T^2$ distributed according to $\Delta^{(U)}(p_T)$ the
veto method is used with the ratio of the integrands of
Eq.~\eqref{eq:yaISRbigINTEGRAND} and
Eq.~\eqref{eq:yaISRsmallINTEGRAND}. For numerical evaluation of Eq.~\eqref{eq:yaISRbigINTEGRAND} we have approximated
\begin{align}
	&\frac{\left(2 \sqrt{r}+1\right) \left(\Pi \left(\frac{2}{1-2 \sqrt{r}}|\frac{4 \sqrt{r}}{2 \sqrt{r}-1}\right)-\Pi \left(\frac{2}{1-2 \sqrt{r}};\frac{\pi }{4}|\frac{4 \sqrt{r}}{2 \sqrt{r}-1}\right)\right)}{\sqrt{1-2 \sqrt{r}}} \nn\\
	&\approx \log \left(2 \sqrt{r}\right) \arctan\left(16.3846 \sqrt[4]{r}\right) \qty[1-0.0000381337 e^{-1.66304 \sqrt{r}} \cos \left(21.1566 \sqrt[4]{r}\right)]\nn\\ &\times\frac{-0.660324 r^2+0.788716 r^{3/2}+0.645556 r+0.00603772 \sqrt{r}+7.591545691777556\cdot 10^{-6}}{-r^2+0.800354 r^{3/2}+1.45553 r+0.00835825 \sqrt{r}+2.475163988222491\cdot 10^{-6}}\nn\\
	&\times P(\sqrt{r},6),
\end{align}
where $P(\sqrt{r},6)$ is a piece wise defined polynomial in $\sqrt{r}$ of sixth degree.

Further, $\xi$ is to be generated according to the integrand of
Eq.~\eqref{eq:xiint-ISR}. Therefore, we overestimate the integrand
\begin{align}
	\sqrt{\frac{\xi }{(\xi -4 r) \left(\xi ^2-4 r\right)}} < \sqrt{\frac{1}{\left(2 \sqrt{r}-4 r\right) \left(2 \xi  \sqrt{r}-4 r\right)}}.
\end{align}
Next, we norm the integrand by providing the factor $\sqrt{r}$ to have
\begin{align}
	\int_{2\sqrt{r}}^1 \dd\xi\; \sqrt{r}\sqrt{\frac{1}{\left(2 \sqrt{r}-4 r\right) \left(2 \xi  \sqrt{r}-4 r\right)}} = 1.
\end{align}
The randomly generated $\xi'$ is obtained by solving
\begin{align}
	\int_{2\sqrt{r}}^{\xi'} \dd\xi\; \sqrt{r}\sqrt{\frac{1}{\left(2 \sqrt{r}-4 r\right) \left(2 \xi  \sqrt{r}-4 r\right)}} = x_\xi\,, 
\end{align}
where $x_\xi$ is random number  generated uniformly between 0 and 1. 
This leads to
\begin{align}
	\xi' = x_\xi^2 \left(1-2 \sqrt{r}\right)+2 \sqrt{r}.
\end{align}
Lastly, we need to keep the generated value only with probability 
\begin{align}
	\frac{\sqrt{\frac{1}{\left(2 \sqrt{r}-4 r\right) \left(2 \xi'  \sqrt{r}-4 r\right)}}}{\sqrt{\frac{\xi' }{(\xi' -4 r) \left((\xi') ^2-4 r\right)}}}.
\end{align}
To do this we check if a new random number $y \in (0,1)$ is smaller than this probability. 
If it is, we keep the generated value, otherwise we generate a new
$\xi$.

The generation of the phase space point, including the choice of the branch cut, then proceeds similarly to what illustrated in the previous section.

\section{Matching with Pythia}
\label{app:matching}

We now provide additional details on how we perform the matching to
the parton shower.  Each event printed in the LHE files is correlated
by a variable \texttt{scalup} that corresponds to the hardness
$k_t^{\rm PWG}$ of the emission.
In App.~\ref{app:genrad}, we have introduced
\begin{equation}
  \kappa_t^{2}=\bar{s} \xi^2 (1-y) \times
  \begin{cases}
   \frac{1}{2} & \qquad \mbox{FSR}, \\
     \frac{1}{2-\xi(1+y)}  & \qquad \mbox{ISR}, \\
  \end{cases}
  \label{eq:ktpwg}
\end{equation}
where $\bar{s}$ is the pre-branching centre-of-mass energy, which here corresponds to the underlying Born centre-of-mass energy, $\xi$ is
twice the energy fraction of the radiated parton in the event frame,
and $y$ is the cosine of the angle between emitter and radiated parton
in the event frame. The variable $\kappa_t^2$ is often dubbed as ``\texttt{scalup}'' and, at
LL~accuracy, it corresponds to the ordering
variable used by all \PYTHIA{} showers.
Thus, one could use \texttt{scalup} as starting scale for the showers.
To achieve this task, for both the \VINCIA{} and simple \PYTHIA{} time- and space-like showers
we need to set \texttt{pTmaxMatch = 1}.

Alternatively, we can start the shower at the maximum kinematical
limit (\texttt{pTmaxMatch = 2}), calculate the transverse momentum
of each emission, and veto emissions harder than the original LHE
event hardness.
To enable the veto, which is performed
by the \texttt{PowhegHooks} and \texttt{PowhegHooksVincia} classes,
which are part of \texttt{Pythia8.3}, we need to set
\begin{verbatim}
POWHEG:veto = 1.
\end{verbatim}
  We use \texttt{scalup} as event hardness, and the transverse
  momentum of every emission is computed using Eq.~\eqref{eq:ktpwg}.\footnote{Notice that we rely on the \texttt{Pythia8.3} definition of $\bar{s}$, i.e.\ the centre-of-mass energy before the last emission, as we have not yet generalised our mappings beyond the first emission.}
  This is achieved via the settings
  \begin{verbatim}
    POWHEG:pThard = 0
    POWHEG:pTdef  = 1.
  \end{verbatim}  
  We have re-implemented the function \texttt{pTpowheg}
  contained in the \texttt{PowhegHooks} and the \texttt{PowhegHooksVincia} classes,
  which are part of \texttt{Pythia8.3}.
  Notice that in case of $g\to gg$ and $g\to q\bar{q}$ final-state splittings,
  we use as $\xi$ the energy fraction of the softer of the two partons.
  
\PYTHIAE{} (dipole and \VINCIA) showers are only formally LL accurate,
so all the above options preserve their logarithmic accuracy.
However, since these showers also capture many NLL effects, providing
e.g.\ the correct NLL DGLAP evolution of initial-state partons~\cite{vanBeekveld:2022ukn}, we believe that the last option for
matching should be used as default. Furthermore it was shown in
Ref.~\cite{Hamilton:2023dwb} that not accounting correctly for the
veto can lead to a breakdown of exponentiation, which indicates failure
at the LL level already.

\section{Real radiation damping}
\label{sec:damp}
In this section we compare three options for the definition of the
singular real contribution $ R^{(s)}$, that we have introduced in
Sec.~\ref{sec:pwgingredients}.
In particular, we consider
\begin{enumerate}
\item no damping, i.e.~$ R^{(s)}=R$,  with $R$ being the whole real cross section; 
\item \texttt{Bornzerodamp} mechanism on;
\item \texttt{Bornzerodamp} and \texttt{hdamp} mechanisms simultaneously activated.
\end{enumerate}
For the latter option, we have modified the implementation of the
damping function $h(k_{\sss \rm T})$ of Eq.~\eqref{eq:hdamp} to be
\begin{equation}
h(k_{\sss \rm T})
  = \frac{Q^2}{\alpha_h k_{\sss \rm T}^2 + Q^2},
\end{equation}
where $\alpha_h$ is a parameter that can be varied. In our study we
considered $\alpha_h=1$. Notice that in the \POWHEGBOXR{}, the \texttt{hdamp} mechanism is applied only for ISR.
However, in our implementation, we use it also for FSR.

    {In this appendix we consider only the photon-exchange contribution to
$e^- p \rightarrow e^- X$ with $E_p = 904.5$~GeV, $E_e = 27.6$~GeV,
and we fix the underlying Born kinematics to be $\xdis=0.116$ and
$Q=57.6$~GeV. Events are required to have $E_{\mathrm{curr}} > Q/10$.}

\begin{figure}[t!]
  \centering\includegraphics[width=0.49\textwidth,
  page=1]{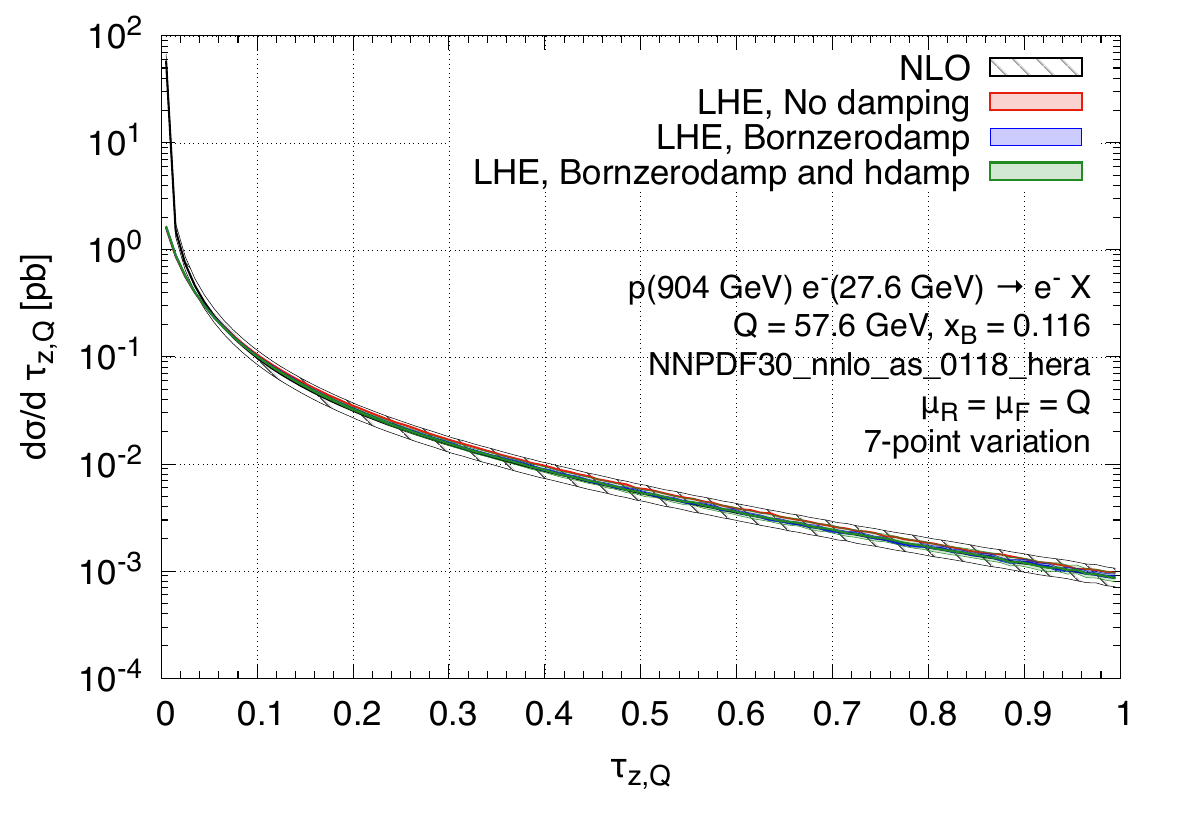}
  \centering\includegraphics[width=0.49\textwidth,
  page=2]{fig-temp/damp-lhef.pdf}
  \caption{ Thrust distribution
  normalised with respect to $Q$ for $Q=57.6\GeV$, $\xdis=0.116$ for the
  photon-exchange contribution to $e^- p \rightarrow e^- X$ at NLO
  (black), and at the LHE level, considering several damping options
  for the definition of the \POWHEG{} cross section: no damping (red),
  with the \texttt{Bornzerodamp} mechanism (blue), and with
  the \texttt{Bornzerodamp} and \texttt{hdamp} mechanisms activated
  simultaneously (green). In the right panel, the ratio with the NLO curve is shown.
  The band in the LHE curves is obtained with the 7-point factorisation- and renormalisation-scale variations.
}
  \label{fig:damptau}
\end{figure}
\begin{figure}[t!]
  \centering\includegraphics[width=0.49\textwidth,
  page=3]{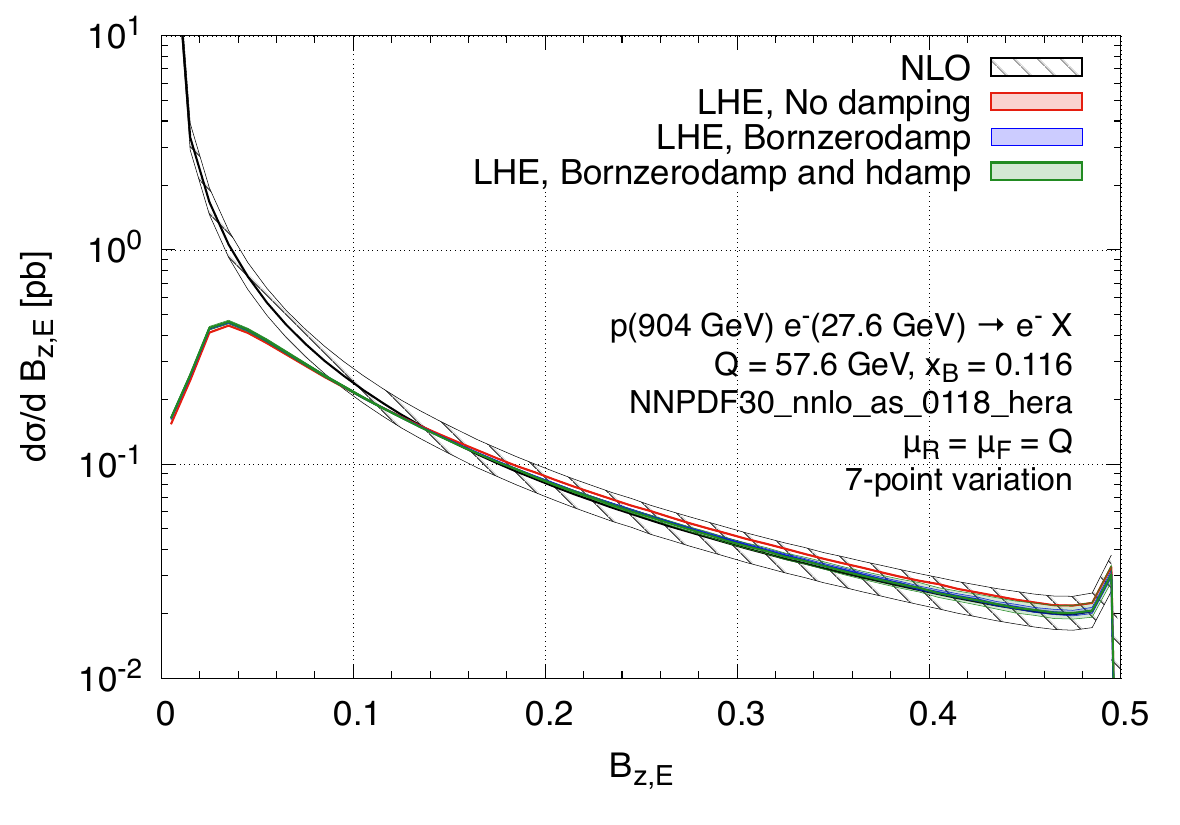} \centering\includegraphics[width=0.49\textwidth,
  page=4]{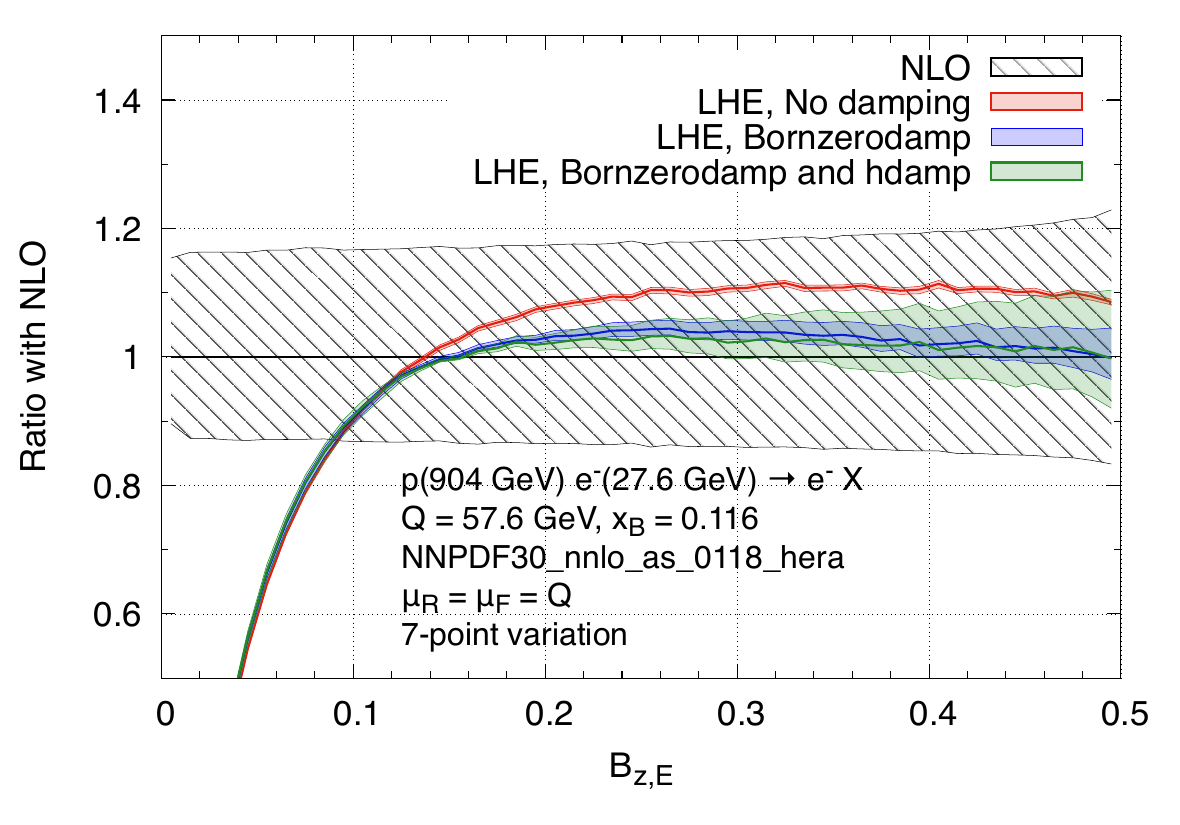} \caption{Same as
  Fig.~\ref{fig:damptau}, but for the broadening distribution
of Eq.~\eqref{eq:ebroadening}.}
  \label{fig:dampbrd}
\end{figure}

In Figs.~\ref{fig:damptau} and~\ref{fig:dampbrd} we compare NLO
predictions with distributions obtained from unshowered LHE events,
produced using these three definitions of the singular contribution
entering the \POWHEG{} cross section $\mathd\sigma_{\sss \rm PWG}$ of
Sec.~\ref{sec:pwgingredients} for $\tau_{\rm z,Q}$ of
Eq.~\eqref{eq:qthrust} and $B_{\rm z,E}$ of Eq.~\eqref{eq:ebroadening}.
For small values of the event shapes, all the LHE level distributions
agree with each other, and the presence of the \POWHEG{}
Sudakov form factor of Eq.~\eqref{eq:SudakovPWG} regulates the divergent behaviour, which
is instead present at NLO.

We observe that if we do not introduce any damping factor, the LHE
distribution overshoots the NLO one by roughly 10\% in the tail.
This enhancement is within the scale-uncertainty band of the
NLO result.
However, we notice the scale-variation band for the LHE curve is almost
absent.
This is due to the fact that the hardest radiation is always generated 
using the transverse momentum as scale entering the emission
probability~\cite{Frixione:2007vw}, so that the factorisation and
renormalisation scale variation affects only the total weight, but not
the differential distribution.
While the NLO cross section is smaller than the LO one
(i.e.\ $\bar{B}/B\sim 0.965 <1$), the increase of the distribution in
the tail, is due to higher-order corrections such as those coming from
the treatment of the running coupling~\cite{Catani:1990rr} in the
squared bracket of~Eq.~\eqref{eq:sigmaPWGnaive}, which can capture the
bulk of NLL corrections arising from subsequent unresolved emissions,
or due to the scale choice in the PDF.
The inclusion of a damping function does instead ensure that such
corrections are not applied for large values of the event shapes. In this case the
central value of the LHE curve aligns with the NLO one.
We also notice that in this case, scale-variation bands are larger, as
the argument of the PDFs and the coupling constant appearing in the
remnant cross section are varied accordingly.
The inclusion of the \texttt{hdamp} mechanism, on top of
the \texttt{Bornzerodamp} one, leaves the central value of the curve
almost unaffected, but increases the size of the uncertainty band in
the tail of the distribution.
In all cases, however, the scale-uncertainty band produced by LHE-level
distributions is much smaller than the NLO one.

\section{Central scale choices}
\label{sec:scale-variations}
\begin{figure}[tb!]
  \centering\includegraphics[width=0.49\textwidth,page=1]{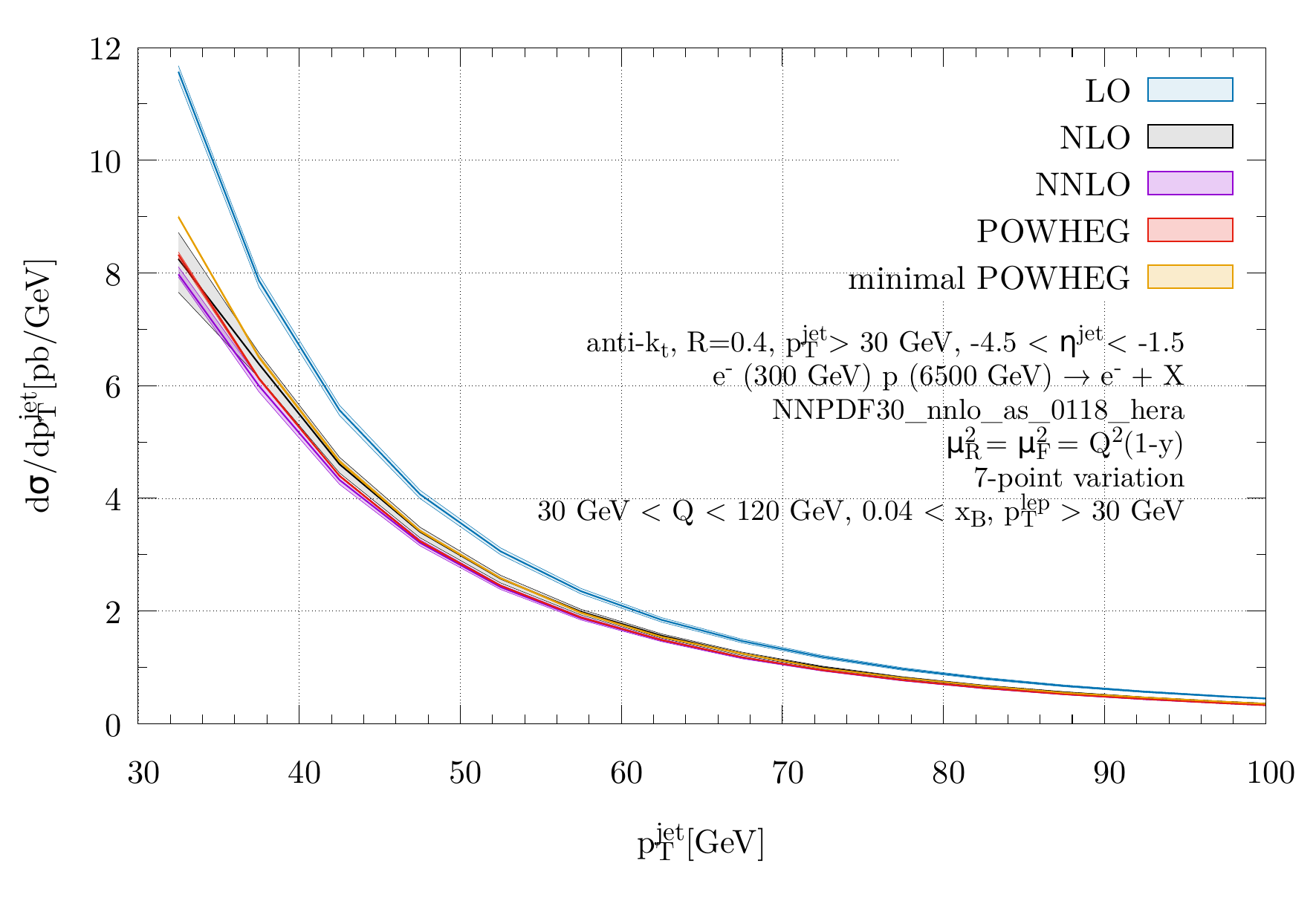}
  \centering\includegraphics[width=0.49\textwidth,page=2]{fig-temp/vbf-plots-ptlep.pdf}
  \caption{The hardest anti-$k_T$ $R=0.4$ jet in the rapidity window
    $-4.5 < \etaj < -1.5$ for events satisfying the cuts of
    Eq.~\eqref{eq:vbfcuts}. We show LO (blue), NLO (grey), NNLO
    (purple), our new DIS implementation showered with \PYTHIAE{}
    (red) and the minimally modified \POWHEG{} implementation with the
    same shower (orange). On the right we show the ratio to the NLO
    prediction. The bands correspond to a 7-point scale variation
    around the central scale defined by $\mu^2=Q^2(1-y)$. }
  \label{fig:vbf-ptj1-ptlep}
\end{figure}
\begin{figure}[tb!]
  \centering\includegraphics[width=0.49\textwidth,page=3]{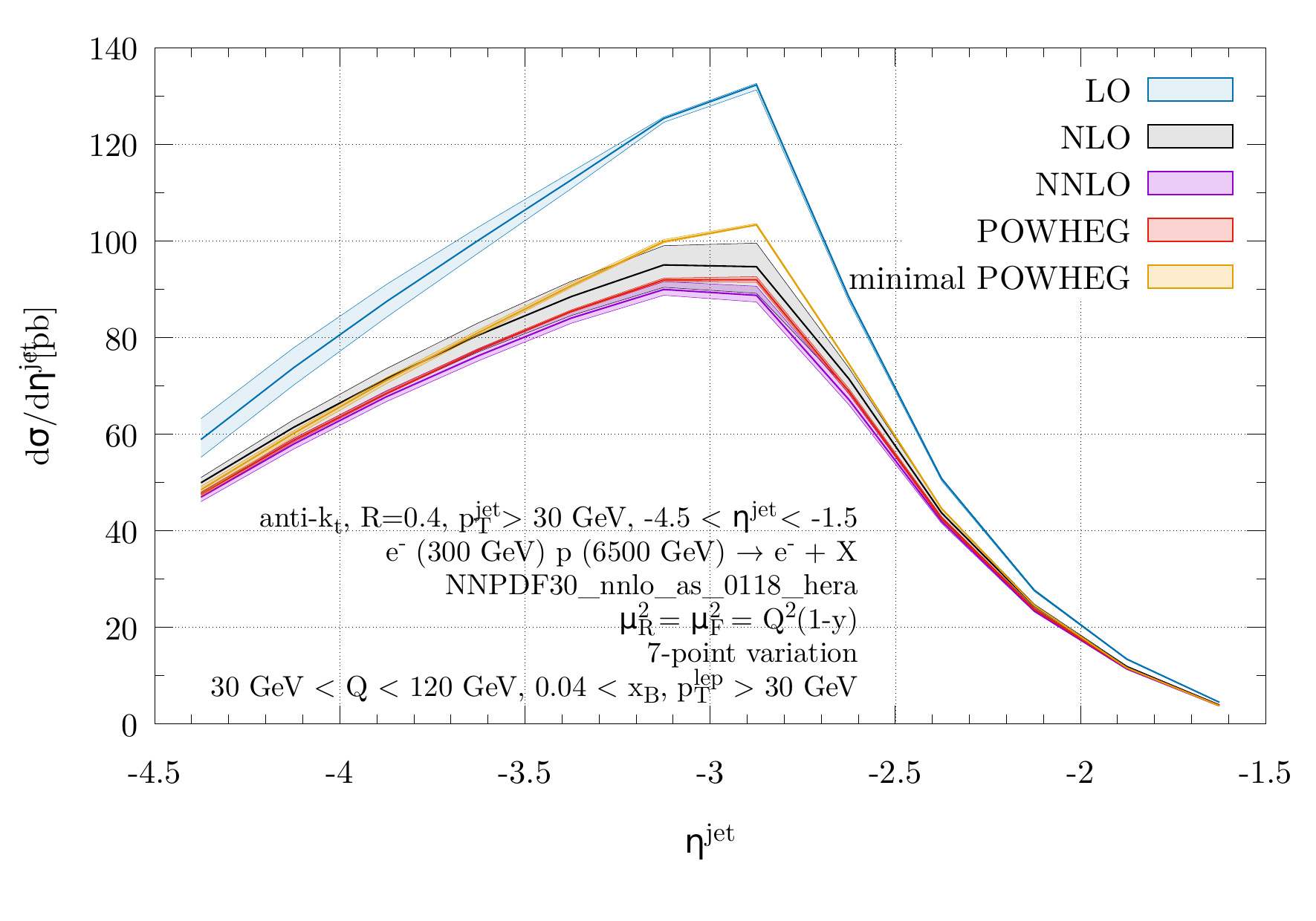}
  \centering\includegraphics[width=0.49\textwidth,page=4]{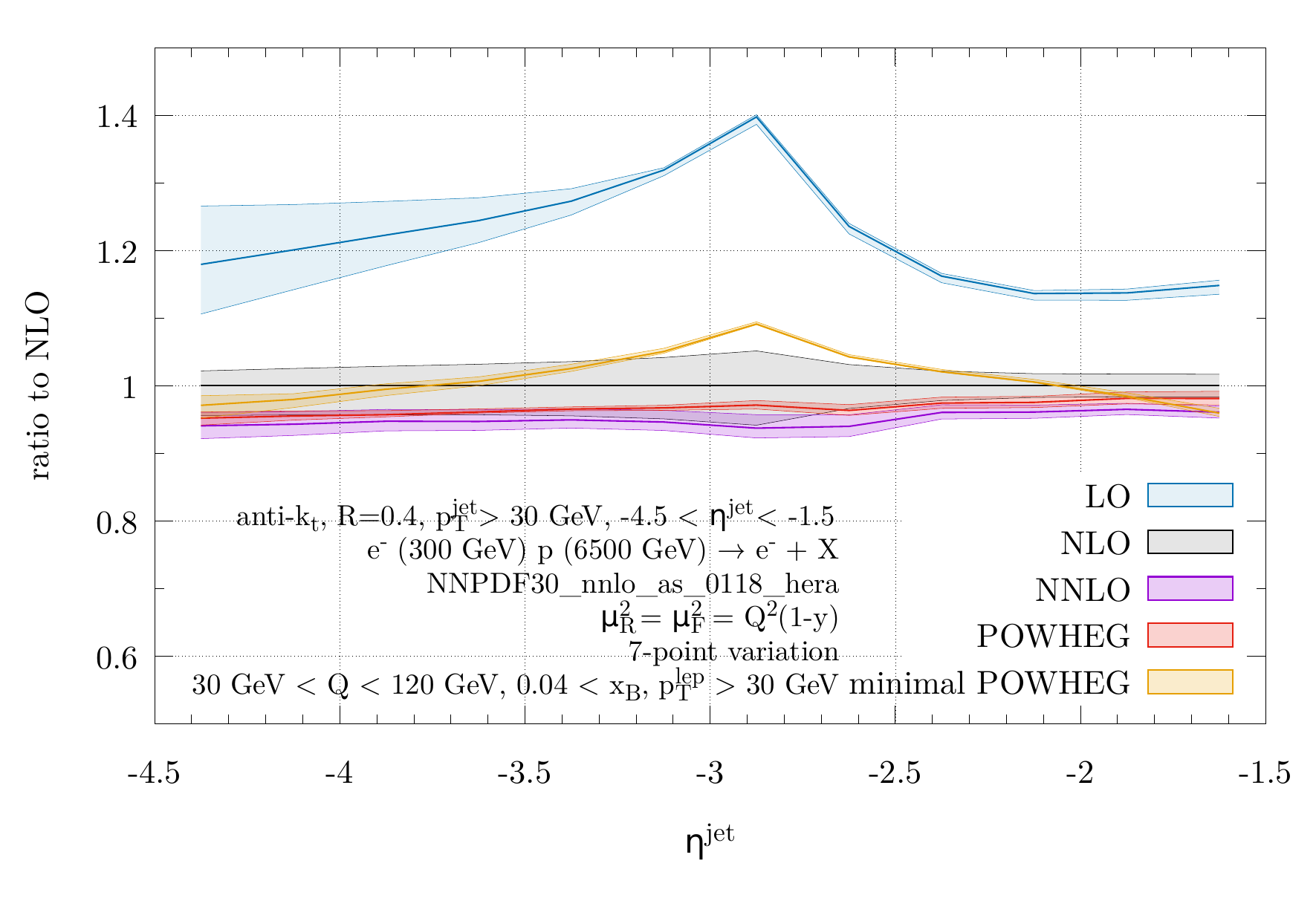}
  \caption{ Same as Fig.~\ref{fig:vbf-ptj1-ptlep}, but now showing the
    rapidity of the hardest jet satisfying $\ptj>30\,\mathrm{GeV}$.}
  \label{fig:vbf-etaj1-ptlep}
\end{figure}
In this appendix we present the results of Sec.~\ref{sec:vbf} for two
different choices for the central value of the renormalisation and
factorisation scales.
This is of interest since a central scale choice of $\mu^2 = Q^2$ is
only well-motivated for inclusive quantities, whereas more exclusive
quantities in general probe different QCD scales. The two alternative
scales that we explore are the following:
\begin{itemize}
\item The transverse momentum of the final-state lepton (in the collision frame), $p_T^{\rm \sss lep}$, which is related to DIS variables via 
  \begin{equation}
    \mu^2 = \left(p_T^{\rm \sss lep}\right)^2 = (1-\ydis) Q^2 < Q^2\,.
  \end{equation}
  At LO, this scale coincides with the transverse momentum of the jet,
  $p_T^{\rm \sss jet}$, and differences between the two scales arise
  from the real radiation corrections.
  Such differences are then formally NNLO.
\item The invariant mass of the recoil system such that 
  \begin{equation}
    \mu^2 = Q^2 \frac{(1-\xdis)}{\xdis}.
  \end{equation}
  This scale is related to the maximum transverse momentum available for the jet.
  This choice can be problematic for $\xdis \to 1$, but we want to include it in our discussion to present an extreme scenario.
\end{itemize}
In the following figures we illustrate the effects of using these two
scales both in the fixed order predictions and the \POWHEG{} results.

\begin{figure}[tb!]
  \centering\includegraphics[width=0.49\textwidth,page=1]{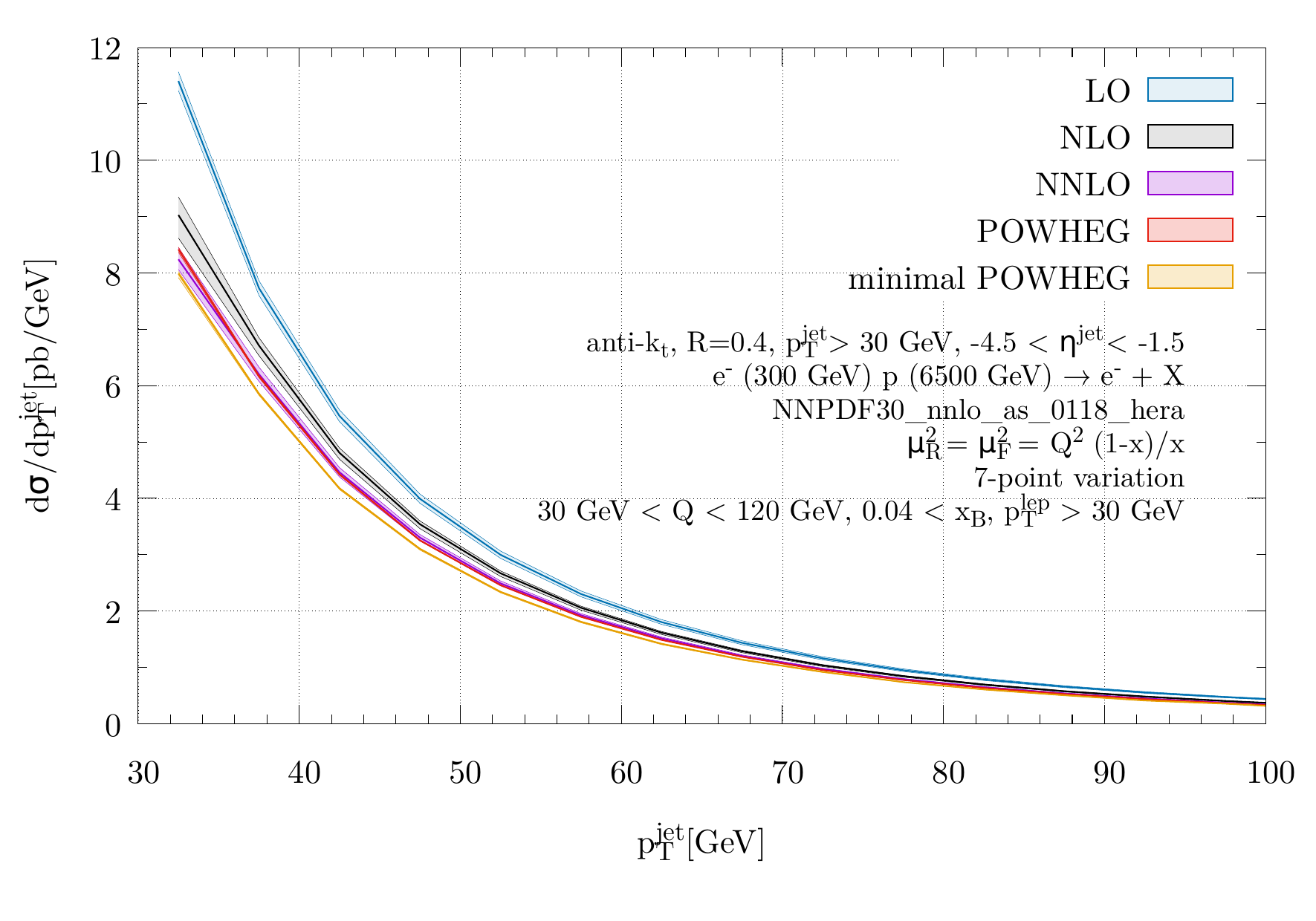}
  \centering\includegraphics[width=0.49\textwidth,page=2]{fig-temp/vbf-plots-Q1mxox.pdf}
  \caption{The hardest anti-$k_T$ $R=0.4$ jet in the rapidity window
    $-4.5 < \etaj < -1.5$ for events satisfying the cuts of
    Eq.~\eqref{eq:vbfcuts}. We show LO (blue), NLO (grey), NNLO
    (purple), our new DIS implementation showered with \PYTHIAE{}
    (red) and the minimally modified \POWHEG{} implementation with the
    same shower (orange). On the right we show the ratio to the NLO
    prediction. The bands correspond to a 7-point scale variation
    around the central scale defined by $\mu^2=\frac{Q^2(1-\xdis)}{\xdis}$. }
  \label{fig:vbf-ptj1-Q1mxox}
\end{figure}
\begin{figure}[tb!]
  \centering\includegraphics[width=0.49\textwidth,page=3]{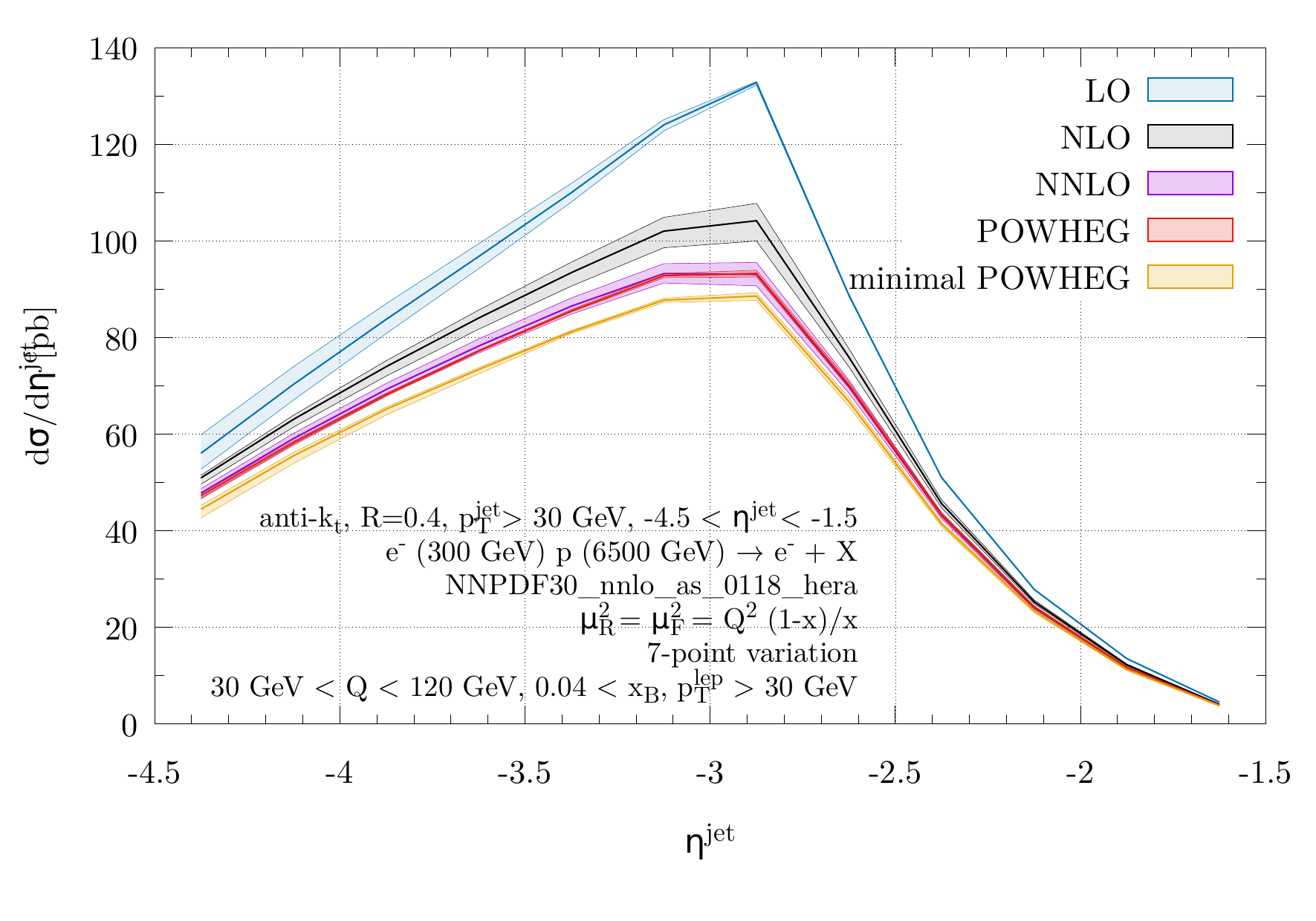}
  \centering\includegraphics[width=0.49\textwidth,page=4]{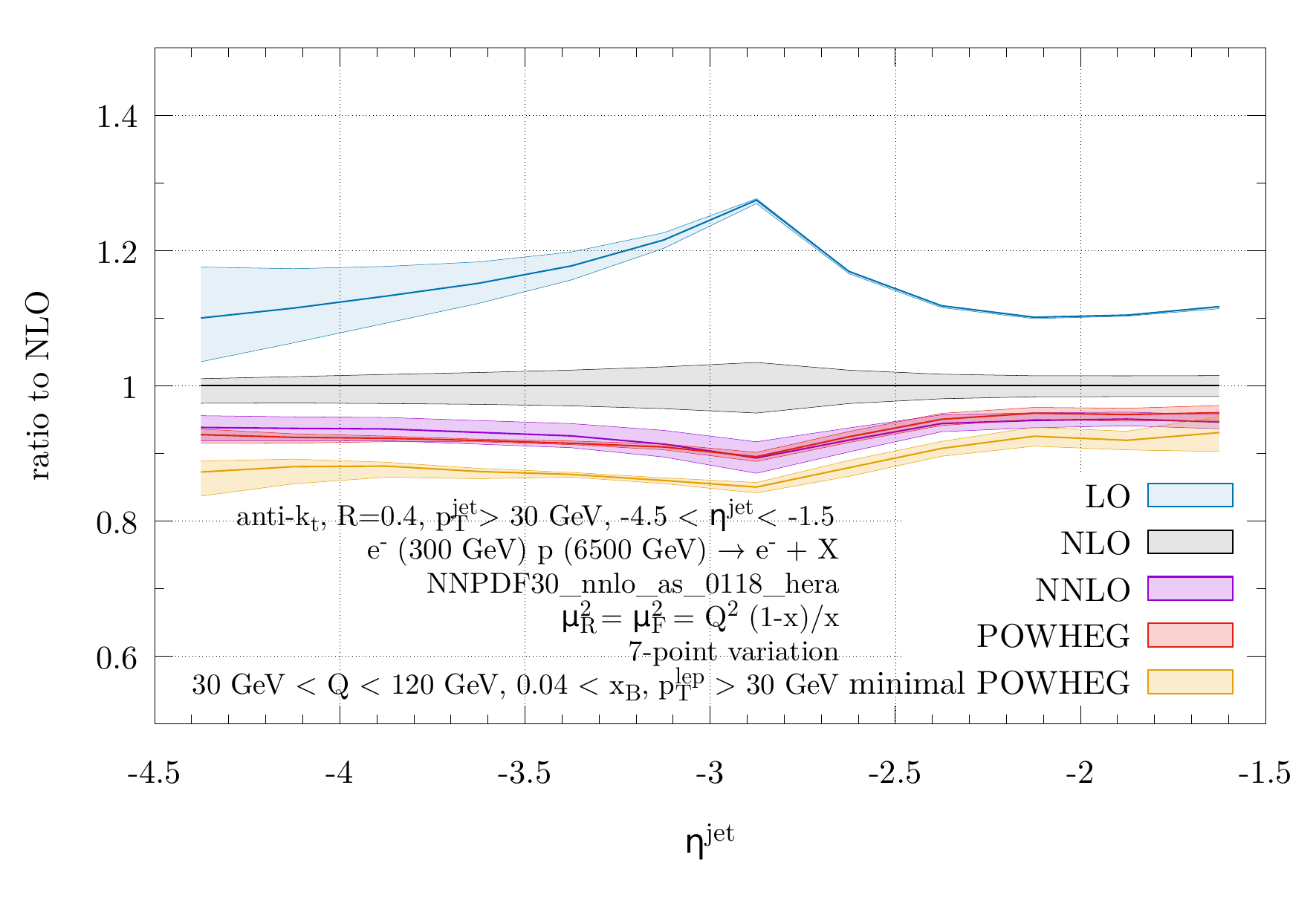}
  \caption{ Same as Fig.~\ref{fig:vbf-ptj1-Q1mxox}, but now showing the
    rapidity of the hardest jet satisfying $\ptj>30\,\mathrm{GeV}$.}
  \label{fig:vbf-etaj1-Q1mxox}
\end{figure}
\begin{figure}[tb!]
  \centering\includegraphics[width=0.49\textwidth,page=5]{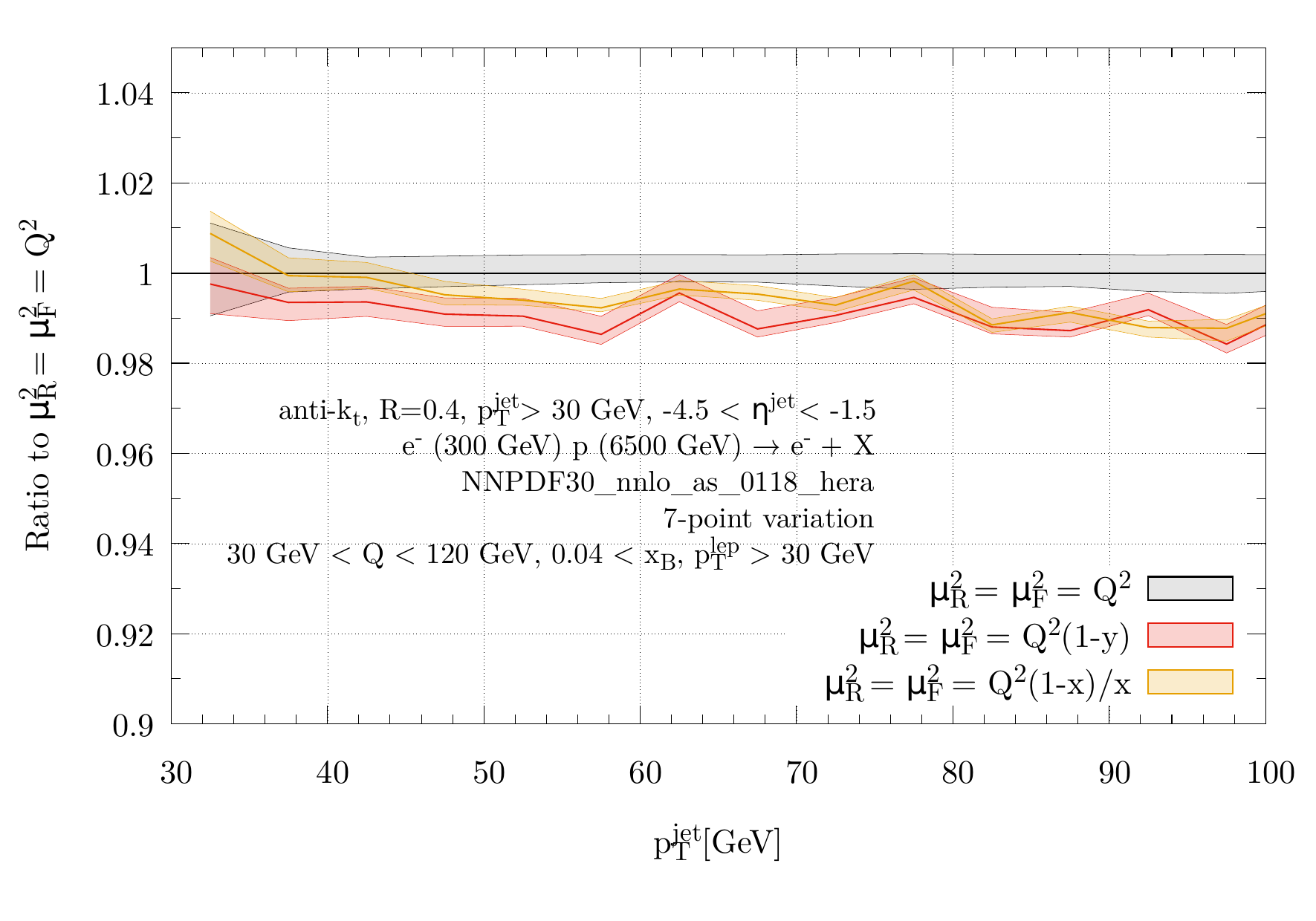}
  \centering\includegraphics[width=0.49\textwidth,page=6]{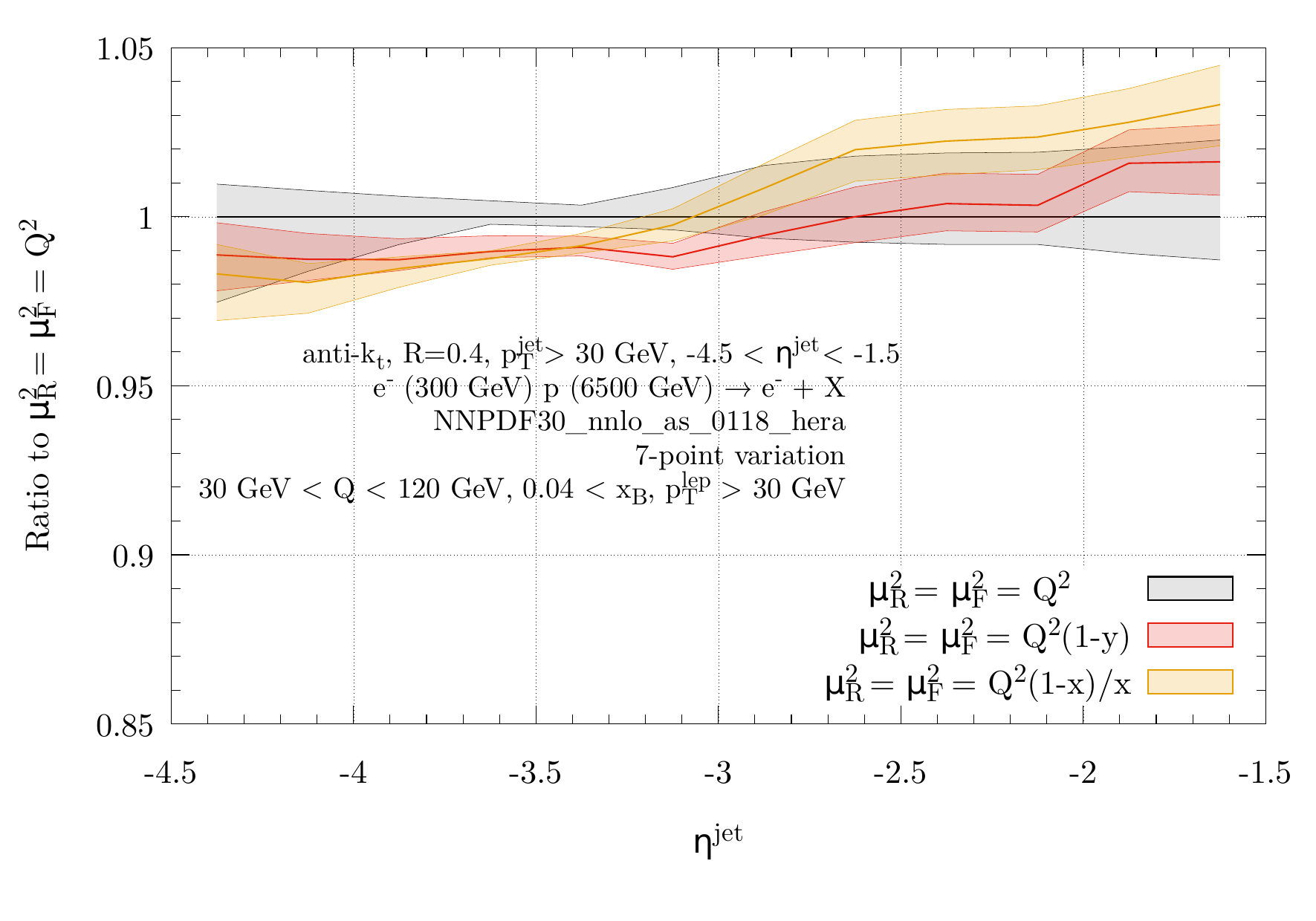}
  \caption{ Same as previous figures but now showing the ratio of the
    \POWHEG{} implementation described in this paper for three
    different central scale choices. In gray we show $\mu^2=Q^2$, in
    red we show $\mu^2 = Q^2(1-y)$, and in yellow we show $\mu^2 =
    \frac{Q^2 (1-\xdis)}{\xdis}$. The plot shows the ratio with
    respect to the $\mu^2=Q^2$ result. }
  \label{fig:vbf-ratio}
\end{figure}
In Figs.~\ref{fig:vbf-ptj1-ptlep}--\ref{fig:vbf-etaj1-ptlep} we show
the results for the first scale, i.e.~the transverse momentum of the
lepton. As can be seen by comparing to the plots of
Figs.~\ref{fig:vbf-ptj1}--\ref{fig:vbf-etaj1} using $\mu^2=Q^2$, in
Figs.~\ref{fig:vbf-ptj1-ptlep}--\ref{fig:vbf-etaj1-ptlep} the pattern
across the various orders is very similar. This is perhaps not a
surprise, since for the setup we study here the values of $\ydis$ that
we probe tend to be small.

For the second scale choice we expect a much larger deviation from
$Q$, since for $\xdis > 0.04$ this scale can get $24$ times larger
than $Q^2$ (and also much smaller although only when $\xdis >
0.5$). In Figs.~\ref{fig:vbf-ptj1-Q1mxox}--\ref{fig:vbf-etaj1-Q1mxox}
we show the results for $\ptj$ and $\etaj$ using this scale. Here it
can be seen that the size of the perturbative corrections seems to
have shifted. In particular the NLO+PS results now sit completely
outside of the scale variation band at NLO -- this is particularly bad
for the minimal \POWHEG{} implementation. Interestingly using the
mappings presented in this paper, there seems to still be very good
agreement with the NNLO prediction.

Finally, in Fig.~\ref{fig:vbf-ratio} we compare all three scale
choices, showing their ratio to the scale choice $\mu=Q$. It is
interesting to note that despite the rather different pattern observed
for the scale defined by $\mu^2=\frac{Q^2(1-\xdis)}{\xdis}$ all three
predictions are in reasonably good agreement with each other. The
scale uncertainties are a bit underestimated as expected, but not
dramatically so.

\bibliographystyle{JHEP}
\bibliography{dis-powheg}

\end{document}